\documentclass{sig-alternate}

\usepackage{amsmath,url,cite,mathrsfs}
\usepackage{amsfonts}
\usepackage{graphicx}
\usepackage{subfigure}

\begin{document}

\title{A Very Efficient Scheme for Estimating Entropy of Data Streams Using Compressed Counting}

\numberofauthors{1}
\author{
\alignauthor
Ping Li\\
       \affaddr{Cornell University}\\
              \affaddr{Ithaca, NY 14850}\\
       \email{pingli@cornell.edu}
}

\maketitle

\noindent \today
\begin{abstract}
\textbf{\em Compressed Counting (CC)} was recently proposed for approximating the $\alpha$th frequency moments of data streams, for $0<\alpha \leq 2$. Under the {\em relaxed strict-Turnstile} model, CC dramatically improves the standard  algorithm based on {\em symmetric stable random projections}, especially as $\alpha\rightarrow 1$.
A direct application of CC is to estimate the entropy, which is an important summary statistic in Web/network measurement and often serves a crucial ``feature'' for data mining.  The R\'enyi entropy and the Tsallis entropy  are functions of the $\alpha$th frequency moments; and both approach the Shannon entropy as $\alpha\rightarrow 1$. A recent theoretical work suggested using the $\alpha$th frequency moment to approximate the Shannon entropy with $\alpha=1+\delta$ and very small $|\delta|$ (e.g., $<10^{-4}$).

In this study, we experiment using CC to estimate frequency moments, R\'enyi entropy, Tsallis entropy, and  Shannon entropy, on real Web crawl data. We demonstrate the variance-bias trade-off in estimating  Shannon entropy and provide practical recommendations. In particular, our experiments enable us to draw some important conclusions:
\begin{itemize}
\item As $\alpha\rightarrow 1$, CC dramatically improves {\em symmetric stable random projections} in estimating frequency moments, R\'enyi entropy, Tsallis entropy, and Shannon entropy. The improvements appear to approach ``infinity.''
\item CC is a highly practical algorithm for estimating Shannon entropy (from either R\'enyi or Tsallis entropy) with $\alpha\approx 1$. Only a very small sample (e.g., 20) is needed to achieve a high accuracy (e.g., $<1\%$ relative errors).
\item Using {\em symmetric stable random projections} and $\alpha = 1+\delta$ with very small $|\delta|$ does not provide a practical algorithm because the required sample size is enormous.
\item If we do need to use {\em symmetric stable random projections} for estimating Shannon entropy, we should exploit the variance-bias trade-off by letting $\alpha$ be away from 1, for much better performance.
\item Even in terms of the best achievable performance in estimating Shannon entropy, CC still considerably improves {\em symmetric stable random projections} by one or two magnitudes, both in terms of the estimation accuracy and the required sample size (storage space). 
\end{itemize}

\end{abstract}
\section{Introduction}

The general theme of ``scaling up for high dimensional data and high speed data streams'' is among the  ``ten challenging problems in data mining research''\cite{Article:ICDM10}. This paper focuses on a very efficient algorithm for estimating the entropy of data streams using a recently developed randomized algorithm called \textbf{\em Compressed Counting (CC)} by Li \cite{Article:Li_CC_v0,Article:Li_CC,Report:Li_CC_oq}. The underlying technique of CC is {\em maximally-skewed stable random projections}. Our experiments on real Web crawl data demonstrate that CC can approximate entropy with very high accuracy. In particular, CC (dramatically) improves {\em symmetric stable random projections} (Indyk \cite{Article:Indyk_JACM06} and Li \cite{Proc:Li_SODA08}) for estimating entropy, under the {\em relaxed strict-Turnstile} model.

\subsection{Data Streams and Relaxed\\ Strict-Turnstile Model}

While traditional machine learning and mining algorithms often assume static data, in reality, data are often constantly updated. Mining data streams\cite{Book:Henzinger_99,Proc:Babcock_PODS02,Proc:Aggarwal_KDD04,Article:Muthukrishnan_05} in (e.g.,) 100 TB scale  databases has become an important area of research, as network data can easily reach that scale\cite{Article:ICDM10}. Search engines are a typical source of data streams (Babcock {\em et.al.} \cite{Proc:Babcock_PODS02}).

We consider the  {\em Turnstile}  stream model (Muthukrishnan \cite{Article:Muthukrishnan_05}). The input  stream $a_t = (i_t, I_t)$, $i_t\in [1,\ D]$ arriving sequentially describes the underlying signal $A$, meaning
\begin{align}\label{eqn_Turnstile}
A_t[i_t] = A_{t-1}[i_t] + I_t,
\end{align} where the increment $I_t$ can be either positive (insertion) or negative (deletion). For example, in an online bookstore, $A_{t-1}[i]$ may record the total number of books that user $i$ has ordered up to time $t-1$ and $I_t$ denotes the number of books that this user orders ($I_t>0$) or cancels ($I_t<0$) at  $t$.

It is often reasonable to assume $A_t[i]\geq 0$, although $I_t$ may be either negative or positive. Restricting $A_t[i]\geq 0$ results in the {\em strict-Turnstile} model, which suffices for describing almost all natural phenomena. For example, in an online store, it is  not possible to cancel orders that do not exist.

\textbf{Compressed Counting (CC)} assumes a {\em relaxed strict-Turnstile} model by only enforcing $A_t[i]\geq0$ at the $t$ one cares about.  At other times $s \neq t$,  CC allows $A_s[i]$ to be arbitrary. This is more general than the {\em strict-Turnstile} model.

\subsection{Moments and Entropies of Data Streams}

The $\alpha$th frequency moment is a fundamental statistic:
\begin{align}
F_{(\alpha)} = \sum_{i=1}^D A_t[i]^\alpha.
 \end{align}
\noindent When $\alpha = 1$, $F_{(1)}$ is the sum of the stream. It is obvious that one can compute $F_{(1)}$ exactly and trivially using a simple counter, because $F_{(1)} = \sum_{i=1}^D A_t[i] = \sum_{s=0}^t I_s$.

$A_t$ is basically a histogram and we can view $p_i = \frac{A_t[i]}{\sum_{i=1}^D A_t[i]}$ as probabilities. An extremely useful (especially in Web and networks\cite{Proc:Zhao_IMC07,Proc:Mei_WSDM08}) summary statistic is the Shannon entropy:
\begin{align}\label{eqn_Shannon}
H = -\sum_{i=1}^D\frac{A_t[i]}{F_{(1)}}\log \frac{A_t[i]}{F_{(1)}}, \hspace{0.2in} \text{where} \ \ F_{(1)} = \sum_{i=1}^D A_t[i].
\end{align}

Various generalizations of the Shannon entropy exist. The R\'enyi entropy\cite{Proc:Renyi_61}, denoted by $H_\alpha$, is defined as
\begin{align}\label{eqn_Renyi}
&H_\alpha =\frac{1}{1-\alpha} \log \frac{\sum_{i=1}^D A_t[i]^\alpha}{\left(\sum_{i=1}^D A_t[i]\right)^\alpha} = \frac{1}{1-\alpha} \log \frac{F_{(\alpha)}}{F_{(1)}^\alpha}.
\end{align}
The Tsallis entropy\cite{Article:Havrda_67,Article:Tsallis_88}, denoted by $T_\alpha$,  is defined as,
\begin{align}\label{eqn_Tsallis}
&T_\alpha = \frac{1}{\alpha -1} \left( 1 - \frac{F_{(\alpha)}}{F_{(1)}^\alpha}\right),
\end{align}
\noindent which was first introduced by Havrda and Charv\'at\cite{Article:Havrda_67} and later popularized by Tsallis\cite{Article:Tsallis_88}.

It is easy to verify that, as $\alpha\rightarrow 1$, both the R\'enyi entropy and Tsallis entropy  converge to the Shannon entropy. Thus $H = H_1 = T_1$ in the limit sense. For this fact, one can also consult \url{http://en.wikipedia.org/wiki/Renyi_entropy}.

Therefore, both the R\'enyi entropy and Tsallis entropy can be computed from the $\alpha$th frequency moment; and one can approximate the Shannon entropy from either $H_\alpha$ or $T_\alpha$ by using $\alpha\approx 1$. In fact, several studies (Zhao {\em et.al.} \cite{Proc:Zhao_IMC07} and Harvey {\em et.al.} \cite{Article:Harvey_entropy_arXiv08,Proc:Harvey_FOCS08})   have used this idea to approximate the Shannon entropy. 

We should mention that \cite{Article:Li_CC} proposed estimating the logarithmic moment, $\sum_{i=1}^D \log A_t[i]$, using $F_{(\alpha)}$ with $\alpha\rightarrow 0$. Their idea is very similar to that in estimating entropy.

\subsection{Challenges in Data Stream Computations}

Because the elements, $A_t[i]$, are time-varying,  a na\'ive counting mechanism requires a system of $D$ counters to compute $F_{(\alpha)}$ exactly (unless $\alpha=1$). This is not always realistic when $D$ is large and the data are  frequently updated at very high rate.  For example, if $A_t[i]$ records activities for each user $i$, identified by his/her IP address, then potentially $D = 2^{64}$ (possibly much larger in the near future).

Due to the huge volume, streaming data are often not (fully) stored, even on disks\cite{Proc:Babcock_PODS02}.  One common strategy is to store only a small ``sample'' the data; and sampling has become an important topic in Web search and data streams\cite{Proc:Henzinger_WWW00,Proc:Bar-Yossef_VLDB00,Article:Henzinger_03}. While some modern databases (e.g., Yahoo!'s 2-petabyte database) and government agencies do store the whole data history, the data analysis  often has to be conducted on a (hopefully) representative small sample of the data. As it is well-understood that  general-purpose simple sampling-based methods often can not give reliable approximation guarantees\cite{Proc:Babcock_PODS02}, developing special-purpose (and one-pass) sampling/sketching techniques in streaming data has become an active area of research.

\subsection{Previous Studies on Approximating Frequency Moments and Entropy}

Pioneered by Alon {\it et.al.}\cite{Proc:Alon_STOC96}, the problem of approximating $F_{(\alpha)}$ in data streams has been heavily studied\cite{Proc:Feigenbaum_FOCS99,Proc:Indyk_FOCS00,Proc:Saks_STOC02,Proc:Kumar_FOCS02,Proc:Indyk_STOC05,Proc:Woodruff_SODA04,Proc:Ganguly_RANDOM07}. The method of {\em symmetric stable random projections} (Indyk \cite{Article:Indyk_JACM06}, Li \cite{Proc:Li_SODA08}) is regarded to be  practical and accurate.

We have mentioned that computing the first moment $F_{(1)}$ in {\em strict-Turnstile} model is trivial using a simple counter. One might naturally speculate that when $\alpha\approx 1$, computing (approximating) $F_{(\alpha)}$ should be also easy. However, none of the previous algorithms including {\em symmetric stable random projections} could capture this intuition. For example, Figure \ref{fig_comp_var_factor} in Section \ref{sec_CC} shows that the performance of {\em symmetric stable random projections} is roughly the same for $\alpha=1$ and $\alpha\approx 1$, even though $\alpha=1$ should be trivial.

\textbf{Compressed Counting (CC)}\cite{Article:Li_CC_v0,Article:Li_CC,Report:Li_CC_oq} was recently proposed to overcome the drawback of previous algorithms at $\alpha\approx 1$. CC improves {\em symmetric stable random projections} uniformly for all $0<\alpha\leq 2$ and the improvement is in a sense ``infinite'' when $\alpha\rightarrow 1$ as shown in Figure \ref{fig_comp_var_factor} in Section \ref{sec_CC}. However, no empirical studies on CC have been reported.

Zhao {\em et.al.}\cite{Proc:Zhao_IMC07} applied {\em symmetric stable random projections} to approximate the Shannon entropy. \cite{Article:Li_CC} cited \cite{Proc:Zhao_IMC07}, as one application of {\em Compressed Counting} (CC).  A nice  theoretical paper in FOCS'08 by Harvey {\em et.al.}\cite{Article:Harvey_entropy_arXiv08,Proc:Harvey_FOCS08} provided the criterion to choose the $\alpha$ so that the Shannon entropy can be approximated with a guaranteed accuracy, using the $\alpha$th frequency moment. \cite{Proc:Harvey_FOCS08} cited both {\em symmetric stable random projections}\cite{Article:Indyk_JACM06,Proc:Li_SODA08} and {\em Compressed Counting}\cite{Article:Li_CC}.

There are other methods for estimating entropy, e.g., \cite{Proc:Guha_SODA06}, which we do not compare with in this study.

\subsection{Summary of Our Contributions}

Our main contribution is the first empirical study of Compressed Counting for estimating entropy.  Some theoretical analysis is also conducted.
\begin{itemize}
\item  We apply Compressed Counting (CC) to compute the R\'enyi entropy, the Tsallis entropy, and the Shannon entropy, on real Web crawl data.
\item We empirically compare CC with {\em symmetric stable random projections} and demonstrate the huge improvement. Thus, our work helps establish CC as a promising practical tool in data stream computations.
\item We provide some theoretical analysis for approximating entropy, for example, the variance-bias trade-off.
\item Our empirical work leads to practical recommendations for various estimators developed in \cite{Article:Li_CC_v0,Article:Li_CC,Report:Li_CC_oq}.
\end{itemize}

For estimating the Shannon entropy, the theoretical work by Harvey {\em et.al.}\cite{Article:Harvey_entropy_arXiv08,Proc:Harvey_FOCS08} used {\em symmetric stable random projections} or CC as a subroutine (a two-stage ``black-box'' approach). That is, they first determined at what $\alpha=1+\delta$ value, $H_\alpha$ (or $T_\alpha$) is close to $H$ within a required accuracy. Then they used this chosen $\alpha$th frequency moment to approximate the Shannon entropy, independent of whether the frequency moments are estimated using CC or {\em symmetric stable random projections}.

In comparisons, we demonstrate that  estimating Shannon entropy is  a variance-bias trade-off; and hence the performance is highly coupled with the underlying estimators. The two-stage ``black-box'' approach   \cite{Article:Harvey_entropy_arXiv08,Proc:Harvey_FOCS08} may have  some theoretical advantage (e.g., simplifying the analysis), while our variance-bias analysis directly reflects the real-world situation and leads to practical recommendations.
\begin{itemize}
\item  \cite{Article:Harvey_entropy_arXiv08,Proc:Harvey_FOCS08} let $\alpha = 1+\delta$ and provided the procedures to compute $\delta$ (or a series of $\delta$'s). If one actually carries out the calculation, their $|\delta|$ is very small (like $10^{-4}$ or smaller). Consequently their theoretically calculated sample size may be (impractically) large, especially when using {\em symmetric stable random projections}.
\item In comparison, we  provide a practical recommendation for estimating Shannon entropy: using CC with $\alpha\approx 0.98\sim 0.99$ and the {\em optimal quantile} estimator. Only a  small sample (e.g., 20) can achieve a high accuracy (e.g., $<1\%$ relative errors).
\item We  demonstrate that due to the variance-bias trade-off, there will be an ``optimal'' $\alpha$ value that could attain the best mean square errors for estimating the Shannon entropy. This optimal $\alpha$  can be quite away from 1  when using {\em symmetric stable random projections}.
\end{itemize}

\subsection{Organization}

Section \ref{sec_entropy} reviews  some applications of entropy. The basic methodologies of CC and various estimators for recovering the $\alpha$th frequency moments are reviewed in Section \ref{sec_CC}. We analyze in Section \ref{sec_entropy_est} the biases and variances in estimating entropies.  Experiments on real Web crawl data are presented in Section \ref{sec_exp}. Finally, Section \ref{sec_conclusion} concludes the paper.

\section{Some Applications of Entropy}\label{sec_entropy}

\subsection{The Shannon Entropy}

The Shannon entropy, $H$ defined in (\ref{eqn_Shannon}), is a fundamental measure of randomness. A recent paper in WSDM'08 (Mei and Church \cite{Proc:Mei_WSDM08}) was devoted to estimating the Shannon entropy of MSN search logs, to help answer some basic problems in Web search, such as,  {\em how big is the web?}

The search logs can be naturally viewed as data streams, although \cite{Proc:Mei_WSDM08} only analyzed several ``snapshots'' of a sample of MSN search logs.  The sample used in \cite{Proc:Mei_WSDM08} contained 10 million <Query, URL,IP> triples; each triple corresponded to a click from a particular IP address on a particular URL for a particular query.  \cite{Proc:Mei_WSDM08} drew their important conclusions on this (hopefully) representative sample. We believe one can (quite easily) apply Compressed Counting (CC) on the same task, on the whole history of MSN (or other search engines) search logs instead of a (static) sample.\\

Using the Shannon entropy as an important ``feature'' for mining anomalies is a widely used technique (e.g., \cite{Proc:Lakhina_SIGCOMM05}). In IMC'07, Zhao {\it et.al.}\cite{Proc:Zhao_IMC07} applied {\em symmetric stable random projections} to estimate the Shannon entropy for all origin-destination (OD) flows in network measurement, for clustering traffic and detecting traffic anomalies.

Detecting anomaly events in real-time (DDoS attacks, network failures, etc.) is highly beneficial in monitoring network performance degradation and service disruptions. Zhao {\it et.al.}\cite{Proc:Zhao_IMC07}  hoped to capture those events in real-time by examining the entropy of every OD flow.  They resorted to approximate algorithms because measuring the Shannon entropy in real-time is not possible on high-speed links due to its memory requirements and high computational cost.\\

\subsection{The R\'enyi Entropy}

The R\'enyi entropy, $H_\alpha$ defined in (\ref{eqn_Renyi}), is a generalization of the classical Shannon entropy $H$. $H_\alpha$ is function of the frequency moment $F_{(\alpha)}$ and approaches $H$ as $\alpha \rightarrow 1$. Thus it is natural to use $H_\alpha$ with $\alpha\approx 1$ to approximate $H$.

The R\'enyi entropy has other applications. It is a diversity index in ecology \cite{Article:Tothmeresz_95,Article:Ricotta_02,Article:Liu_06}. It is used for analyzing expander graphs\cite{Article:Hoory_06} and other applications, e.g. \cite{Article:Zyczkowski_03}.

\subsection{The Tsallis Entropy}

The Tsallis entropy, $T_{\alpha}$  defined in (\ref{eqn_Tsallis}), is another generalization of the Shannon entropy $H$. Since $T_{\alpha}\rightarrow H$ as $\alpha\rightarrow 1$, the Tsallis entropy provides another algorithm for approximating the Shannon entropy.

The Tsallis entropy is widely used in statistical physics and  mechanics. Interested readers may consult the link \\ \url{www.cscs.umich.edu/~crshalizi/notabene/tsallis.html}.

\section{Review Compressed Counting (CC)}\label{sec_CC}

Compressed Counting (CC) assumes the {\em relaxed strict-Turnstile} data stream model. Its underlying technique is based on {\em maximally-skewed stable random projections}.

\subsection{Maximally-Skewed Stable Distributions}

A random variable
$Z$ follows a maximally-skewed $\alpha$-stable distribution if the Fourier transform of its density  is\cite{Book:Zolotarev_86}
\begin{align}\notag
{\mathscr{F}}_Z(t) &= \text{E}\exp\left(\sqrt{-1}Zt\right)\\\notag
&= \exp\left(-F|t|^\alpha\left(1-\sqrt{-1}\beta\text{sign}(t)\tan\left(\frac{\pi\alpha}{2}\right)\right)\right),
\end{align}
where $0<\alpha \leq 2$, $F>0$, and $\beta = 1$. We denote $Z\sim S(\alpha,\beta=1,F)$.
The skewness parameter $\beta$ for general stable distributions ranges in $[-1,1]$; but CC uses $\beta = 1$, i.e., maximally-skewed. Previously, the method of {\em symmetric stable random projections}\cite{Article:Indyk_JACM06,Proc:Li_SODA08} used $\beta = 0$.

Consider two independent variables, $Z_1, Z_2 \sim S(\alpha, \beta=1,1)$. For any non-negative constants $C_1$ and $C_2$, the ``$\alpha$-stability'' follows from  properties of Fourier transforms:
\begin{align}\notag
Z = C_1Z_1 + C_2Z_2 \sim S\left(\alpha, \beta=1, C_1^\alpha + C_2^\alpha\right).
\end{align}

Note that if $\beta = 0$, then the above stability holds for any constants $C_1$ and $C_2$. This is why {\em symmetric stable random projections}\cite{Article:Indyk_JACM06,Proc:Li_SODA08} can work on general data but CC only works on non-negative data  (i.e., {\em relaxed strict-Turnstile model}). Since we are interested in the entropy, the non-negativity constraint is natural, because the probability should be non-negative.

\subsection{Random Projections}

Conceptually, one can generate a  matrix $\mathbf{R}\in\mathbb{R}^{D\times k}$ and multiply it with the data stream $A_t$, i.e., $X = \mathbf{R}^\text{T} A_t \in\mathbb{R}^k$. The resultant vector $X$ is only of length $k$. The entries of $\mathbf{R}$, $r_{ij}$,  are i.i.d. samples of a stable distribution $S(\alpha,\beta=1,1)$.

By property of Fourier transforms, the entries of $X$, $x_j$ $j = 1$ to $k$,  are i.i.d. samples of a stable distribution
{\small\begin{align}\notag
x_j =& \left[\mathbf{R}^\text{T}A_t\right]_j = \sum_{i=1}^D r_{ij} A_t[i]\\
 \sim& S\left(\alpha,\beta=1,F_{(\alpha)} = \sum_{i=1}^D A_t[i]^\alpha\right),
\end{align}}
\noindent whose scale parameter $F_{(\alpha)}$ is exactly the $\alpha$th frequency moment of $A_t$.

Therefore, CC boils down to a statistical estimation problem. If we can estimate the scale parameter from $k$ samples, we can then estimate the frequency moments and  entropies.

For real implementations, one should conduct $\mathbf{R}^\text{T}A_t$ incrementally. This is possible because the {\em Turnstile} model (\ref{eqn_Turnstile}) is  a linear updating model. That is, for every incoming $a_t = (i_t, I_t)$, we update $x_j \leftarrow x_j + r_{i_tj} I_t$ for $j = 1$ to $k$.  Entries of $\mathbf{R}$ are generated on-demand as necessary.

\subsection{The Efficiency in Processing Time}

Ganguly and Cormode \cite{Proc:Ganguly_RANDOM07} commented that, when $k$ is large, generating entries of $\mathbf{R}$ on-demand and multiplications $r_{i_tj} I_t$, $j = 1$ to $k$, can be prohibitive when  data arrive at very high rate.  This can be a drawback of {\em stable random projections}. An easy ``fix'' is to use $k$ as small as possible.

At the same $k$, all procedures of CC and {\em symmetric stable random projections} are the same except the entries in $\mathbf{R}$ follow different distributions. Thus, both methods have the same efficiency in processing time at the same $k$. However, since CC is much more accurate especially when $\alpha\approx 1$, it requires a much smaller  $k$ for reaching a specified level of accuracy. For example, while using {\em symmetric stable random projections} with $k = 10000$ is prohibitive, using CC with $k = 20$ only may be practically feasible.  Therefore, CC in a sense naturally provides a solution to the problem of processing efficiency. 

\subsection{Three Statistical Estimators for CC}

In this study, we consider three estimators from \cite{Article:Li_CC_v0,Article:Li_CC,Report:Li_CC_oq}, which are promising for good performance near $\alpha=1$.

Recall CC boils down to estimating the scale parameter $F_{(\alpha)}$ from $k$ i.i.d. samples $x_j \sim S\left(\alpha,\beta=1, F_{(\alpha)}\right)$.

\subsubsection{The Geometric Mean Estimator}
\vspace{-0.1in}
\begin{align}\label{eqn_F_gm}
&\hat{F}_{(\alpha),gm} = \frac{\prod_{j=1}^k |x_j|^{\alpha/k}} {D_{gm}}
\end{align}

\begin{align}\notag
&D_{gm} =  \left(\cos^k\left(\frac{\kappa(\alpha)\pi}{2k}\right)/\cos \left(\frac{\kappa(\alpha)\pi}{2}\right)\right)\\\notag
&\hspace{0.5in}\times  \left[\frac{2}{\pi}\sin\left(\frac{\pi\alpha}{2k}\right)
\Gamma\left(1-\frac{1}{k}\right)\Gamma\left(\frac{\alpha}{k}\right)\right]^k, \\\notag
&\kappa(\alpha) = \alpha, \ \ \  \  \text{if} \  \ \alpha<1, \   \ \kappa(\alpha) = 2-\alpha \  \ \text{if} \  \ \alpha>1.
\end{align}

This estimator is strictly unbiased, i.e., $\text{E}\left(\hat{F}_{(\alpha),gm}\right) = F_{(\alpha),gm}$,  and its asymptotic (i.e., as $k\rightarrow \infty$) variance is
{\scriptsize \begin{align}\label{eqn_F_gm_var}
\text{Var}\left(\hat{F}_{(\alpha),gm}\right) 
=&\left\{
\begin{array}{l}
\frac{F_{(\alpha)}^2}{k}\frac{\pi^2}{6}\left(1-\alpha^2\right)+O\left(\frac{1}{k^2}\right), \ \ \   \alpha<1\\\\
\frac{F_{(\alpha)}^2}{k}\frac{\pi^2}{6}(\alpha-1)(5-\alpha)+O\left(\frac{1}{k^2}\right),    \alpha>1
\end{array}
\right.
\end{align}}
\noindent As $\alpha\rightarrow 1$, the asymptotic variance  approaches zero.

The geometric mean estimator is  important for theoretical analysis. For example, \cite{Article:Li_CC} showed that when $\alpha=1\pm\Delta\rightarrow 1$ (i.e., $\Delta\rightarrow0$), the ``constant'' $G$ in its sample complexity bound $ k = O\left(\frac{G}{\epsilon^2}\right)$ approaches $G\rightarrow \epsilon$ at the rate of $\sqrt{\Delta}$. That is, as $\alpha\rightarrow 1$, the complexity becomes $k = O\left(1/\epsilon\right)$ instead of  $O\left(1/\epsilon^2\right)$. Note that $O\left(1/\epsilon^2\right)$ is the well-known large-deviation bound for {\em symmetric stable random projections}. 
The sample complexity bound determines the sample size $k$ needed for achieving a relative accuracy within a $1\pm\epsilon$ factor of the truth.

In many theory papers, the ``constants'' in tail bounds are often ignored. The geometric mean estimator for CC demonstrates that in special cases the ``constants'' may be so small that they should not be treated as ``constants'' any more. 

\subsubsection{The Harmonic Mean Estimator}

\begin{align}\label{eqn_F_hm}
\hat{F}_{(\alpha),hm} = \frac{k\frac{\cos\left(\frac{\alpha\pi}{2}\right)}{\Gamma(1+\alpha)}}{\sum_{j=1}^k|x_j|^{-\alpha}}
\left(1- \frac{1}{k}\left(\frac{2\Gamma^2(1+\alpha)}{\Gamma(1+2\alpha)}-1\right) \right),
\end{align}
which is asymptotically unbiased and has variance
{\small\begin{align}\label{eqn_F_hm_var}
\text{Var}\left(\hat{F}_{(\alpha),hm}\right) = \frac{F^{2}_{(\alpha)}}{k}\left(\frac{2\Gamma^2(1+\alpha)}{\Gamma(1+2\alpha)}-1\right) + O\left(\frac{1}{k^2}\right).
\end{align}}

$\hat{F}_{(\alpha),hm}$ is defined only for $\alpha<1$ and is considerably more accurate than the geometric mean estimator $\hat{F}_{(\alpha),gm}$.

\subsubsection{The Optimal Quantile Estimator}

\begin{align}\label{eqn_F_oq}
\hat{F}_{(\alpha),oq} = \left(\frac{q^*\text{-Quantile}\{|x_j|, j = 1, 2, ..., k\}}{W_\alpha}\right)^\alpha.
\end{align}
where
\begin{align}
W_\alpha = q^*\text{-Quantile}\{|S(\alpha,\beta=1,1)|\}.
\end{align}

To compute $\hat{F}_{(\alpha),oq}$, one sorts $|x_j|$, $j =1$ to $k$ and uses the $q^*$th smallest, i.e., $q^*\text{-Quantile}\{|x_j|, j = 1, 2, ..., k\}$.  $q^*$ is chosen to minimize the asymptotic variance.

\cite{Report:Li_CC_oq} provides the values for $q^*$, $W_\alpha$, as well as the asymptotic variances. For convenience, we tabulate the values for $\alpha \in [0.8, \  1.2]$ in Table \ref{tab_oq}. The last column contains the asymptotic variances (with $F_{(\alpha)}=1$) without the $\frac{1}{k}$ factor.

\begin{table}[h]
\caption{\small
 }
\begin{center}{\scriptsize
\begin{tabular}{l l l l}
\hline \hline
$\alpha$ &$q^*$  &$W_\alpha$  &Var\\\hline
0.80 &0.108 &2.256365  &0.15465894  \\
0.90 &0.101 &5.400842 \  &0.04116676 \\
0.95 &0.098 &11.74773 &0.01059831    \\
0.98  & 0.0944 &30.82616 & 0.001724739  \\
0.989 & 0.0941 &56.86694 &0.0005243589 \\
1.011 & 0.8904 &58.83961  &0.0005554749 \\
1.02 & 0.8799 &32.76892  &0.001901498  \\
1.05 & 0.855 &13.61799   &0.01298757 \\
1.10 & 0.827 &7.206345   &0.05717725 \\
1.20 & 0.799 &4.011459   &0.2516604 \\
\hline\hline
\end{tabular}
}
\end{center}
\label{tab_oq}
\end{table}

Compared with the geometric mean and harmonic mean estimators, $\hat{F}_{(\alpha),gm}$ and $\hat{F}_{(\alpha),hm}$, the optimal quantile estimator $\hat{F}_{(\alpha),oq}$ has some noticeable advantages:
\begin{itemize}
\item When the sample size $k$ is not too small (e.g., $k\geq 50$), $\hat{F}_{(\alpha),oq}$ is more accurate then $\hat{F}_{(\alpha),gm}$, especially for $\alpha>1$. It is also more accurate than $\hat{F}_{(\alpha),hm}$, when $\alpha$ is close to 1. Our experiments will verify this point.
\item $\hat{F}_{(\alpha),oq}$ is computationally more efficient because both $\hat{F}_{(\alpha),gm}$ and $\hat{F}_{(\alpha),hm}$ require $k$ fractional power operations, which are expensive.
\end{itemize}

The drawbacks of the optimal quantile estimator are:
\begin{itemize}
\item For small samples (e.g., $k\leq20$), $\hat{F}_{(\alpha),oq}$ exhibits bad behaviors when $\alpha>1$.
\item Its theoretical analysis, e.g., variances and tail bounds, is based on the density function of skewed stable distributions, which do not have closed-forms.
\item The  parameters, $q^*$ and $W_\alpha$, are obtained from the numerically-computed density functions. \cite{Report:Li_CC_oq} provided $q^*$ and $W_\alpha$ values for $\alpha\geq 1.011$ and $\alpha\leq 0.989$.
\end{itemize}

\subsubsection{The Geometric Mean Estimator for Symmetric Stable Random Projections}

For {\em symmetric stable random projections}, the following geometric mean estimator is close to be statistically optimal when $\alpha\approx 1$ \cite{Proc:Li_SODA08}:
\begin{align}\label{eqn_F_gm_sym}
&\hat{F}_{(\alpha),gm,sym} = \frac{\prod_{j=1}^k |z_j|^{\alpha/k}} { \left[\frac{2}{\pi}\sin\left(\frac{\pi\alpha}{2k}\right)\Gamma\left(1-\frac{1}{k}\right)\Gamma\left(\frac{\alpha}{k}\right)\right]^k}\\\label{eqn_F_gm_sym_var}
&\text{Var}\left(\hat{F}_{(\alpha),gm,sym}\right)
\frac{F_{(\alpha)}^2}{k}\frac{\pi^2}{12}\left(2+\alpha^2\right)+O\left(\frac{1}{k^2}\right).
\end{align}
\noindent where $z_j \sim S\left(\alpha,\beta=0,F_{(\alpha)}\right)$.

Therefore, we only compare CC with this estimator, which was explicitly used  in \cite{Article:Harvey_entropy_arXiv08,Proc:Harvey_FOCS08}  for the task of residual moment estimation for the general Turnstile model. 

\subsubsection{Comparisons of Asymptotic Variances}

\begin{figure}[h]
\begin{center}
\includegraphics[width = 2.3in]{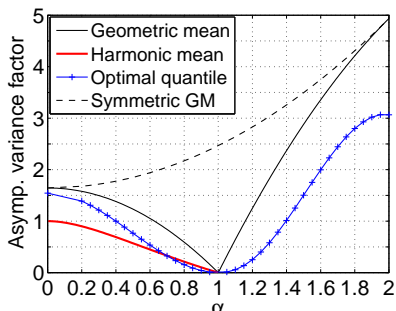}
\end{center}
\vspace{-0.3in}
\caption{Let $\hat{F}$ be an estimator of $F$ with asymptotic variance $\text{Var}\left(\hat{F}\right) = V\frac{F^2}{k} + O\left(\frac{1}{k^2}\right)$. We plot the $V$ values for the {\em geometric mean} estimator,  the {\em harmonic mean} estimator (for $\alpha<1$), and the {\em optimal quantile} estimator, along with the $V$ values for the {\em geometric mean} estimator for {\em symmetric stable random projections} in \cite{Proc:Li_SODA08} (``symmetric GM'').
}\label{fig_comp_var_factor}
\end{figure}

Figure \ref{fig_comp_var_factor} compares the variances of the three estimators for CC, as well as the geometric mean estimator for {\em symmetric stable random projections}.

\subsection{Sampling from Maximally-Skewed Stable Random Distributions}

The standard procedure for sampling from skewed stable distributions is based on the Chambers-Mallows-Stuck method\cite{Article:Chambers_JASA76}. One first generates an exponential random variable with mean 1, $W \sim \exp(1)$,  and a uniform random variable $U \sim uniform \left(-\frac{\pi}{2}, \frac{\pi}{2}\right)$, then,
\begin{align}\notag
Z &= \frac{\sin\left(\alpha(U+\rho)\right)}{\left[\cos U \cos\left(\rho \alpha\right)
\right]^{1/\alpha}} \left[\frac{\cos\left( U - \alpha(U + \rho)\right)}{W}
\right]^{\frac{1-\alpha}{\alpha}}\\\label{eqn_sampling_skewed}
& \sim S(\alpha,\beta=1,1),
\end{align}
where $\rho = \frac{\pi}{2}$ when $\alpha<1$ and $\rho = \frac{\pi}{2}\frac{2-\alpha}{\alpha}$ when $\alpha>1$. \\


Sampling from symmetric ($\beta = 0$) stable distributions uses the same procedure with $\rho = 0$. Thus, the only difference is the $\cos^{1/\alpha}\left(\rho\alpha\right)$ term, which is  a constant and can be removed out of the sampling procedure and put back to the estimates in the end, which in fact also provides better numerical stability when $\alpha\rightarrow 1$. Note  that the estimators (\ref{eqn_F_gm}) and (\ref{eqn_F_hm}) already contain $\cos\left(\rho\alpha\right)$ in the numerators. Thus, we can sample $Z^\prime=Z\cos^{1/\alpha}\left(\rho\alpha\right)  \sim S\left(\alpha,\beta,\cos\left(\rho\alpha\right)\right)$ instead of $Z = S\left(\alpha,\beta,1\right)$ and evaluate (\ref{eqn_F_gm}) (\ref{eqn_F_hm}) without $\cos\left(\rho\alpha\right)$.

\section{Estimating Entropies Using CC}\label{sec_entropy_est}

The basic procedure is to first estimate the $\alpha$th frequency moment $F_{(\alpha)}$ using CC and then compute various entropies using the estimated $F_{(\alpha)}$. Here we use $\hat{F}_{(\alpha)}$ to denote a generic estimator of $\hat{F}_{(\alpha)}$, which could be $\hat{F}_{(\alpha),gm}$,  $\hat{F}_{(\alpha),hm}$,  $\hat{F}_{(\alpha),oq}$,  or $\hat{F}_{(\alpha),gm,sym}$.

In the following subsections, we analyze the variances and biases in estimating the R\'enyi entropy $H_\alpha$, the Tsallis entropy $T_\alpha$, and the Shannon entropy $H$.

\subsection{R\'enyi Entropy}

We denote a generic estimator of $H_\alpha$ by $\hat{H}_{\alpha}$:
\begin{align}\label{eqn_Renyi_est}
&\hat{H}_\alpha  = \frac{1}{1-\alpha} \log \frac{\hat{F}_{(\alpha)}}{F_{(1)}^\alpha},
\end{align}
which becomes $\hat{H}_{\alpha,gm}$,  $\hat{H}_{\alpha,hm}$,  $\hat{H}_{\alpha,oq}$,  and $\hat{H}_{\alpha,gm,sym}$, respectively, when $\hat{F}_{(\alpha)}$ becomes $\hat{F}_{(\alpha),gm}$,  $\hat{F}_{(\alpha),hm}$,  $\hat{F}_{(\alpha),oq}$,  or $\hat{F}_{(\alpha),gm,sym}$. Since $F_{(1)}$ can be computed exactly and trivially using a simple counter, we assume it is a constant.

Since $\hat{F}_{(\alpha)}$ is unbiased or asymptotically unbiased, $\hat{H}_{\alpha}$ is also asymptotically unbiased. The asymptotic variance of $\hat{H}_{\alpha}$ can be computed by Taylor expansions (the so-called ``delta method'' in statistics):
\begin{align}\notag
\text{Var}\left( \hat{H}_\alpha \right) =& \frac{1}{(1-\alpha)^2} \text{Var}\left( \log \left(\hat{F}_{(\alpha)}\right)\right)\\\notag
=& \frac{1}{(1-\alpha)^2} \text{Var}\left(\hat{F}_{(\alpha)}\right)\left( \frac{\partial \log F_{(\alpha)}}{\partial F_{(\alpha)}}  \right)^2 + O\left(\frac{1}{k^2}\right)\\\label{eqn_Renyi_est_var}
=& \frac{1}{(1-\alpha)^2} \frac{1}{F_{(\alpha)}^2} \text{Var}\left(\hat{F}_{(\alpha)}\right) + O\left(\frac{1}{k^2}\right).
\end{align}

\subsection{Tsallis Entropy}
The generic estimator for the Tsallis entropy $T_\alpha$ would be
\begin{align}\label{eqn_Tsallis_est}
&\hat{T}_\alpha = \frac{1}{\alpha -1} \left( 1 - \frac{\hat{F}_{(\alpha)}}{F_{(1)}^\alpha}\right),
\end{align}
which is asymptotically unbiased and has variance
\begin{align}\label{eqn_Tsallis_est_var}
&\text{Var}\left(\hat{T}_\alpha\right) = \frac{1}{(\alpha -1)^2} \frac{1}{F_{(1)}^{2\alpha}}
 \text{Var}\left(\hat{F}_{(\alpha)}\right) + O\left(\frac{1}{k^2}\right).
\end{align}

\subsection{Shannon Entropy}

We use $\hat{H}_{\alpha,R}$ and $\hat{H}_{\alpha,T}$ to denote the estimators for Shannon entropy using the estimated $\hat{H}_{\alpha}$ and $\hat{T}_{\alpha}$, respectively.

The variances remain unchanged, i.e.,
\begin{align}\label{eqn_entropy_var}
\text{Var}\left(\hat{H}_{\alpha,R}\right) = \text{Var}\left(\hat{H}_{\alpha}\right), \hspace{0.05in} \text{Var}\left(\hat{H}_{\alpha,T}\right) = \text{Var}\left(\hat{T}_{\alpha}\right).
\end{align}

However, $\hat{H}_{\alpha,R}$ and $\hat{H}_{\alpha,T}$ are no longer unbiased, even asymptotically (unless $\alpha\rightarrow 1$). The biases would be
\begin{align}\label{eqn_entropy_R_bias}
&\text{Bias}\left(\hat{H}_{\alpha,R}\right) = \text{E}\left( \hat{H}_{\alpha,R} - H\right) =  H_\alpha - H + O\left(\frac{1}{k}\right), \\\label{eqn_entropy_T_bias}
&\text{Bias}\left(\hat{H}_{\alpha,T}\right) = \text{E}\left( \hat{T}_{\alpha,R} - H\right) =  T_\alpha - H + O\left(\frac{1}{k}\right).
\end{align}
The $O\left(\frac{1}{k}\right)$  biases arise from the estimation biases in $\hat{H}_\alpha$ and $\hat{T}_\alpha$ and diminish  quickly as $k$ increases. In fact, there are standard statistics procedures to reduce the $O\left(\frac{1}{k}\right)$ bias to $O\left(\frac{1}{k^2}\right)$.  However, the ``intrinsic biases,'' $H_\alpha - H$ and $T_\alpha - H$, can not be removed by increasing $k$; they can only be reduced by letting $\alpha$ close to 1.

The total error is usually measured by the mean square error: MSE = Bias$^2$ + Var. Clearly, there is a variance-bias trade-off in estimating $H$ using $H_\alpha$ or $T_\alpha$.  For a particular data stream, at each sample size $k$, there will be an optimal $\alpha$ to attain the smallest MSE. The optimal $\alpha$ is data-dependent and hence some prior knowledge of the data is needed in order to determine it. The prior knowledge may be accumulated during the data stream process. Alternatively, we could seek an estimator that is very accurate near $\alpha=1$ to alleviate the variance-bias affect.

\section{Experiments}\label{sec_exp}

The goal of the experimental study is to demonstrate the effectiveness of Compressed Counting (CC) for estimating entropies and to determine a good strategy for estimating the Shannon entropy.  In particular, we focus on the estimation accuracy and would like to verify the formulas for (asymptotic) variances in (\ref{eqn_Renyi_est_var}) and (\ref{eqn_Tsallis_est_var}).  

\subsection{Data}

Since the estimation accuracy is what we are interested in, we can simply use static data instead of real data streams. This is because the projected data vector $X = \mathbf{R}^\text{T} A_t$ is the same, regardless whether it is computed at once  (i.e., static) or incrementally (i.e., dynamic). As we have commented, the processing and storage cost of CC is the same as the cost of {\em symmetric stable random projections} at the same sample size $k$. Therefore, to compare these two methods, it suffices to compare their estimation accuracies.

Ten English words are selected from a chunk of Web crawl data with $D = 2^{16} = 65536$ pages: THE, A, THIS, HAVE, FUN, FRIDAY, NAME, BUSINESS, RICE, and TWIST. The words are selected fairly randomly, except that we make sure they cover a whole range of sparsity, from function words (e.g., A, THE), to  common words (e.g., FRIDAY) to rare words (e.g., TWIST).

Thus, as summarized in Table \ref{tab_data},  our data set consists of ten vectors of length $D = 65536$ and the entries are the numbers of word occurrences in each document.

Table \ref{tab_data} indicates that the R\'enyi entropy $H_\alpha$ provides a much better approximation to the Shannon entropy $H$, than the Tsallis entropy $T_\alpha$ does. On the other hand, if the purpose is to find a summary statistic that is different from the Shannon entropy (i.e., sensitive to $\alpha$), then the Tsallis entropy may be more suitable.

\begin{table}[h]
\caption{\small  The data set consists of 10 English words selected from a chunk of $D=65536$ Web pages, forming 10 vectors of length $D$ whose values are the word occurrences. The table lists their numbers of non-zeros (sparsity), the Shannon entropy $H$, the R\'enyi entropy $H_\alpha$ and the Tsallis entropy $T_\alpha$ (for $\alpha = 0.95$ and 1.05).
 }
\begin{center}{\scriptsize\tiny
\begin{tabular}{l l l l l l l}
\hline \hline\\
Word &Nonzero  & $H$ &$H_{0.95}$  &$H_{1.05}$   &$T_{0.95}$  &$T_{1.05}$ \\  \\\hline
TWIST &274 &5.4873 &5.4962 &5.4781 &6.3256 &4.7919\\
RICE &490 &5.4474 &5.4997 &5.3937 &6.3302 &4.7276\\
FRIDAY &2237 &7.0487 &7.1039 &6.9901 &8.5292 &5.8993 \\
FUN &3076 & 7.6519   & 7.6821 &    7.6196  &   9.3660  &   6.3361\\
BUSINESS &8284 &8.3995 &8.4412 &8.3566 &10.502 &6.8305\\
NAME & 9423 &8.5162 &9.5677 &8.4618 &10.696 &6.8996\\
HAVE & 17522  &8.9782 &9.0228 & 8.9335 & 11.402 & 7.2050\\
THIS & 27695  &9.3893 &9.4370 &9.3416 &12.059 &7.4634 \\
A    & 39063  &9.5463  &9.5981  &9.4950  &12.318   &7.5592\\
THE  & 42754  & 9.4231 &9.4828  &9.3641  &12.133  &7.4775\\
\hline\hline
\end{tabular}
}
\end{center}
\label{tab_data}
\end{table}
\begin{figure}[h]
\begin{center}\mbox{
{\includegraphics[width=1.75in]{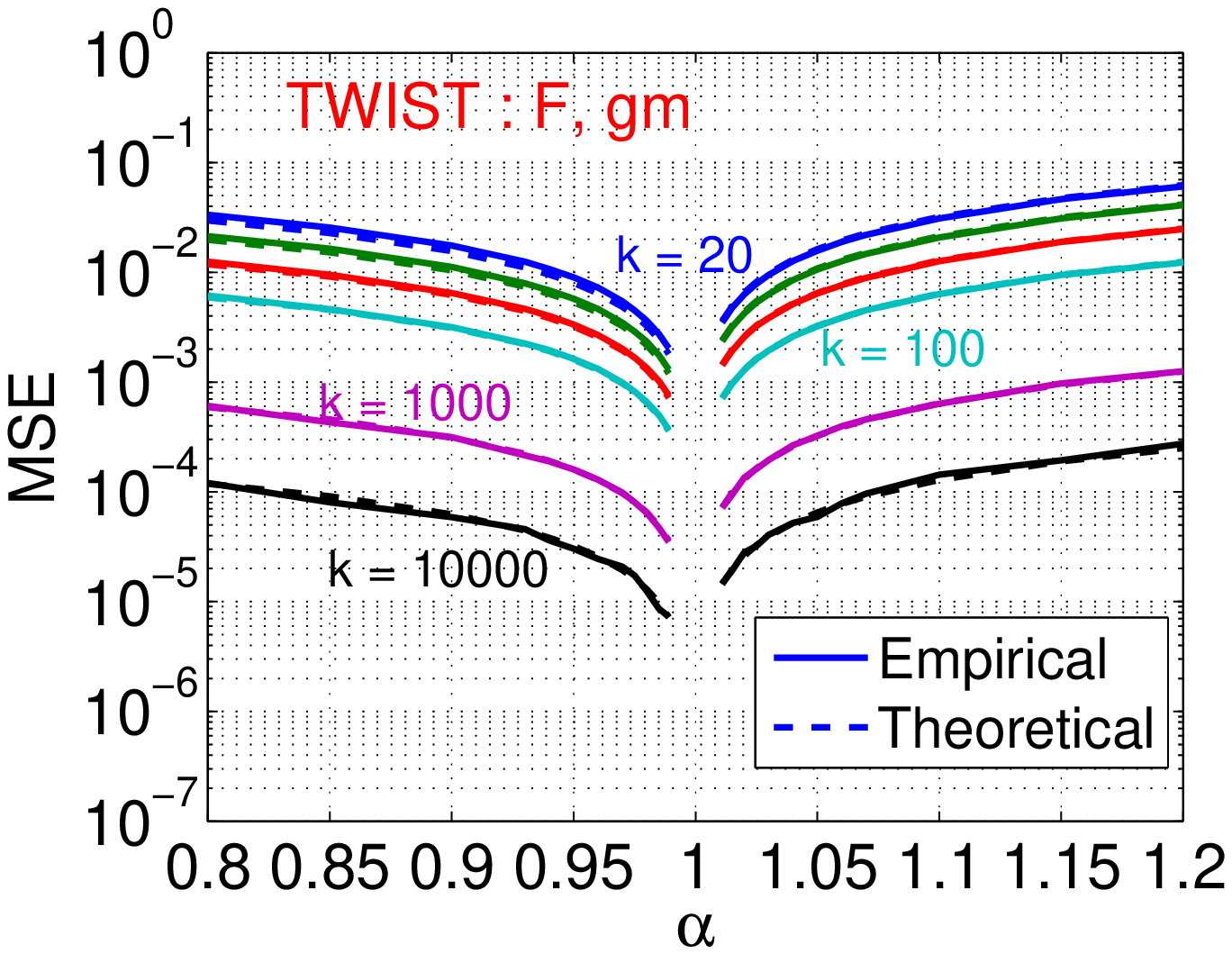}} \hspace{-0.1in}
{\includegraphics[width=1.75in]{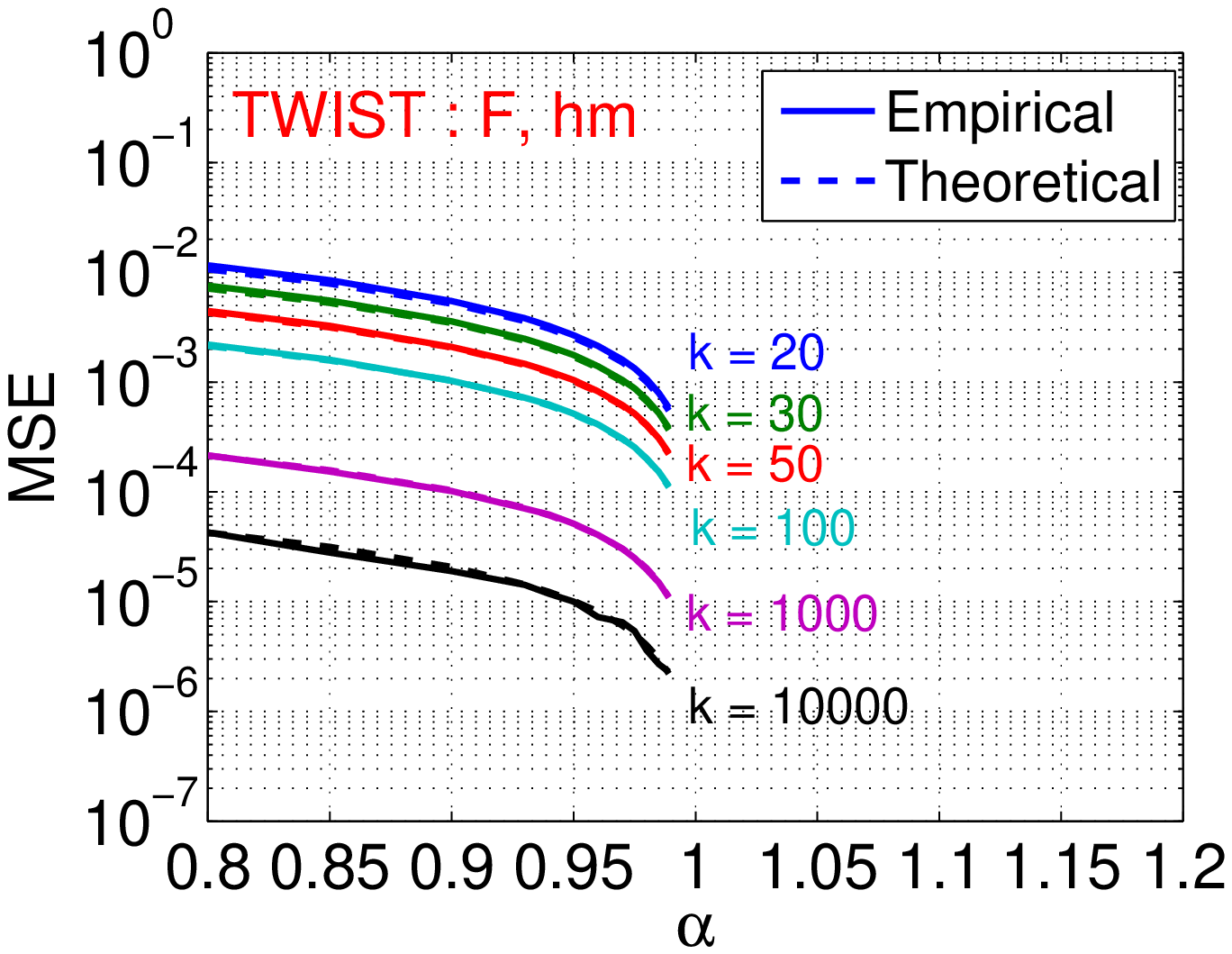}}}\\
\mbox{
{\includegraphics[width=1.75in]{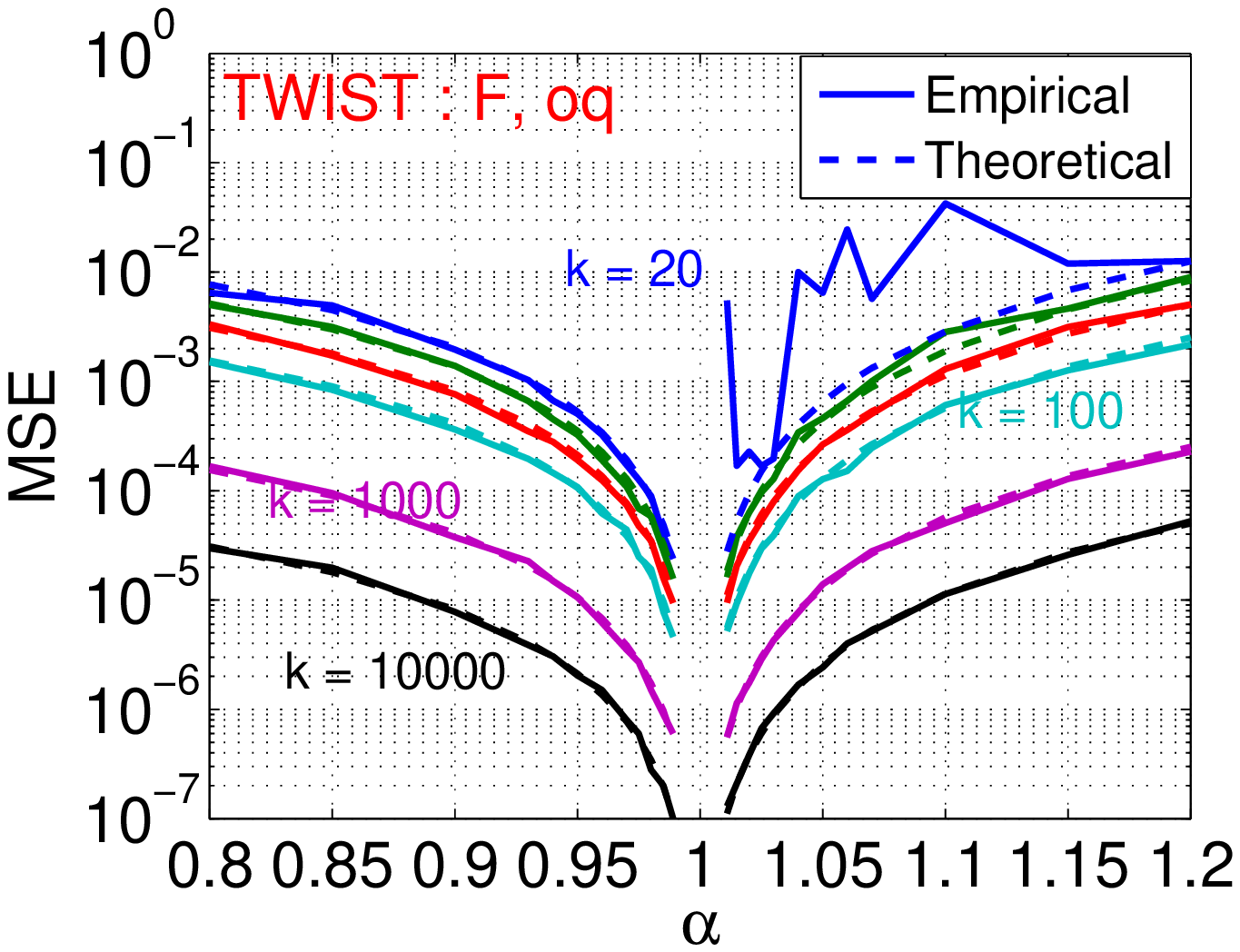}} \hspace{-0.1in}
{\includegraphics[width=1.75in]{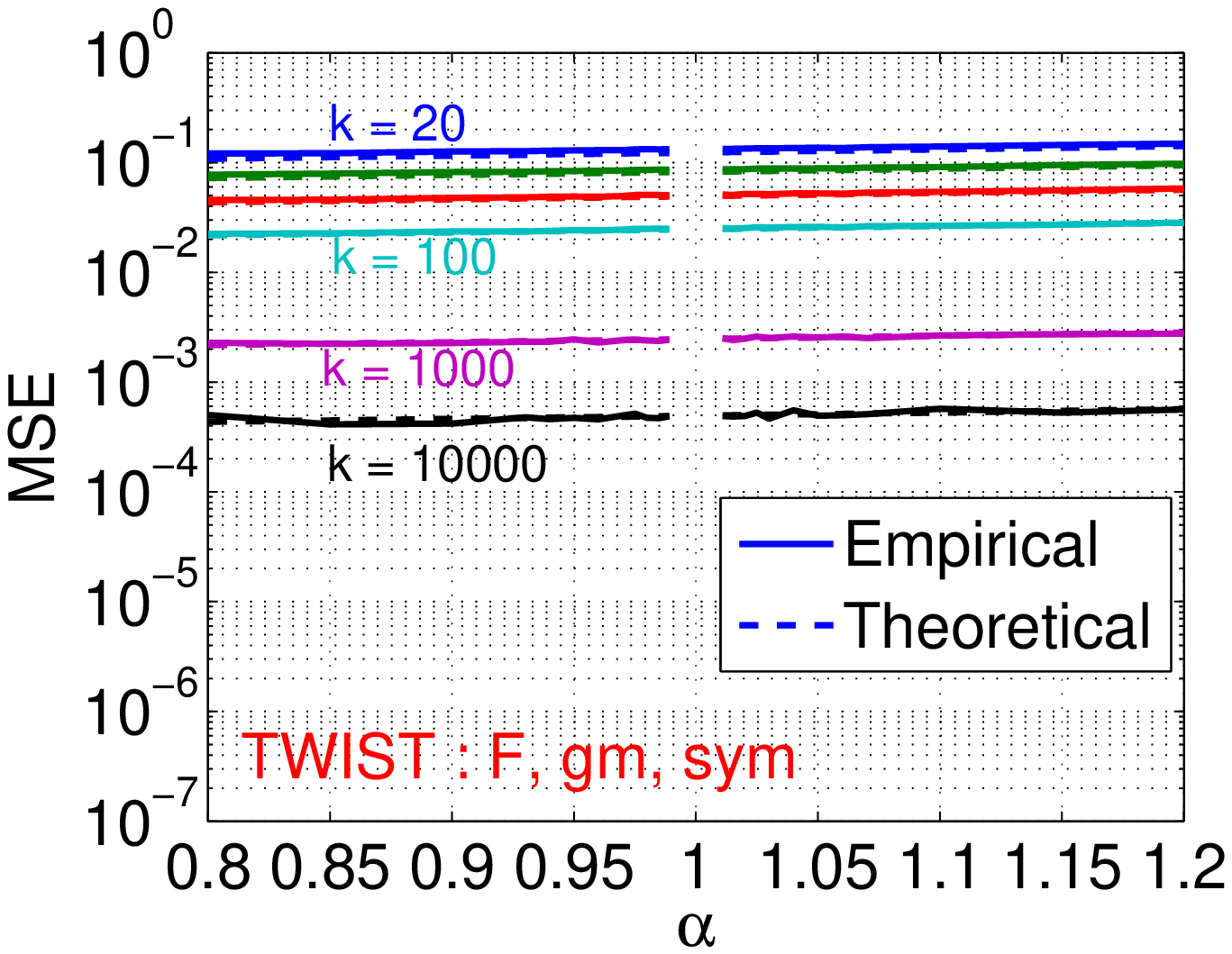}}
}
\end{center}
\vspace{-0.15in}
\caption{Frequency moments, $F_{(\alpha)}$, for TWIST.  Solid curves are   empirical mean square errors (MSEs) and dashed curves are theoretical asymptotic variances in (\ref{eqn_F_gm_var}), (\ref{eqn_F_hm_var}), (\ref{eqn_F_gm_sym_var}), and Table \ref{tab_oq}.  ``F,gm'' stands for the geometric mean estimator $\hat{F}_{(\alpha),gm}$  (\ref{eqn_F_gm}), ``F,hm'' for the harmonic mean estimator $\hat{F}_{(\alpha),hm}$   (\ref{eqn_F_hm}), ``F,oq'' for the optimal quantile estimator $\hat{F}_{(\alpha),oq}$ (\ref{eqn_F_oq}), and ``F,gm,sym'' for the geometric mean estimator $F_{(\alpha),gm,sym}$ (\ref{eqn_F_gm_sym}) in {\em symmetric stable random projections}. }\label{fig_twist_F}
\end{figure}

\subsection{Results}

The results for estimating frequency moments, R\'enyi entropy, Tsallis entropy, and Shannon entropy are presented in the following subsections, in terms of the normalized (i.e., relative) mean square errors (MSEs), e.g., $\frac{\text{MSE}\left(\hat{F}_{(\alpha)}\right)}{F_{(\alpha)}^2}$, $\frac{\text{MSE}\left(\hat{H}_{\alpha}\right)}{H_{\alpha}^2}$, etc. After normalization, we observe that the results are quite similar across different words. To avoid boring the readers, not all words are selected for the presentation. However, we provides the experimental results for all 10 words, in estimating Shannon entropy.

In our experiments, the sample size $k$ ranges from $20$ to $10^4$. We choose $0.8\leq \alpha\leq 0.989$ and $1.011 \leq \alpha\leq 1.2$. This is because \cite{Report:Li_CC_oq} only provided the optimal quantile estimator for $\alpha\geq 1.011$ and $\alpha\leq 0.989$. For the geometric mean and harmonic mean estimators, we actually had no problem of using (e.g.,) $\alpha = 1-10^{-4}$ or $\alpha = 1+10^{-4}$.

\subsubsection{Estimating Frequency Moments}

\begin{figure}[h]
\begin{center}\mbox{
{\includegraphics[width=1.75in]{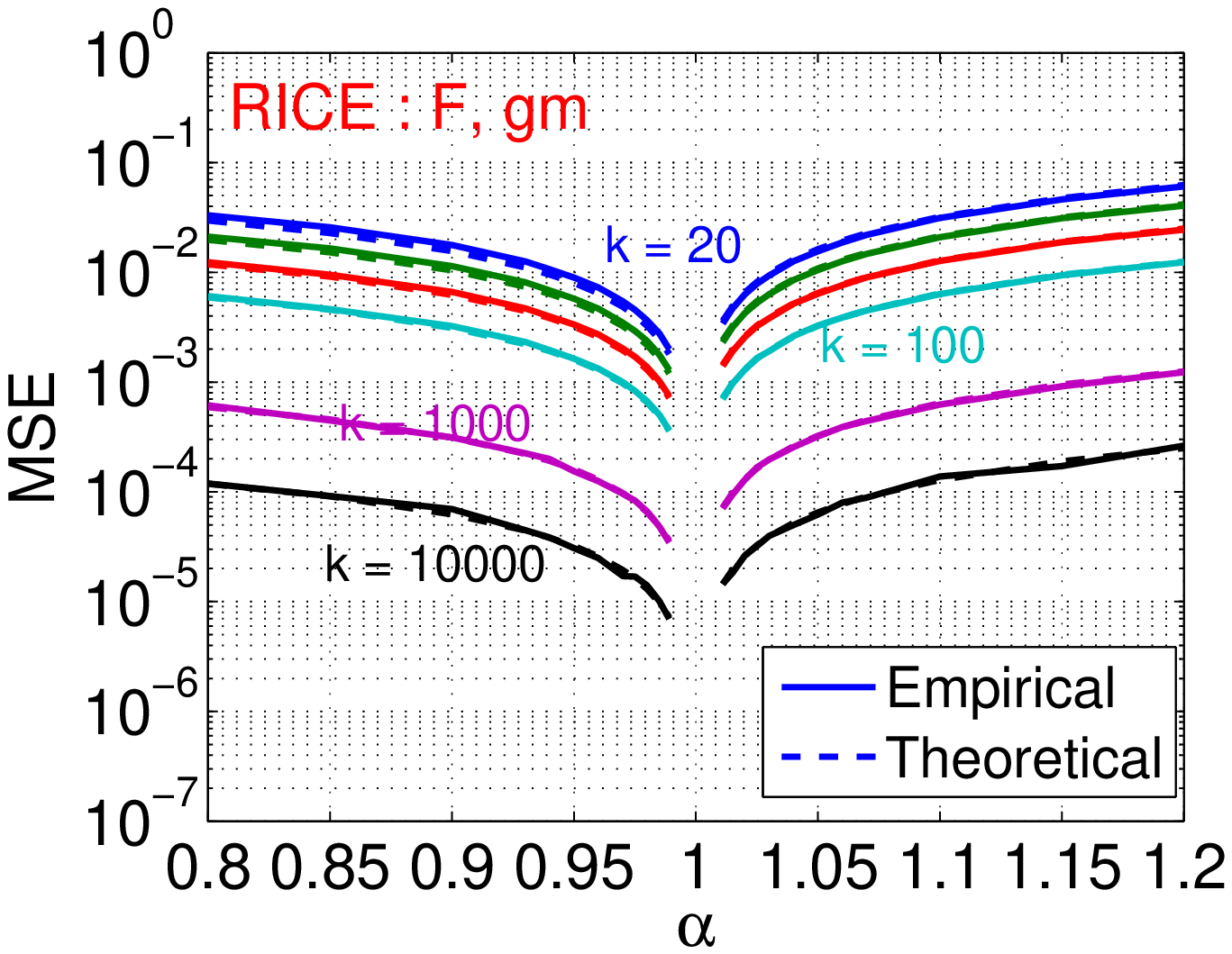}} \hspace{-0.1in}
{\includegraphics[width=1.75in]{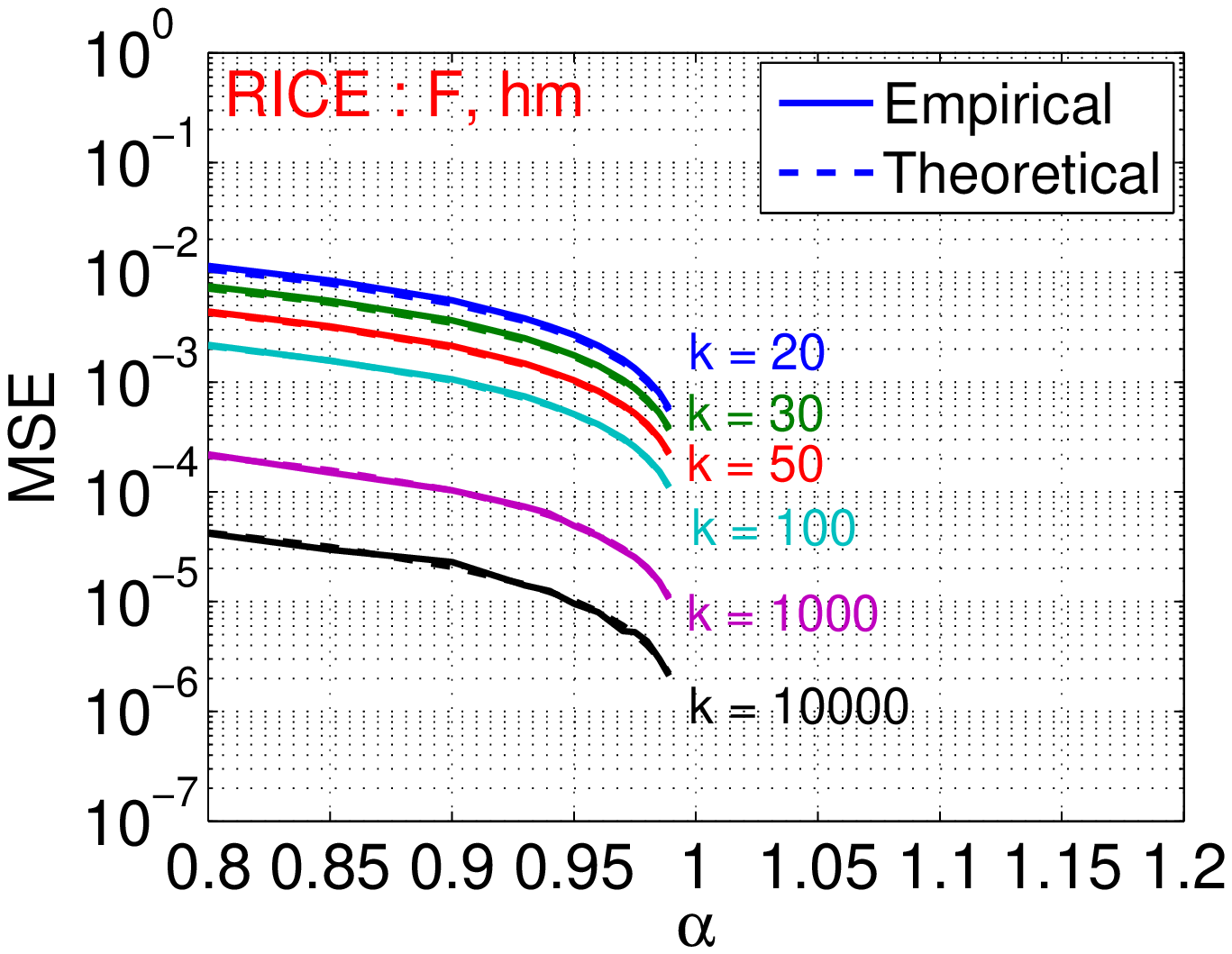}}}\\
\mbox{
{\includegraphics[width=1.75in]{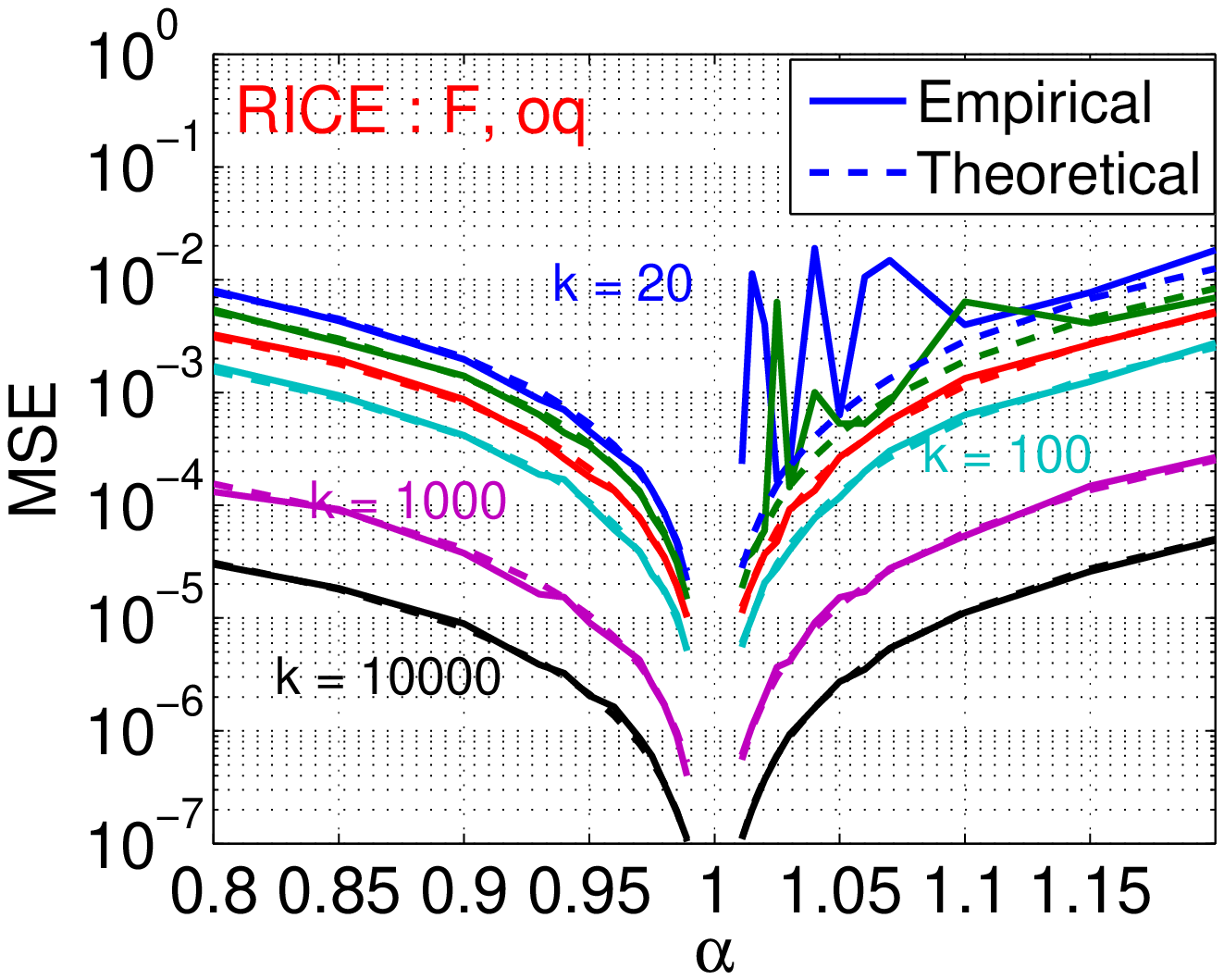}} \hspace{-0.1in}
{\includegraphics[width=1.75in]{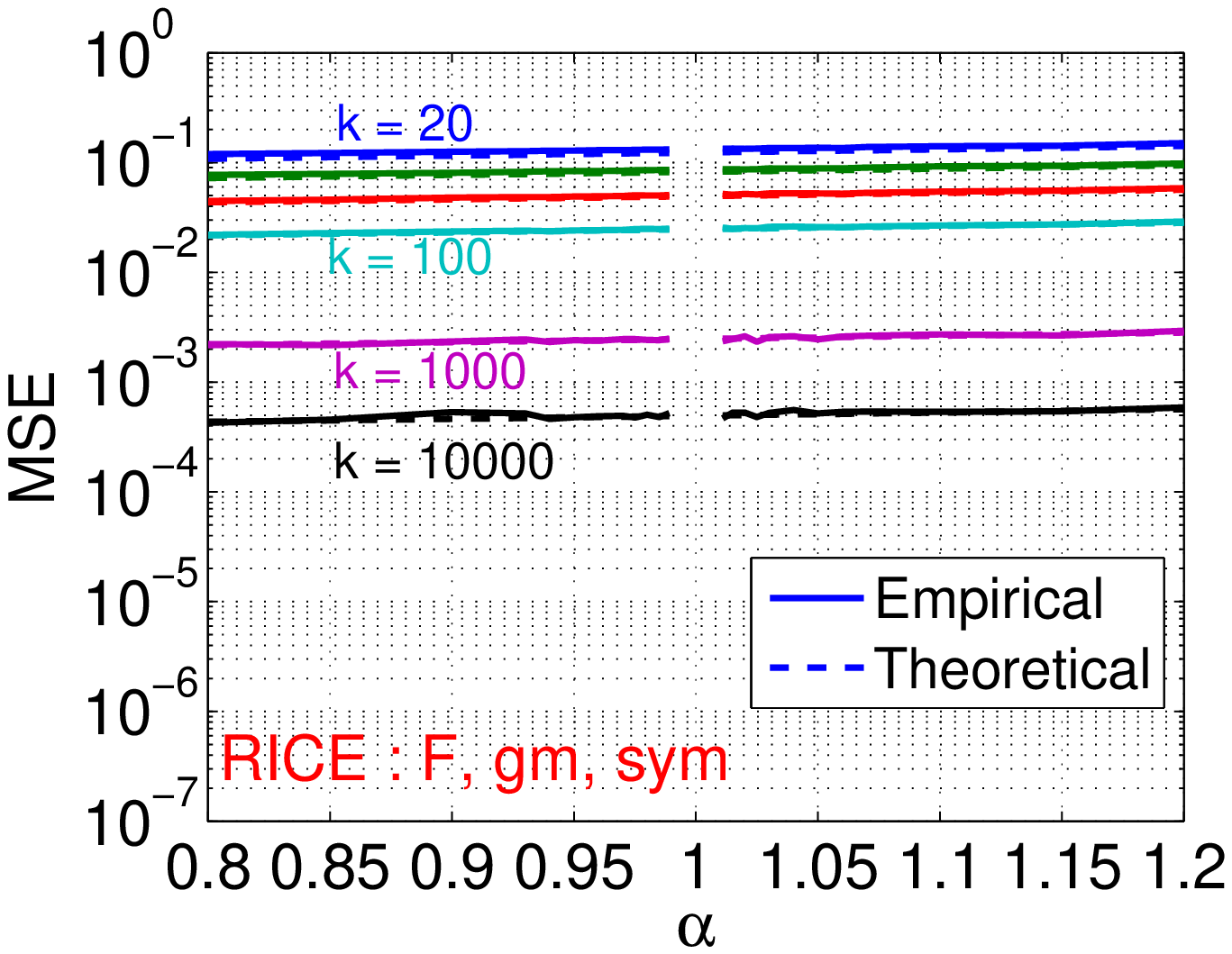}}
}
\end{center}
\vspace{-0.15in}
\caption{ Frequency moments, $F_{(\alpha)}$, for RICE. See the caption of Figure \ref{fig_twist_F} for more explanations.}\label{fig_rice_F}
\end{figure}

\begin{figure}[h]
\begin{center}\mbox{
{\includegraphics[width=1.75in]{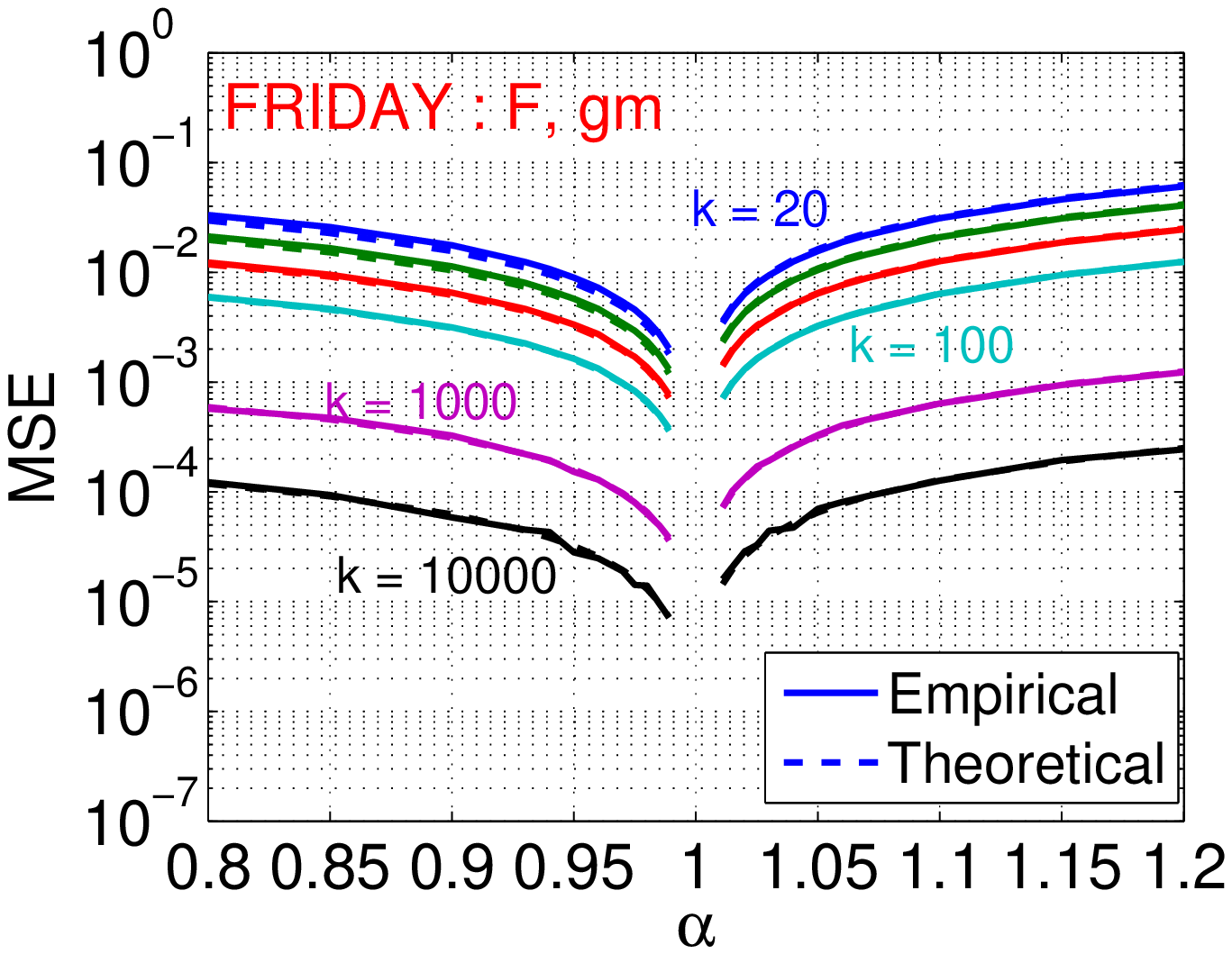}} \hspace{-0.1in}
{\includegraphics[width=1.75in]{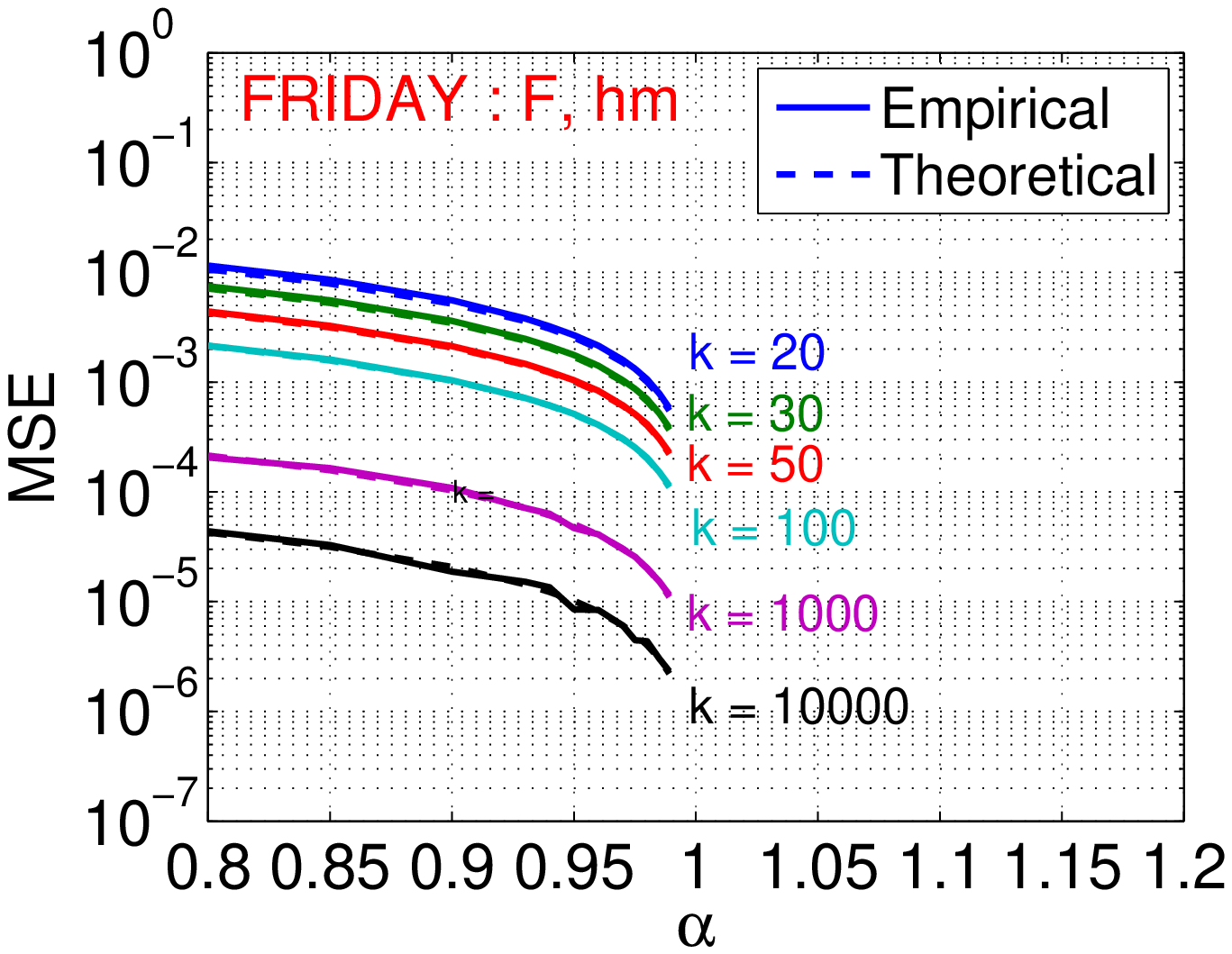}}}\\
\mbox{
{\includegraphics[width=1.75in]{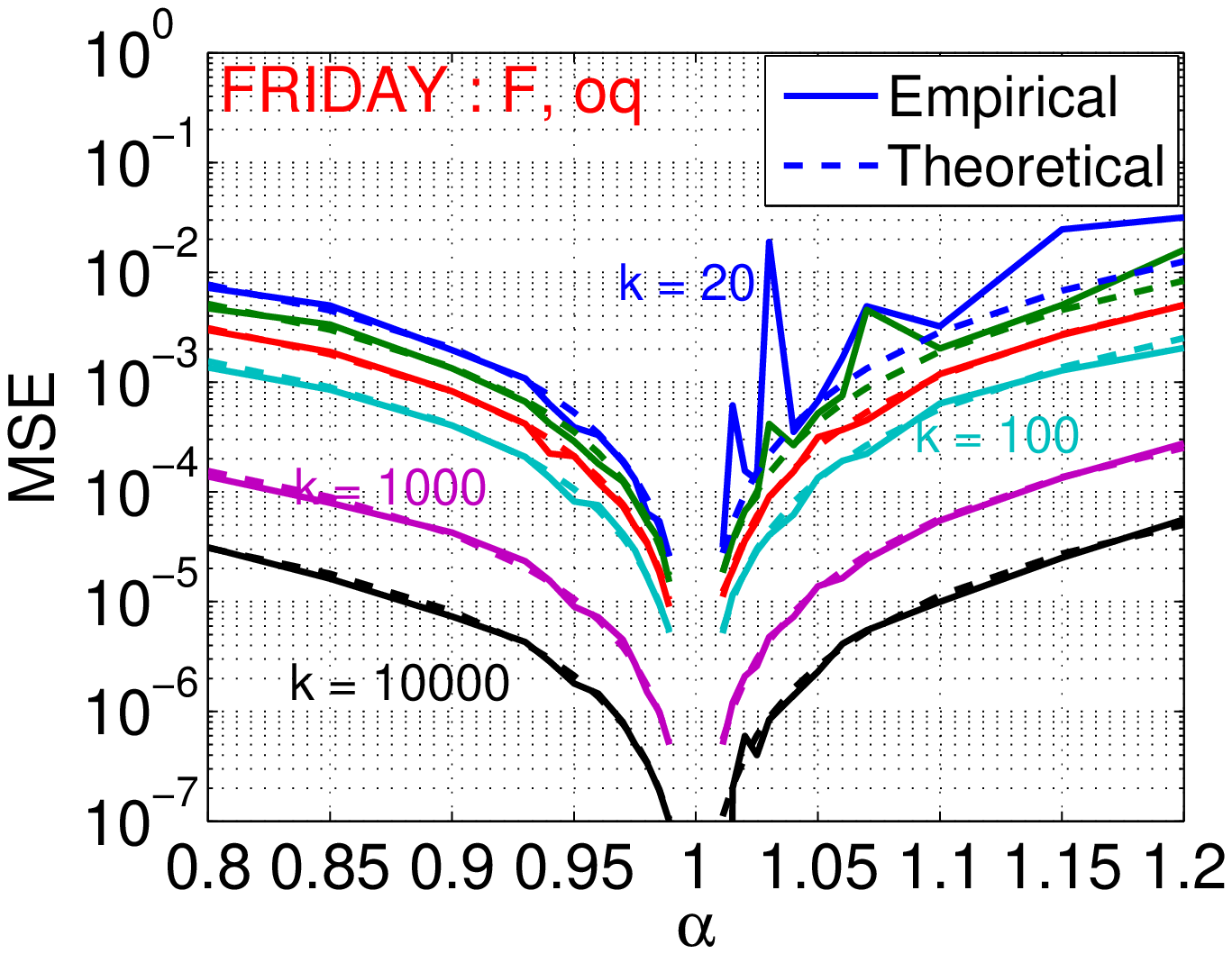}} \hspace{-0.1in}
{\includegraphics[width=1.75in]{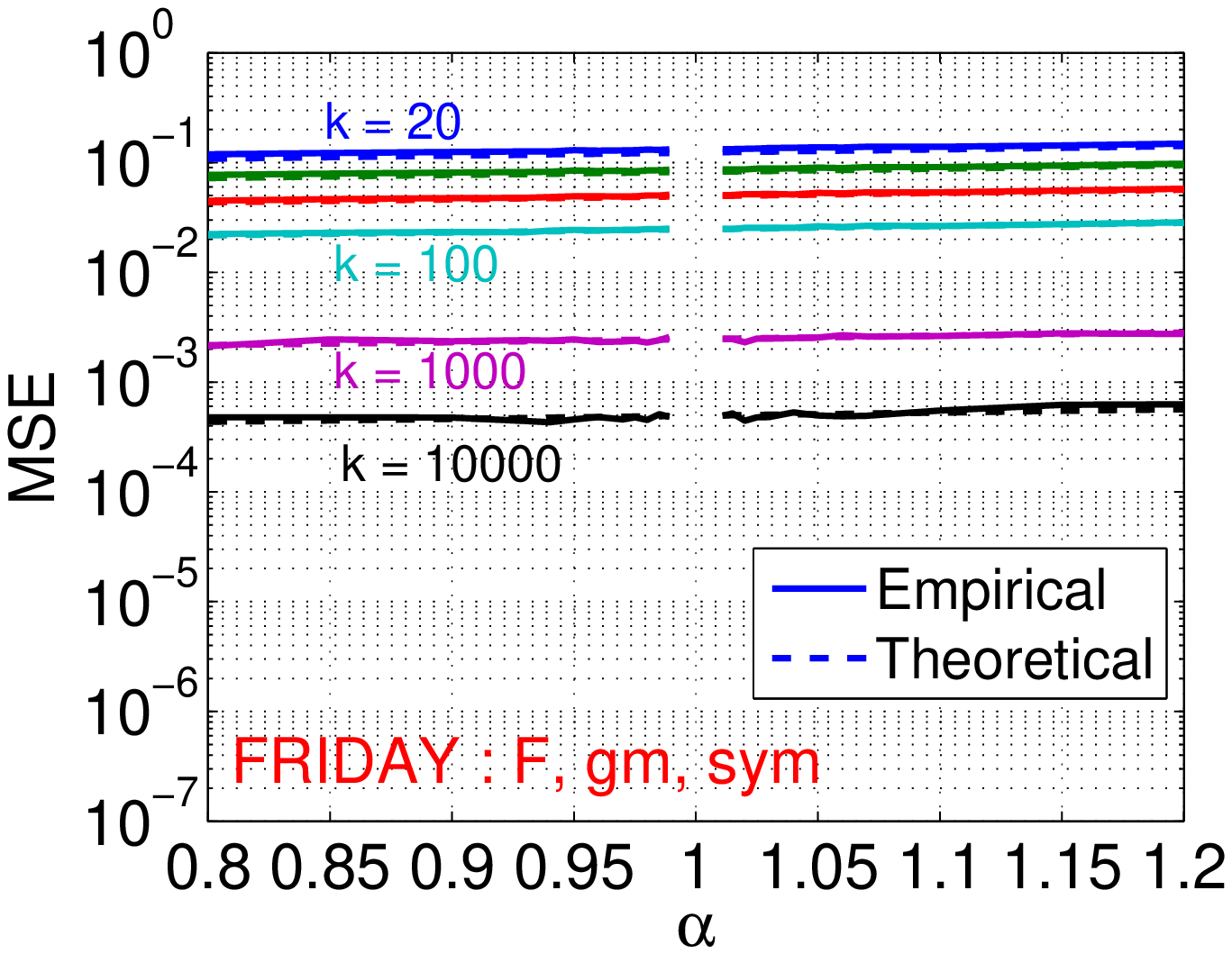}}\vspace{-0.1in}
}
\end{center}
\vspace{-0.15in}
\caption{ Frequency moments, $F_{(\alpha)}$, for FRIDAY. }\label{fig_friday_F}
\end{figure}

Figure \ref{fig_twist_F}, Figure \ref{fig_rice_F}, and Figure \ref{fig_friday_F} provide the MSEs for estimating the $\alpha$th  frequency moments, $F_{(\alpha)}$, for TWIST, RICE, and FRIDAY, respectively.
\begin{itemize}
\item The errors of the three estimators for CC decrease (to zero, potentially) as $\alpha\rightarrow 1$, while the errors of {\em symmetric stable random projections} do not vary much near $\alpha=1$. The improvement of CC is enormous as $\alpha\rightarrow 1$. For example, when $k = 20$ and $\alpha = 0.989$, the MSE of CC using the optimal quantile estimator is about $10^{-5}$ while the MSE of {\em symmetric stable random projections} is about $10^{-1}$, a 10000-fold error reduction.
\item The optimal quantile estimator $\hat{F}_{(\alpha),oq}$ is in general more accurate than the geometric mean and harmonic mean estimators near $\alpha =1$. However, for small $k$ (e.g., 20) and $\alpha>1$, $\hat{F}_{(\alpha),oq}$ exhibits some bad behaviors, which disappear when $k\geq 50$ (or even $k \geq 30$).
\item The theoretical asymptotic variances in (\ref{eqn_F_gm_var}), (\ref{eqn_F_hm_var}), (\ref{eqn_F_gm_sym_var}), and Table \ref{tab_oq} are accurate.
\end{itemize}

\newpage

\vspace{-0.2in}

\vspace{0.2in}
\subsubsection{Estimating R\'enyi Entropy}

\begin{figure}[h]
\begin{center}\mbox{
{\includegraphics[width=1.75in]{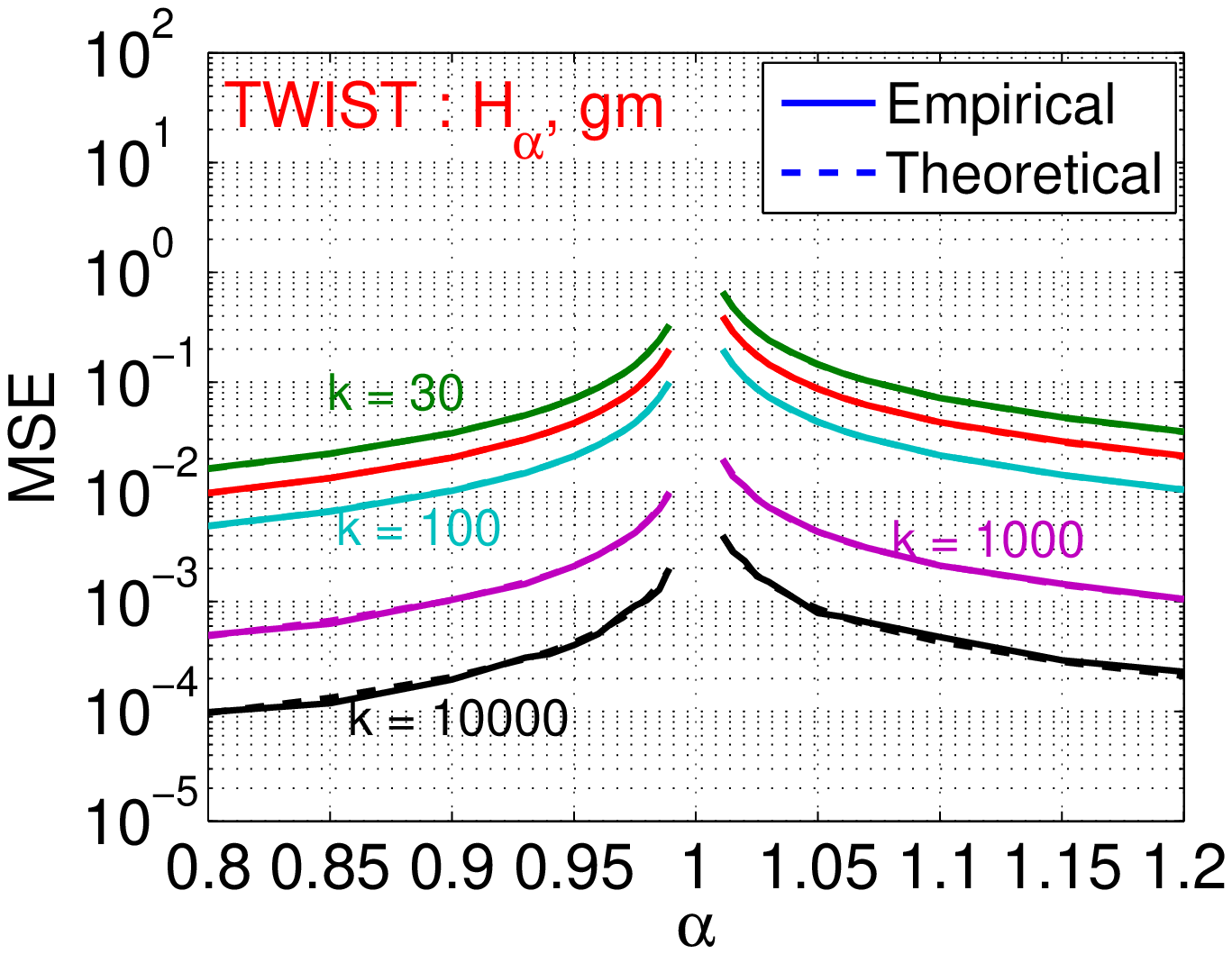}} \hspace{-0.1in}
{\includegraphics[width=1.75in]{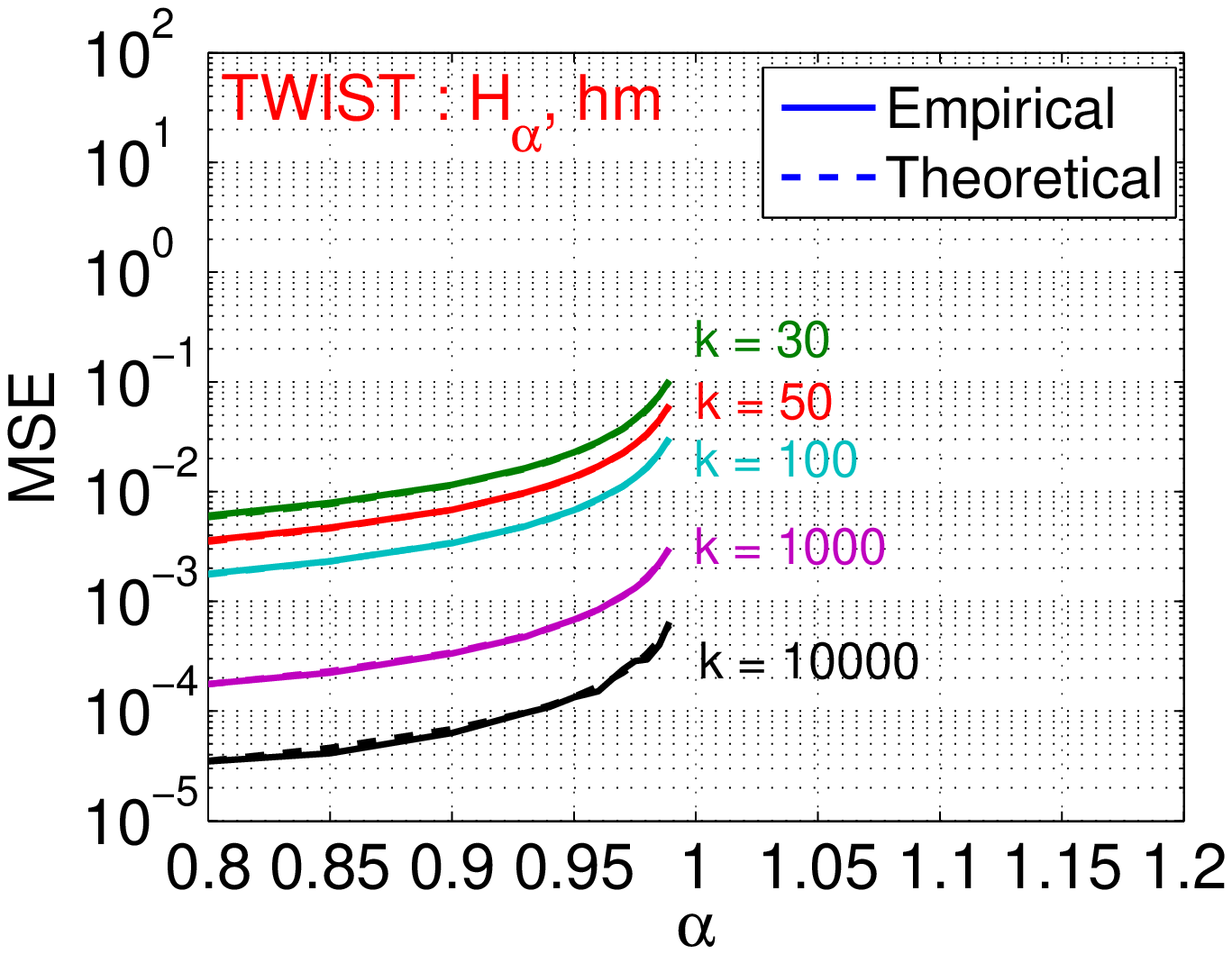}}}\\
\mbox{
{\includegraphics[width=1.75in]{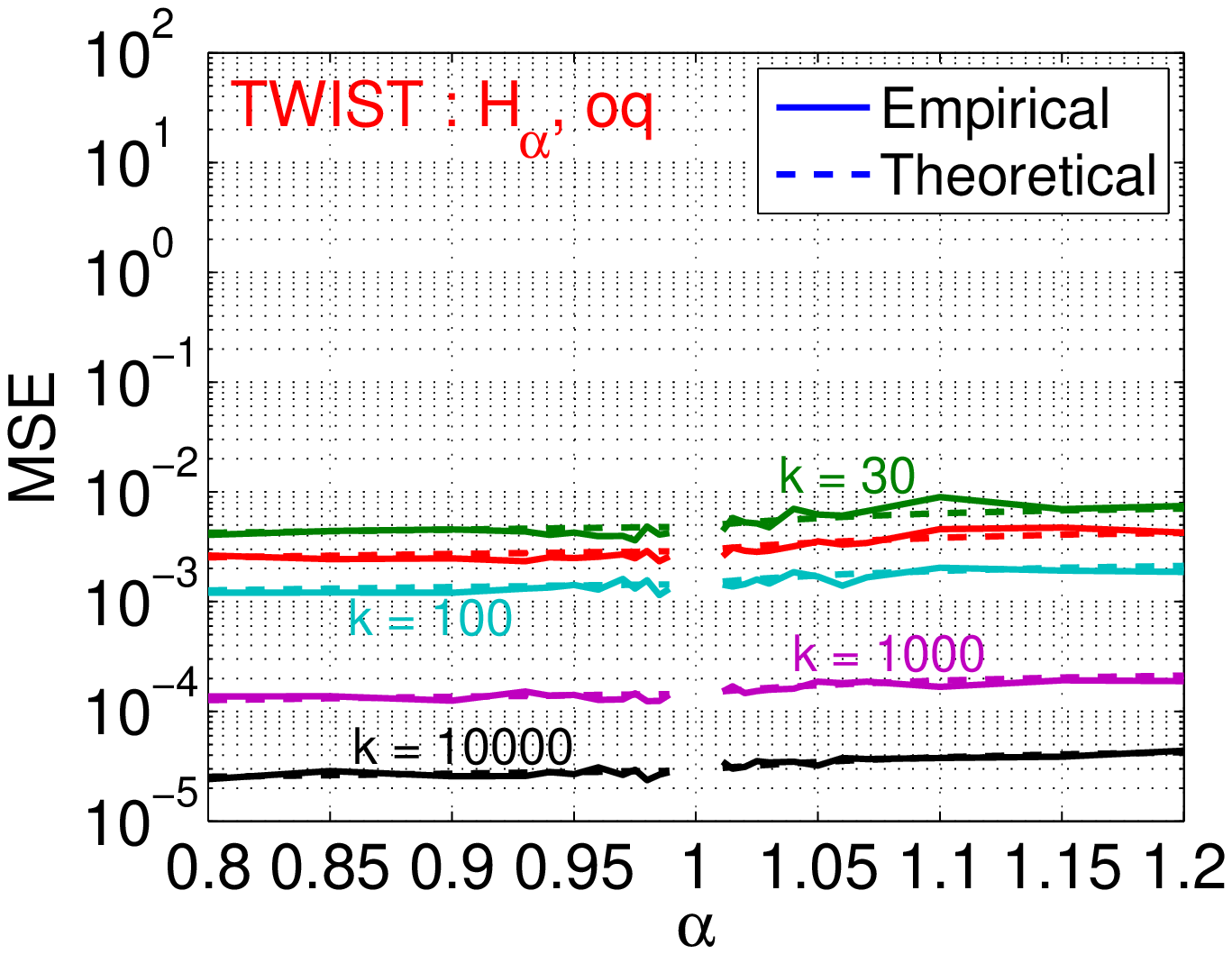}} \hspace{-0.1in}
{\includegraphics[width=1.75in]{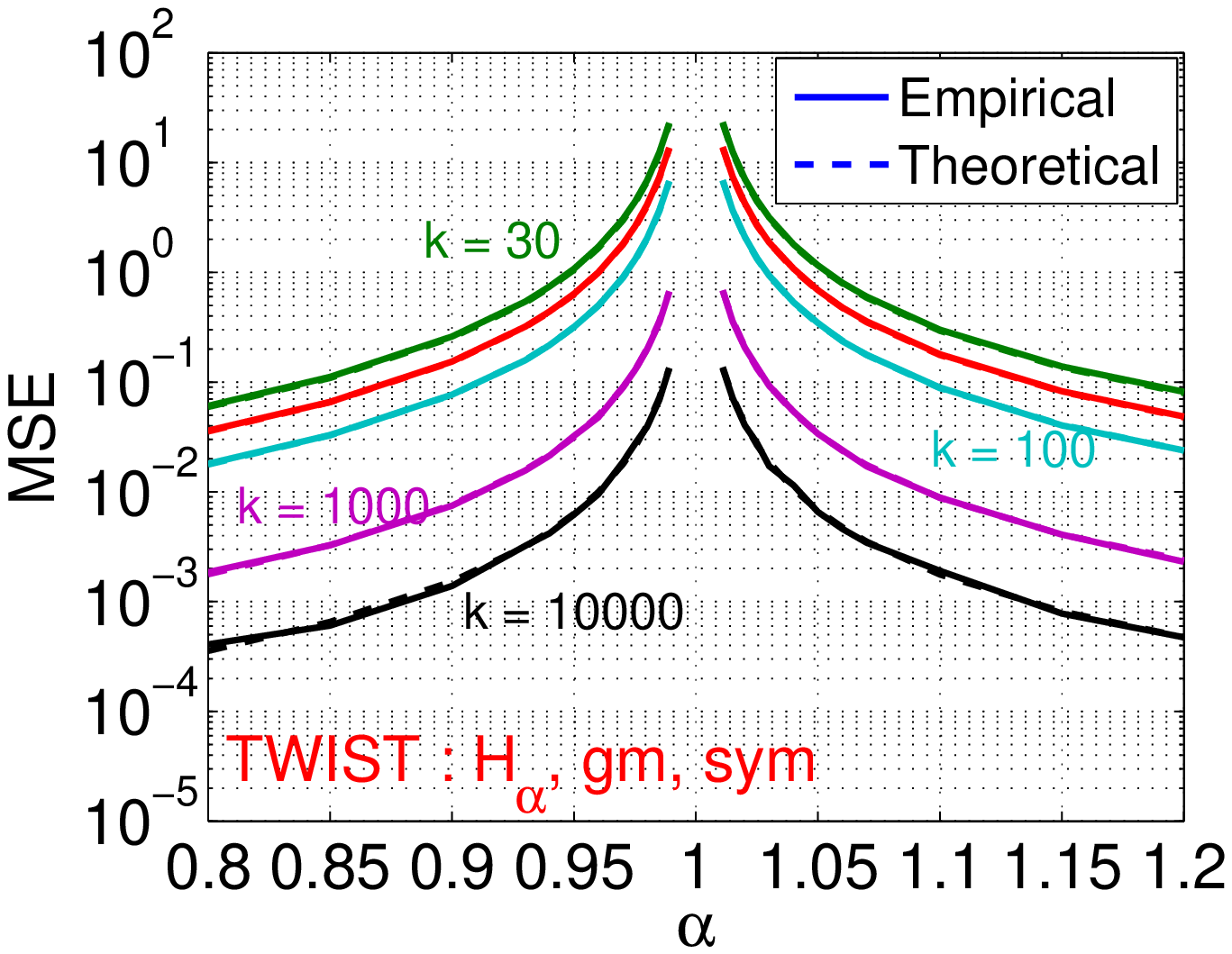}}
}
\end{center}
\vspace{-0.15in}
\caption{R\'eny entropy, $H_{\alpha}$, for TWIST. The theoretical variances (dashed) are computed from (\ref{eqn_Renyi_est_var}).  }\label{fig_twist_H}
\end{figure}

Figure \ref{fig_twist_H} plots the MSEs for estimating the R\'eny entropy for TWIST,  with the curves for $k = 20$ removed.
The figure illustrates that: (1) CC improves {\em symmetric stable rand projections} enormously when $\alpha\rightarrow 1$; (2) The generic variance formula (\ref{eqn_Renyi_est_var}) is accurate.

\vspace{0.2in}

\subsubsection{Estimating Tsallis Entropy}

\begin{figure}[h]
\begin{center}\mbox{
{\includegraphics[width=1.75in]{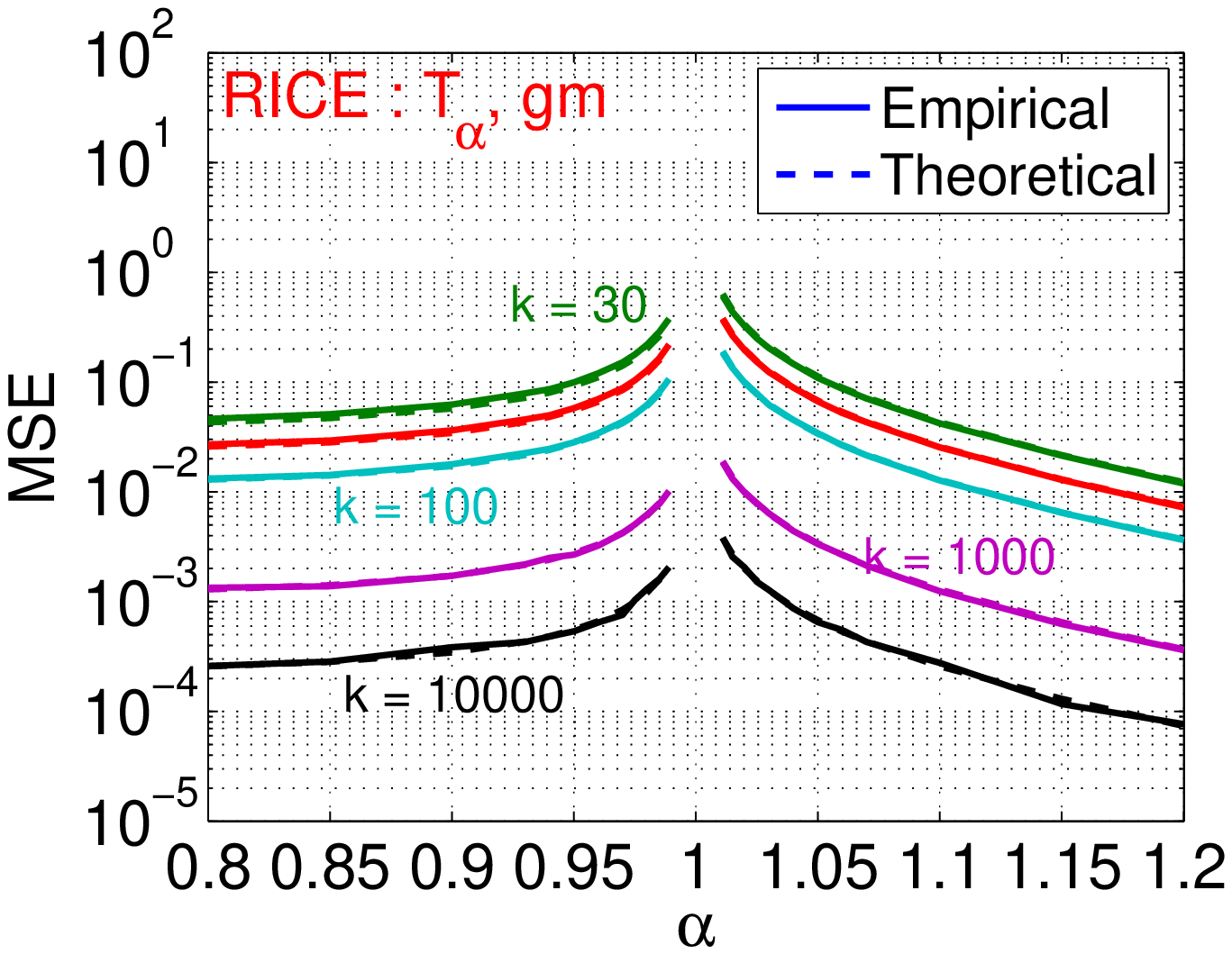}} \hspace{-0.1in}
{\includegraphics[width=1.75in]{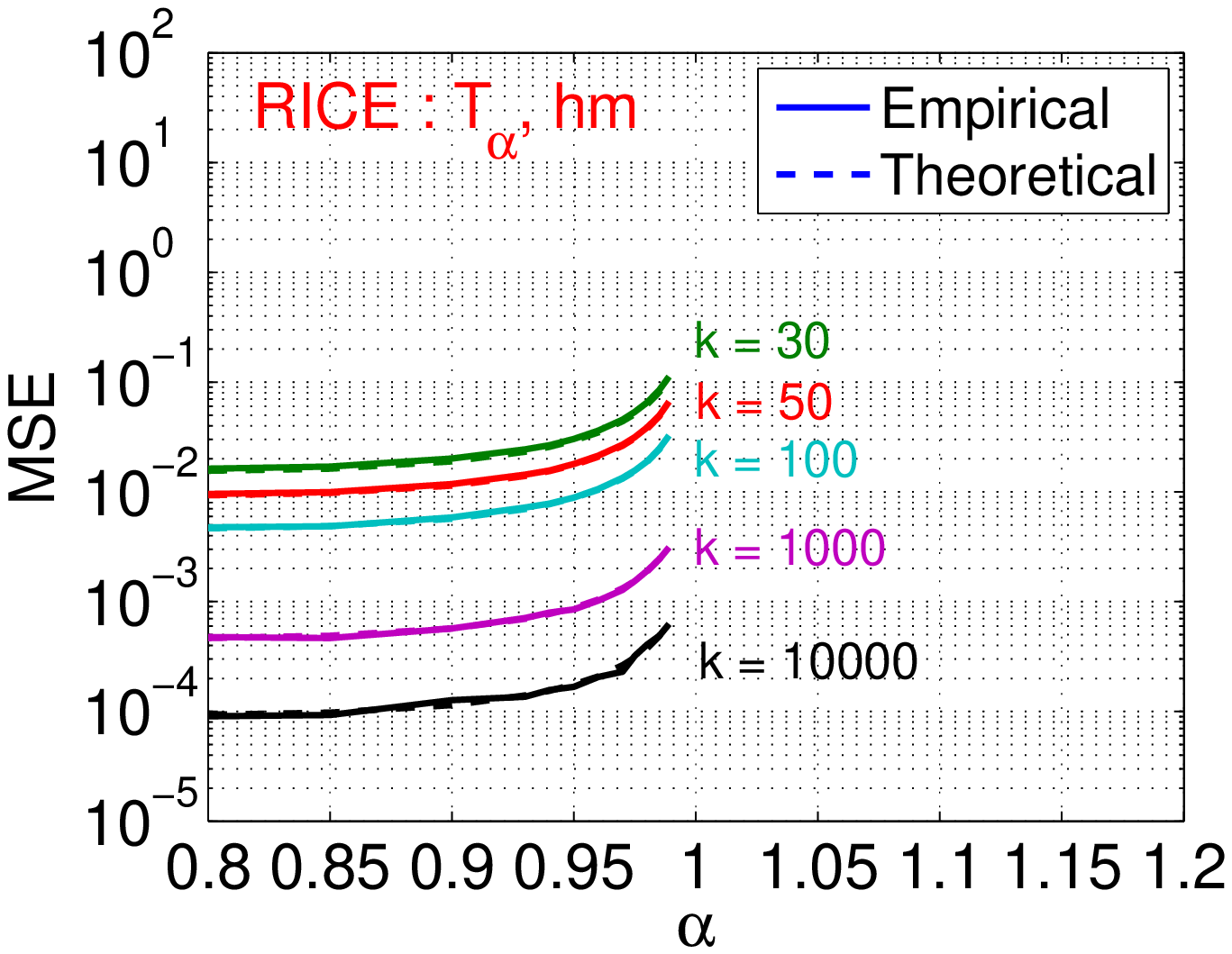}}}\\
\mbox{
{\includegraphics[width=1.75in]{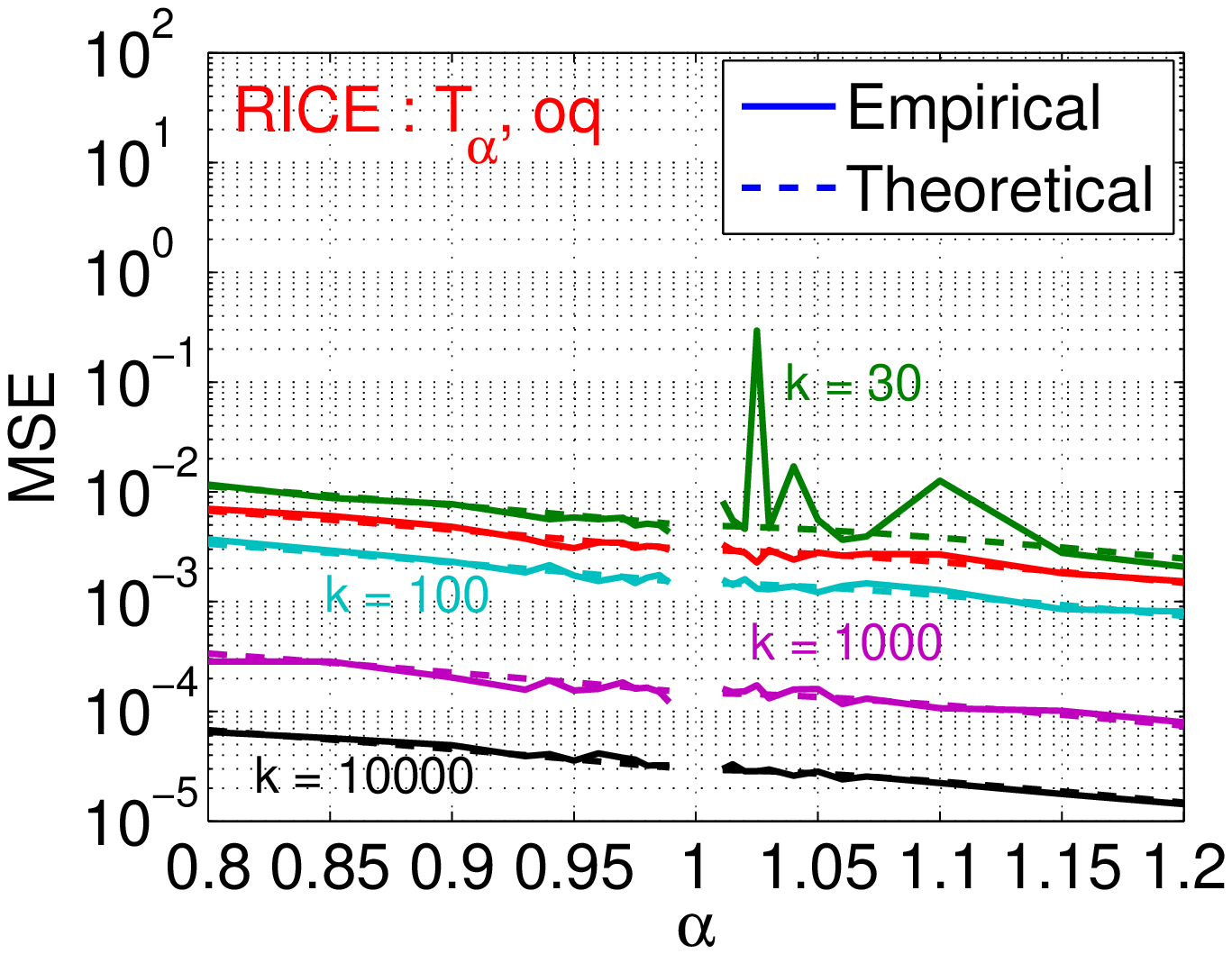}} \hspace{-0.1in}
{\includegraphics[width=1.75in]{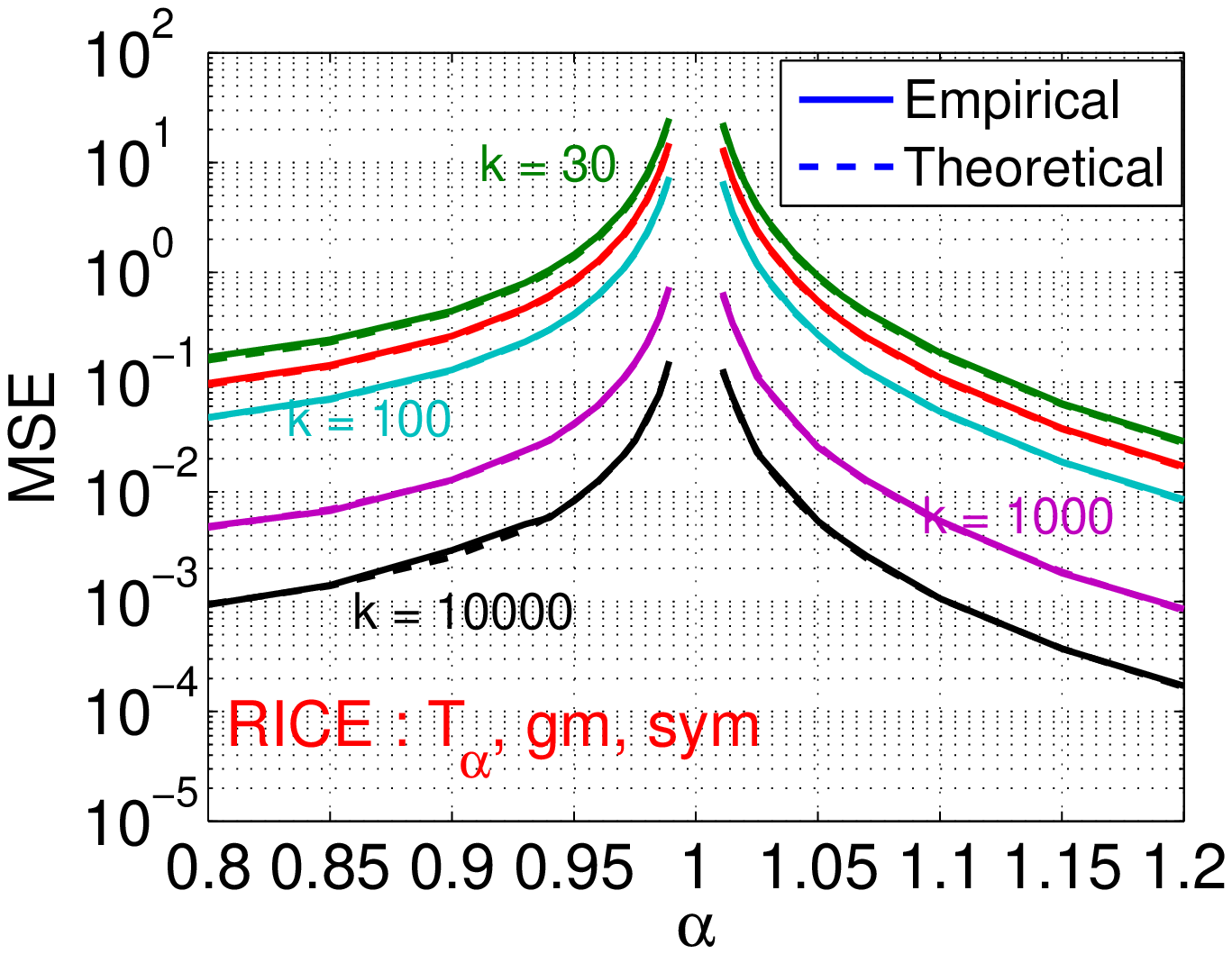}}
}
\end{center}
\vspace{-0.15in}
\caption{Tsallis entropy, $H_{\alpha}$, for RICE. The theoretical variances (dashed) are computed from (\ref{eqn_Tsallis_est_var}). }\label{fig_rice_T}
\end{figure}

Figure \ref{fig_rice_T} plots the MSEs for estimating the Tsallis entropy for RICE, illustrating that: (1) CC improves {\em symmetric stable rand projections} enormously when $\alpha\rightarrow 1$; (2) The generic variance formula (\ref{eqn_Tsallis_est_var}) is accurate.

\vspace{0.2in}
\subsubsection{Estimating Shannon Entropy\\ from R\'enyi Entropy}

\begin{figure}[h]
\begin{center}\mbox{
{\includegraphics[width=1.75in]{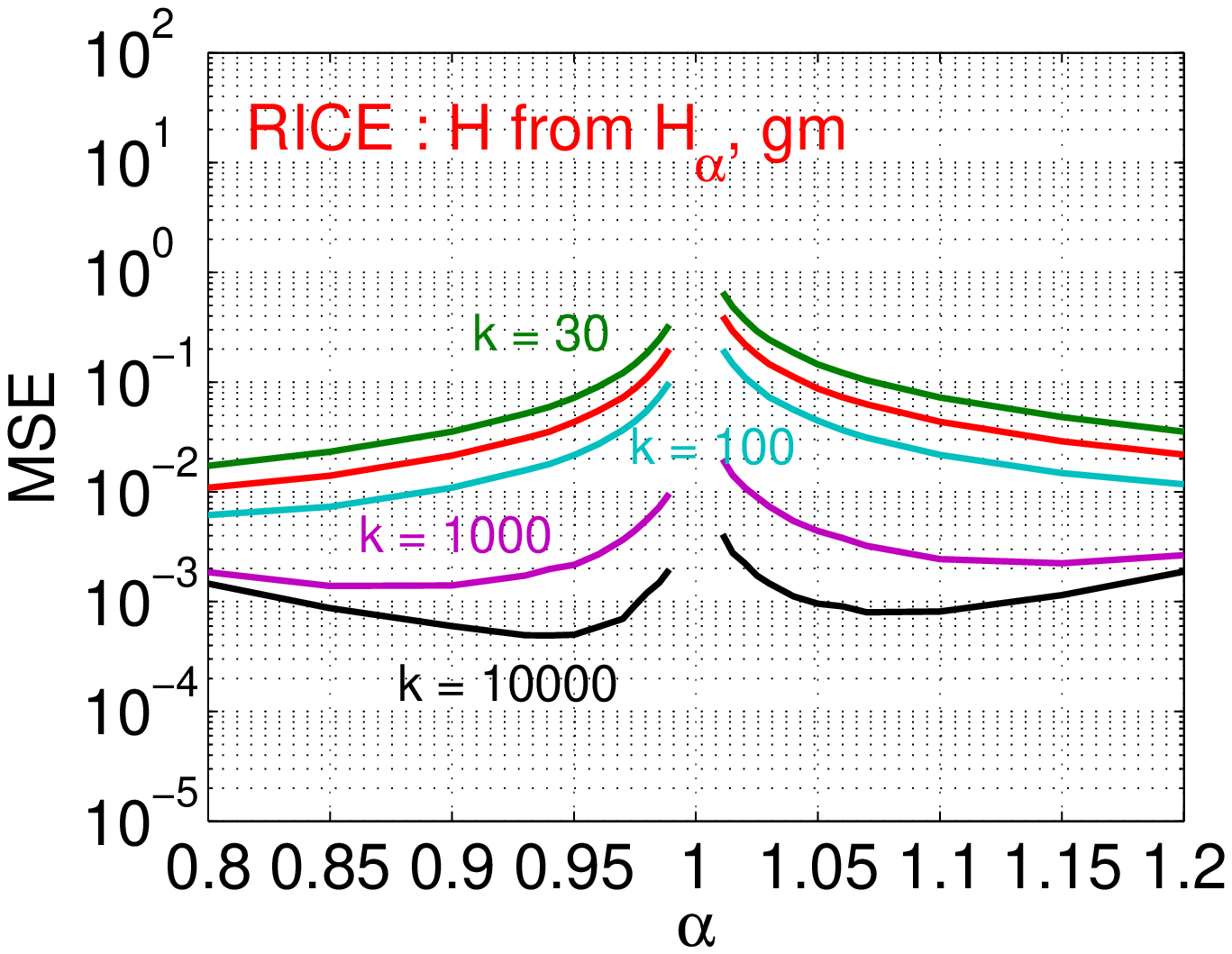}} \hspace{-0.1in}
{\includegraphics[width=1.75in]{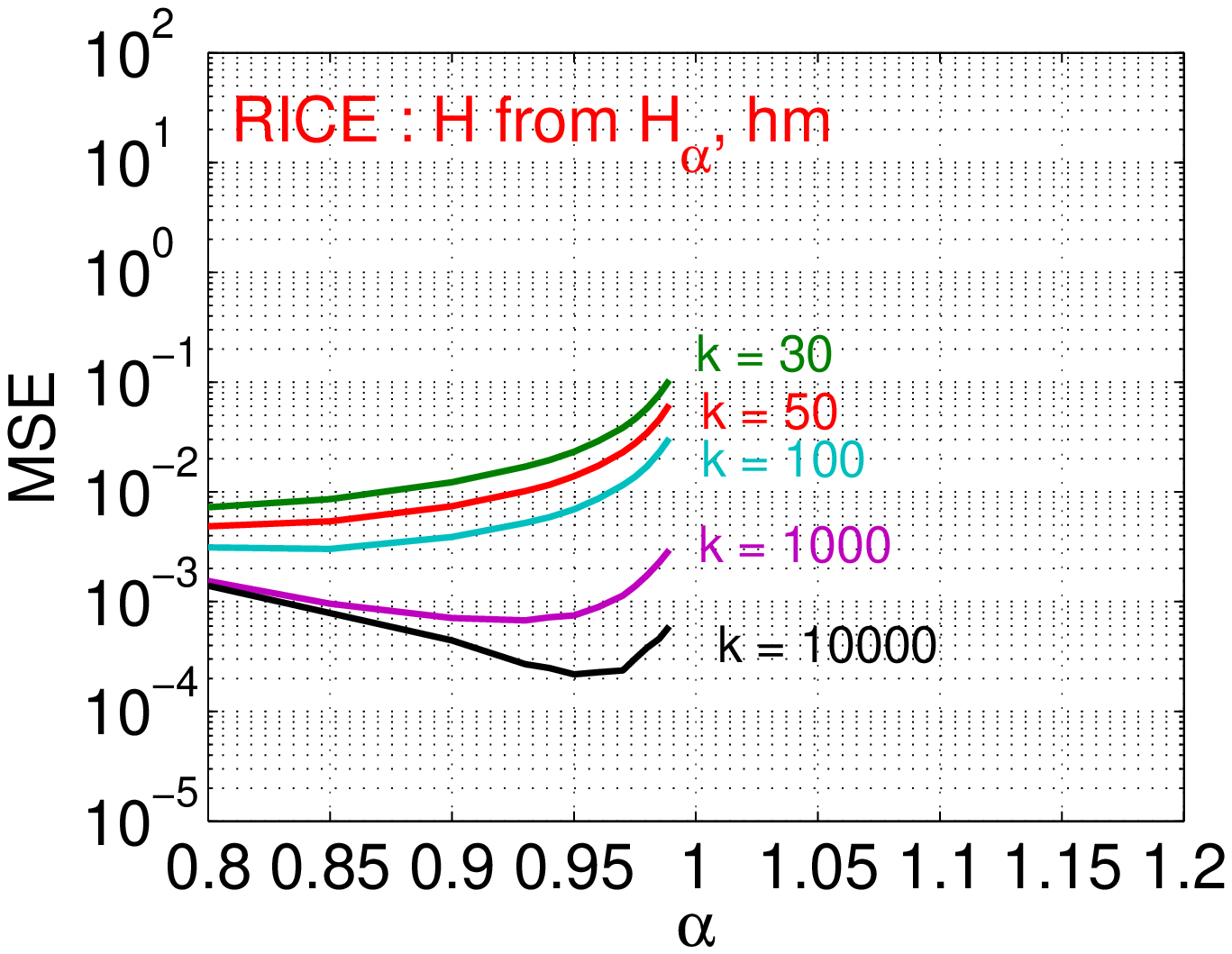}}}\vspace{-0.1in}
\mbox{
{\includegraphics[width=1.75in]{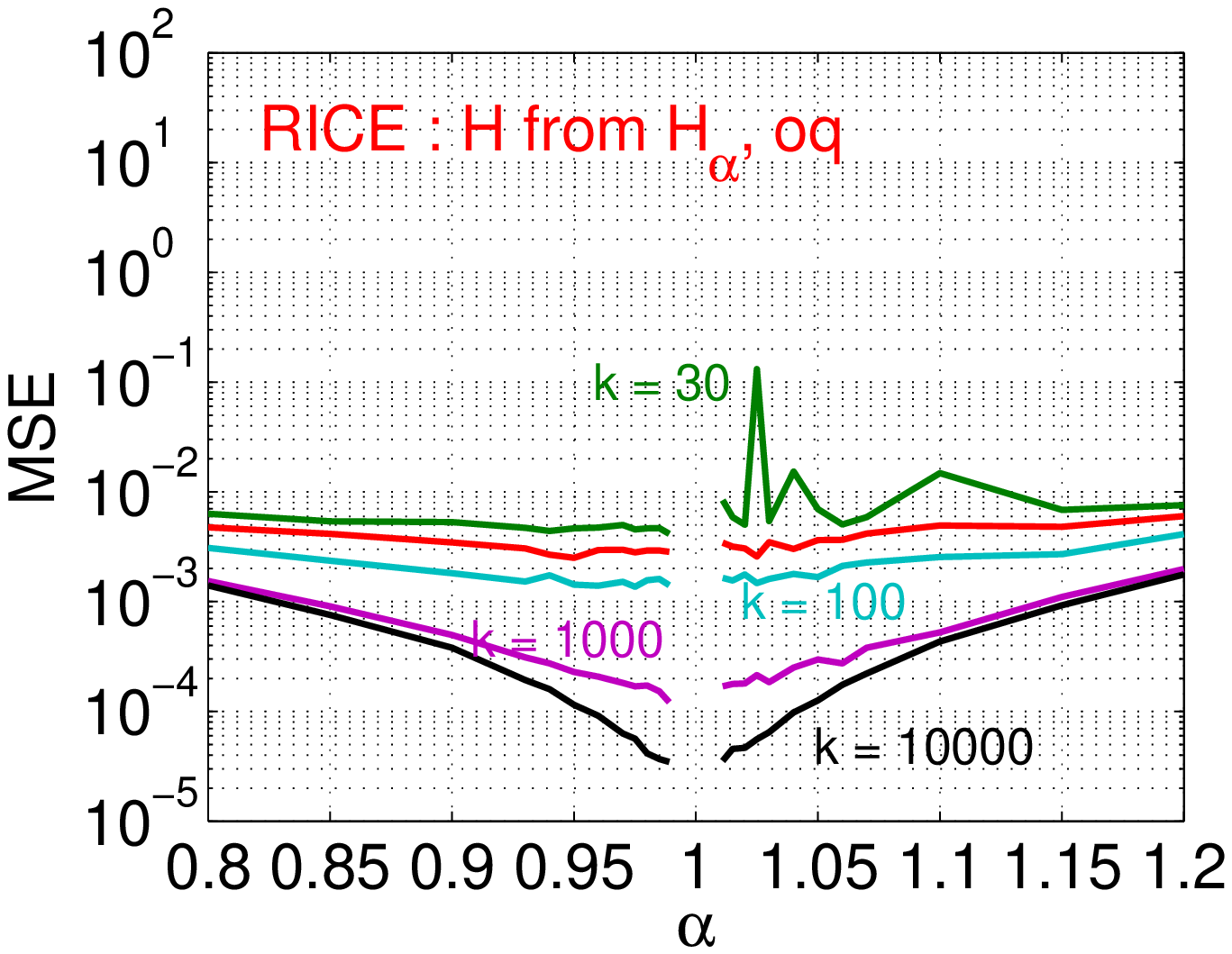}} \hspace{-0.1in}
{\includegraphics[width=1.75in]{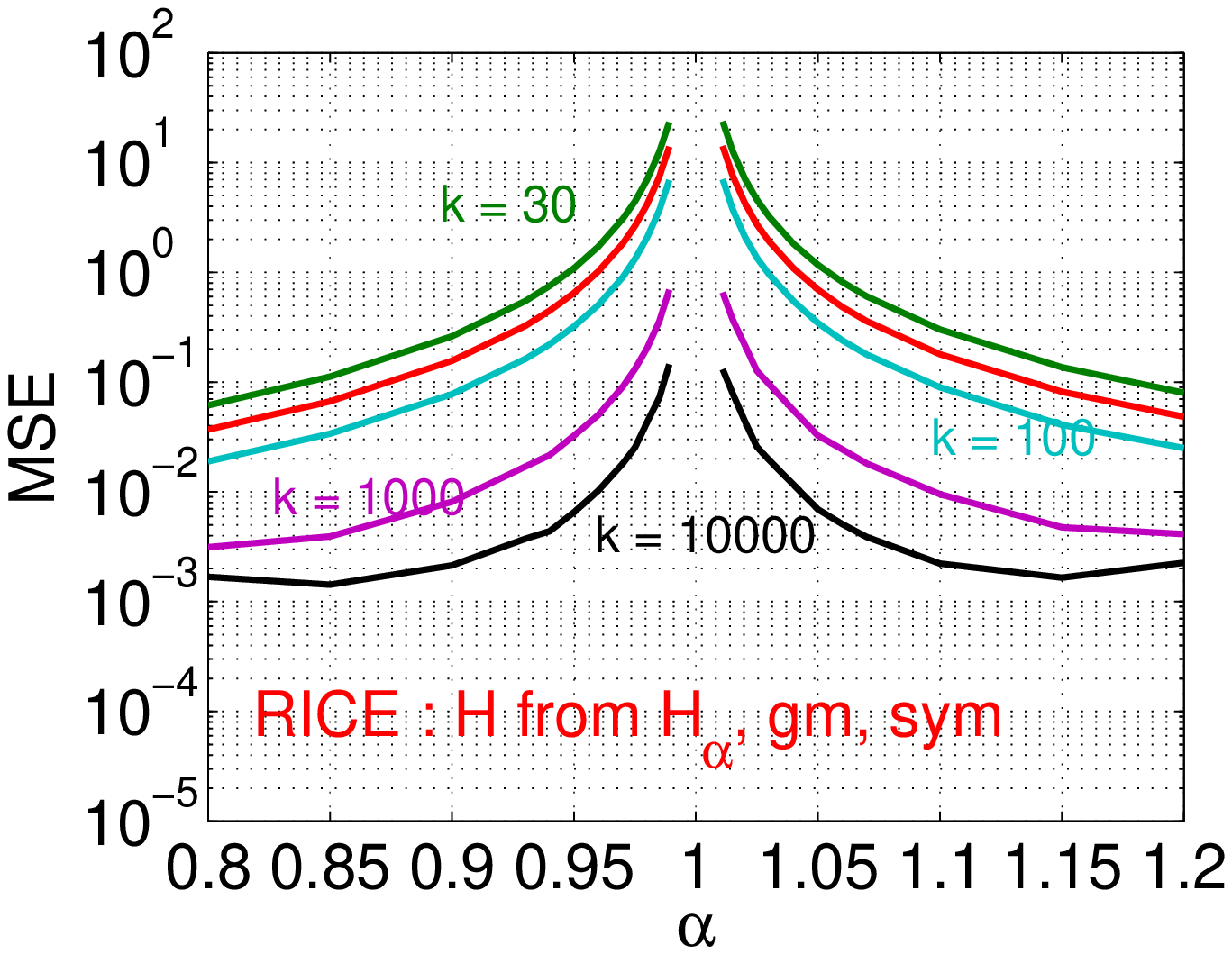}}\vspace{-0.1in}
}
\end{center}
\vspace{-0.25in}
\caption{Shannon entropy, $H$, estimated from R\'enyi entropy, $H_{\alpha}$, for RICE. Curves are the mean square errors (MSEs). }\label{fig_rice_HR}
\end{figure}

Figure \ref{fig_rice_HR} illustrates the MSEs from estimating the Shannon entropy using the R\'enyi entropy, for RICE.
\begin{itemize}
\item Using {\em symmetric stable random projections} with $\alpha=1+\delta$ and very small $|\delta|$ is not a good strategy and not practically feasible because the required sample size is enormous. For example, using $|\delta|\approx 0.01$, we need $k = 10000$ in order to achieve a relative MSE of $1\%$. 
\item There is clearly a variance-bias trade-off, especially for the geometric mean and harmonic mean estimator.  That is, for each $k$, there is an ``optimal'' $\alpha$ which achieves the smallest MSE.
\item Using the optimal quantile estimator does not show a strong variance-bias trade-off, because its has very small variance near $\alpha=1$ and its MSEs are mainly dominated by the (intrinsic) biases, $H_\alpha - H$.
\item The improvement of CC over {\em symmetric stable random projections} is very large when $\alpha$ is close 1. When $\alpha$ is away from 1, the improvement becomes less obvious because the MSEs are dominated by the biases.
\item
Using the optimal quantile estimator with $\alpha$ very close to 1 (preferably  $\alpha<1$) is our recommended procedure for estimating Shannon entropy from R\'enyi entropy.
\end{itemize}

For a fixed $\alpha$ and $k$, we can see that CC improves {\em symmetric stable random projections} enormously when $\alpha\rightarrow 1$.  If we follow the theoretical suggestion of \cite{Article:Harvey_entropy_arXiv08,Proc:Harvey_FOCS08} by using (e.g.) $\alpha = 1+10^{-4}$, then the improvement of CC over {\em symmetric stable random projections} will be enormous.

As a practical recommendation, we do not suggest letting $\alpha$ too close to 1 when using {\em symmetric stable random projections}. Instead, one should take advantage of the variance-bias trade-off by using $\alpha$ away from 1. There will be an ``optimal'' $\alpha$ that attains the smallest mean square error (MSE), at each $k$.

As illustrated in  Figure \ref{fig_rice_HR}, CC is not affected much by the variance-bias trade-off and it is preferable to choose $\alpha$  close to 1 when using the optimal quantile estimator. Therefore, we will present the comparisons mainly in terms of the minimum MSEs (i.e., best achievable performance), which we believe actually heavyily favors {\em symmetric stable random projections}.

\clearpage

\begin{figure}[h]
\begin{center}\mbox{
{\includegraphics[width=1.75in]{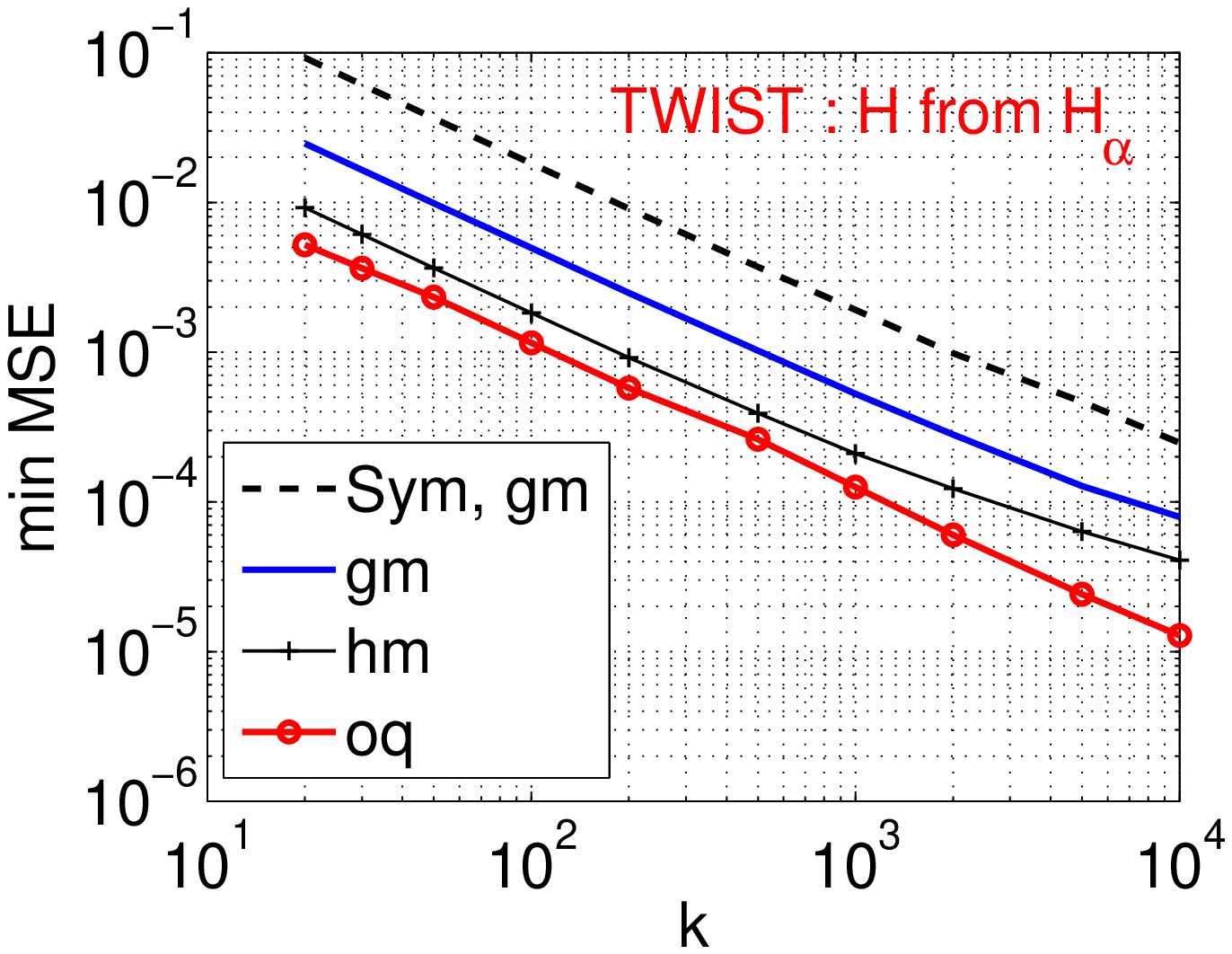}} \hspace{-0.1in}
{\includegraphics[width=1.75in]{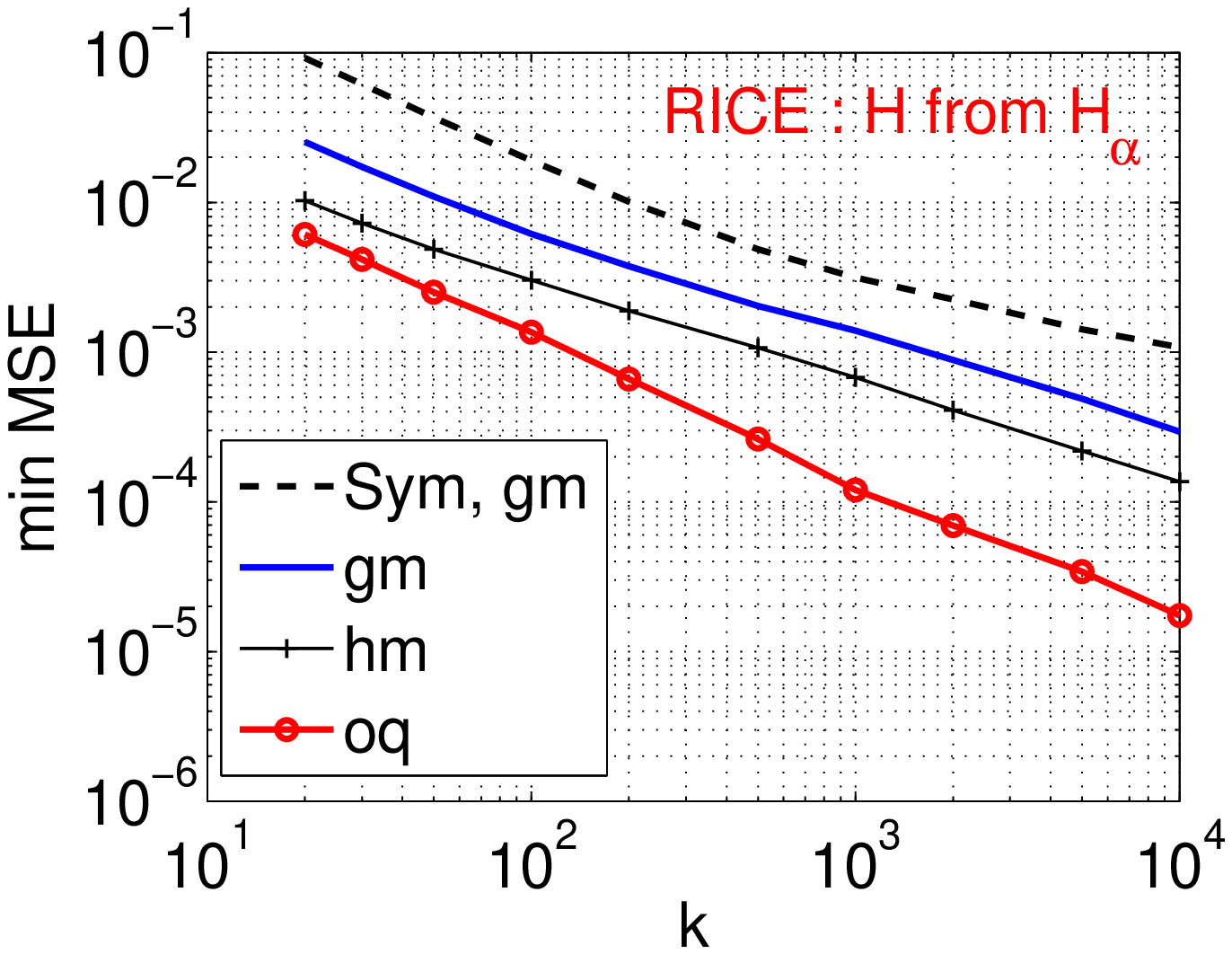}}}
\mbox{
{\includegraphics[width=1.75in]{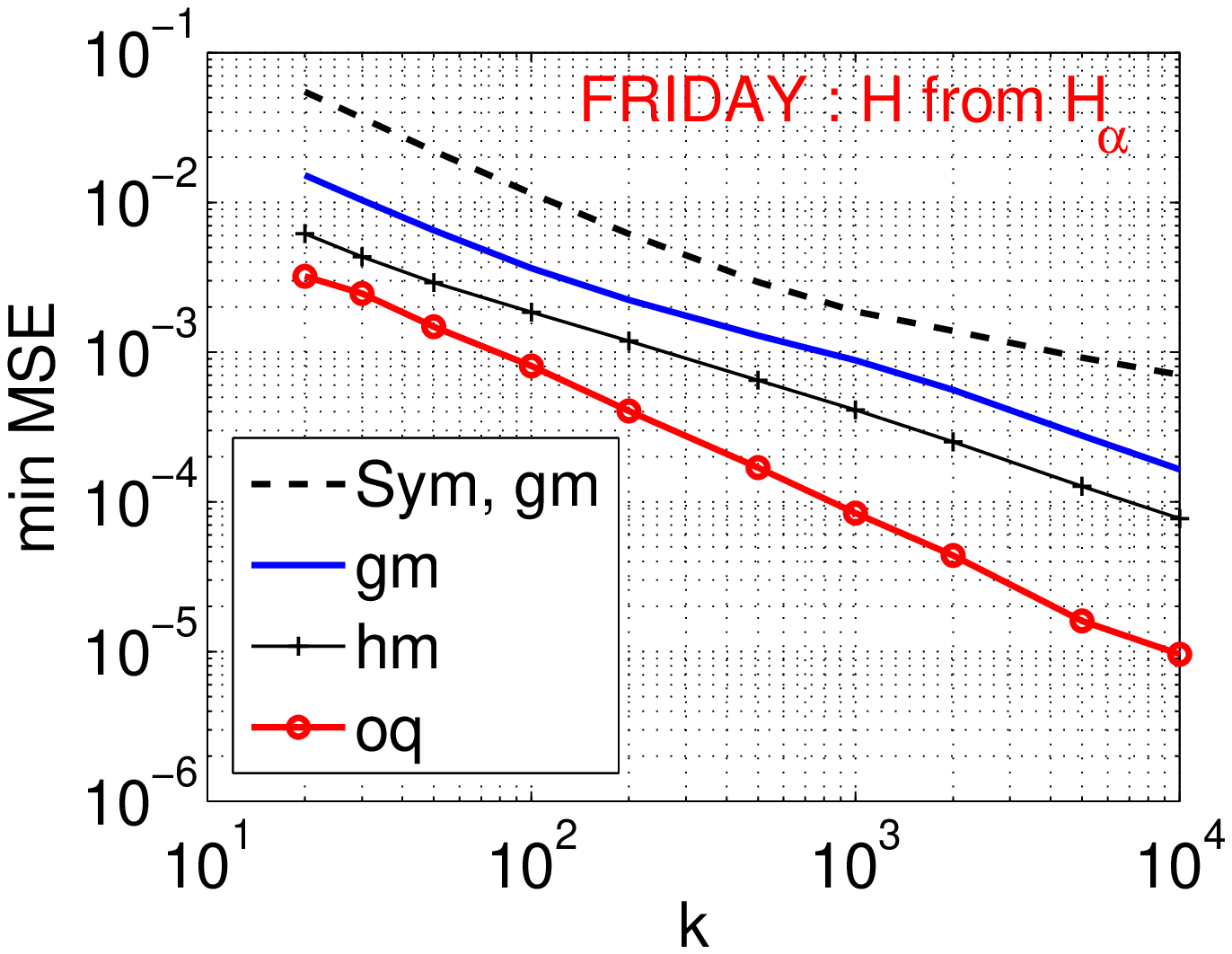}} \hspace{-0.1in}
{\includegraphics[width=1.75in]{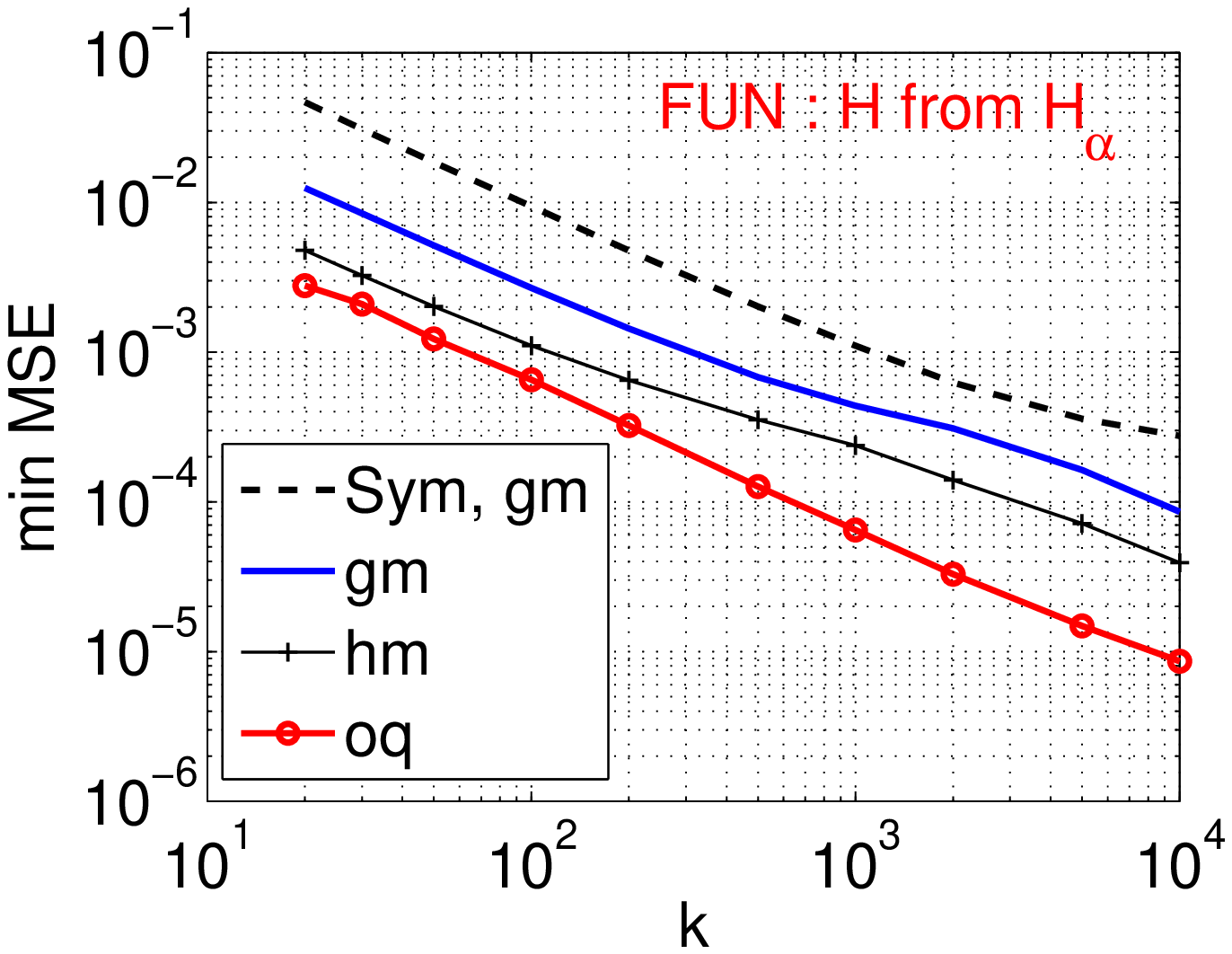}}}
\mbox{
{\includegraphics[width=1.75in]{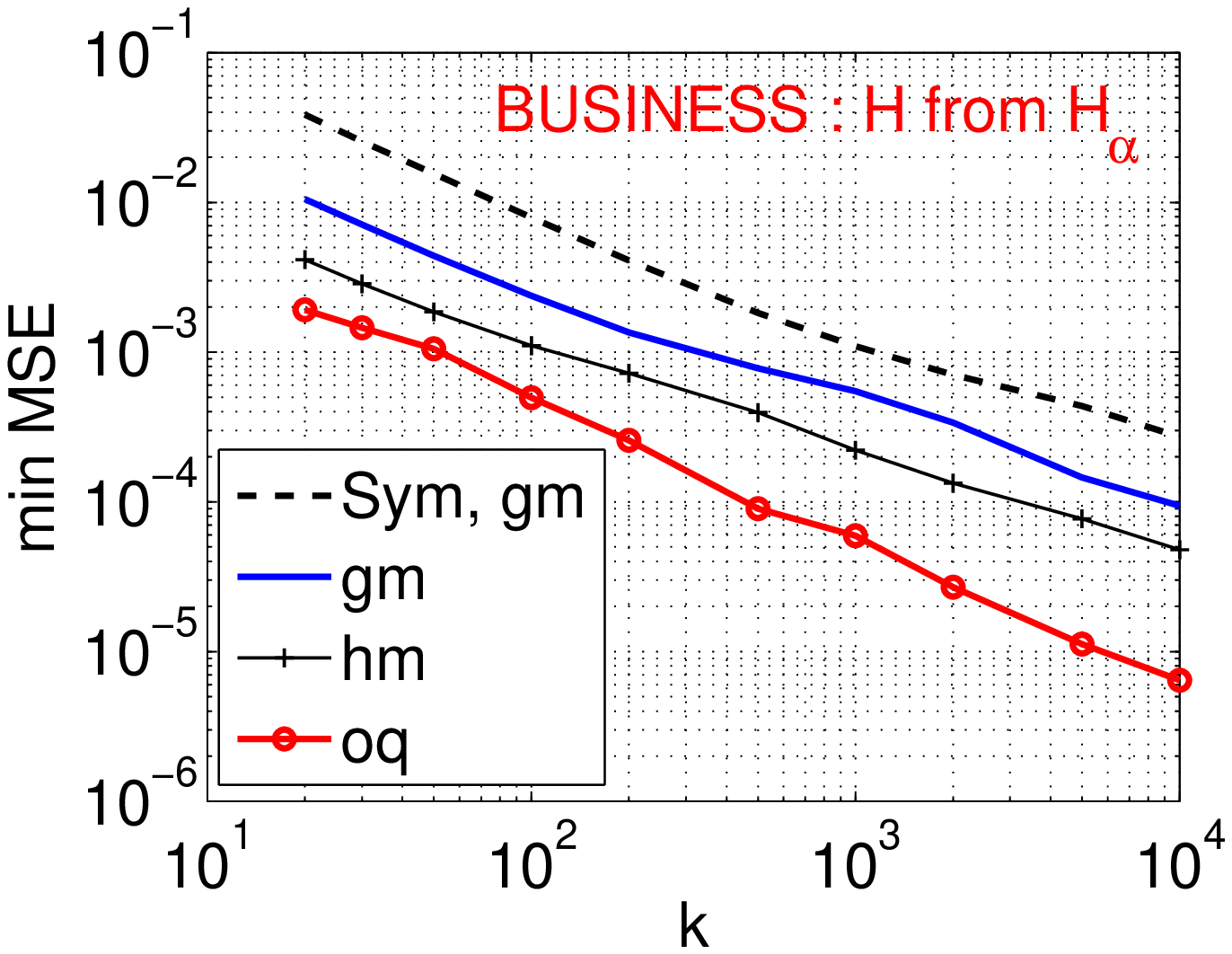}} \hspace{-0.1in}
{\includegraphics[width=1.75in]{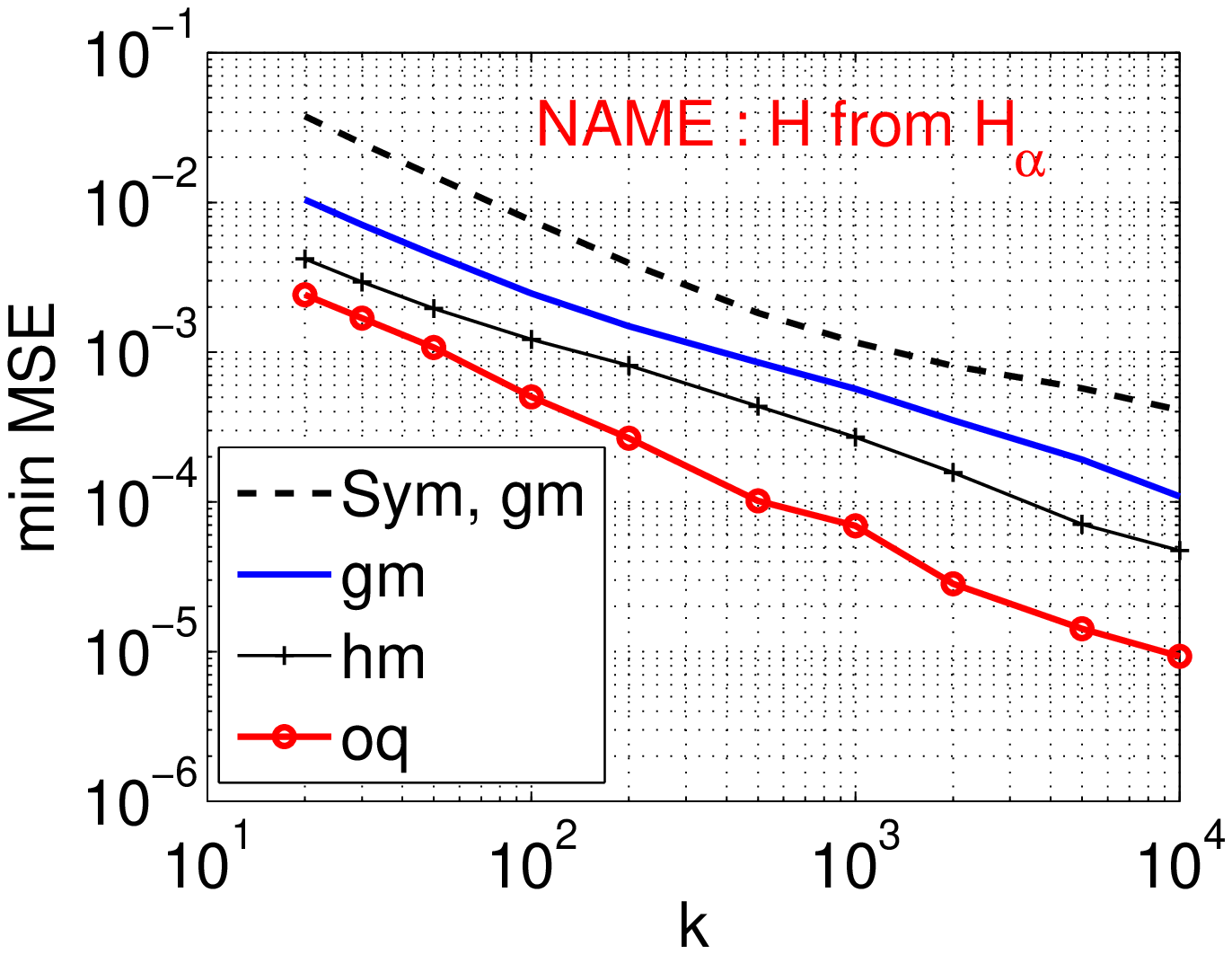}}}
\mbox{
{\includegraphics[width=1.75in]{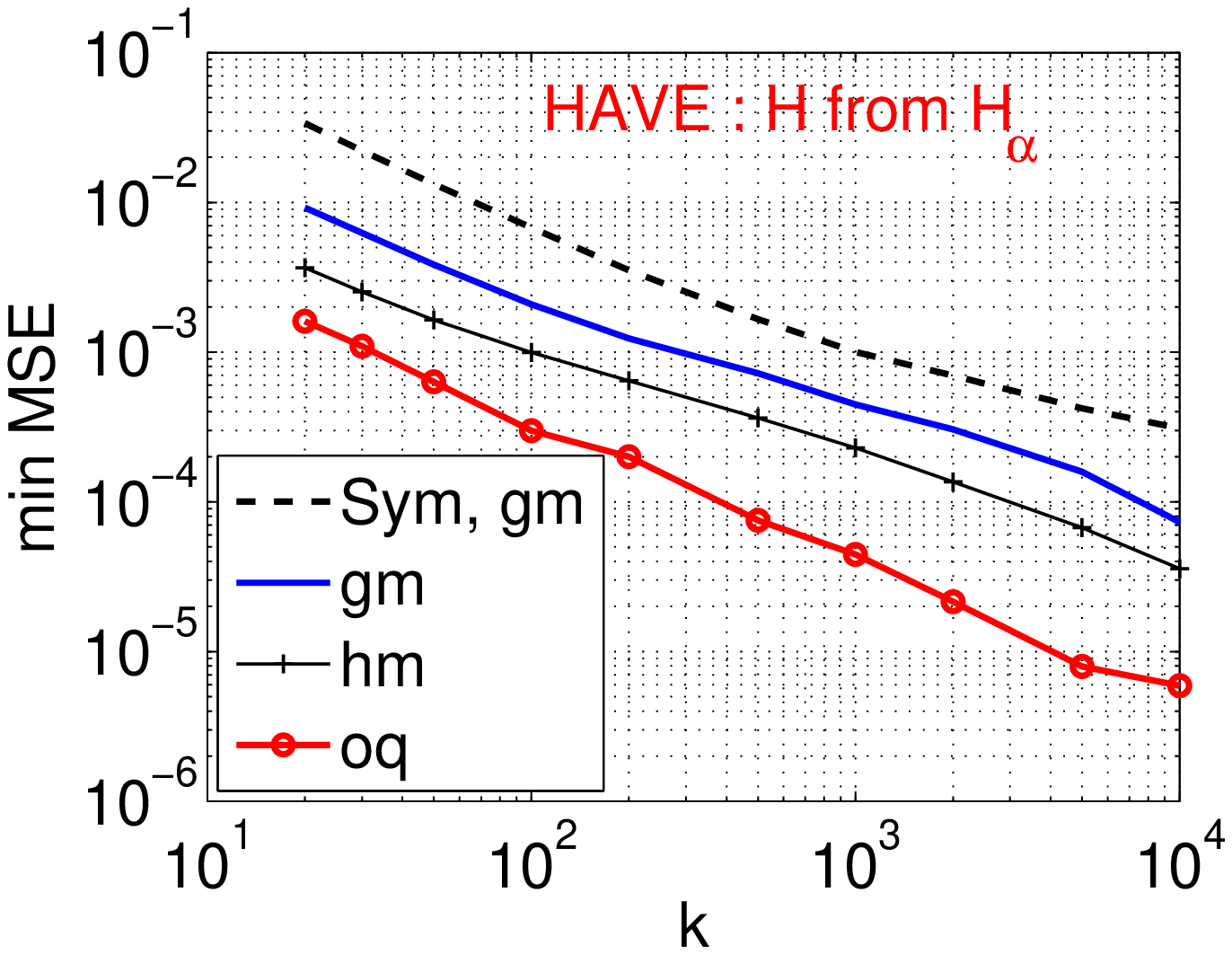}} \hspace{-0.1in}
{\includegraphics[width=1.75in]{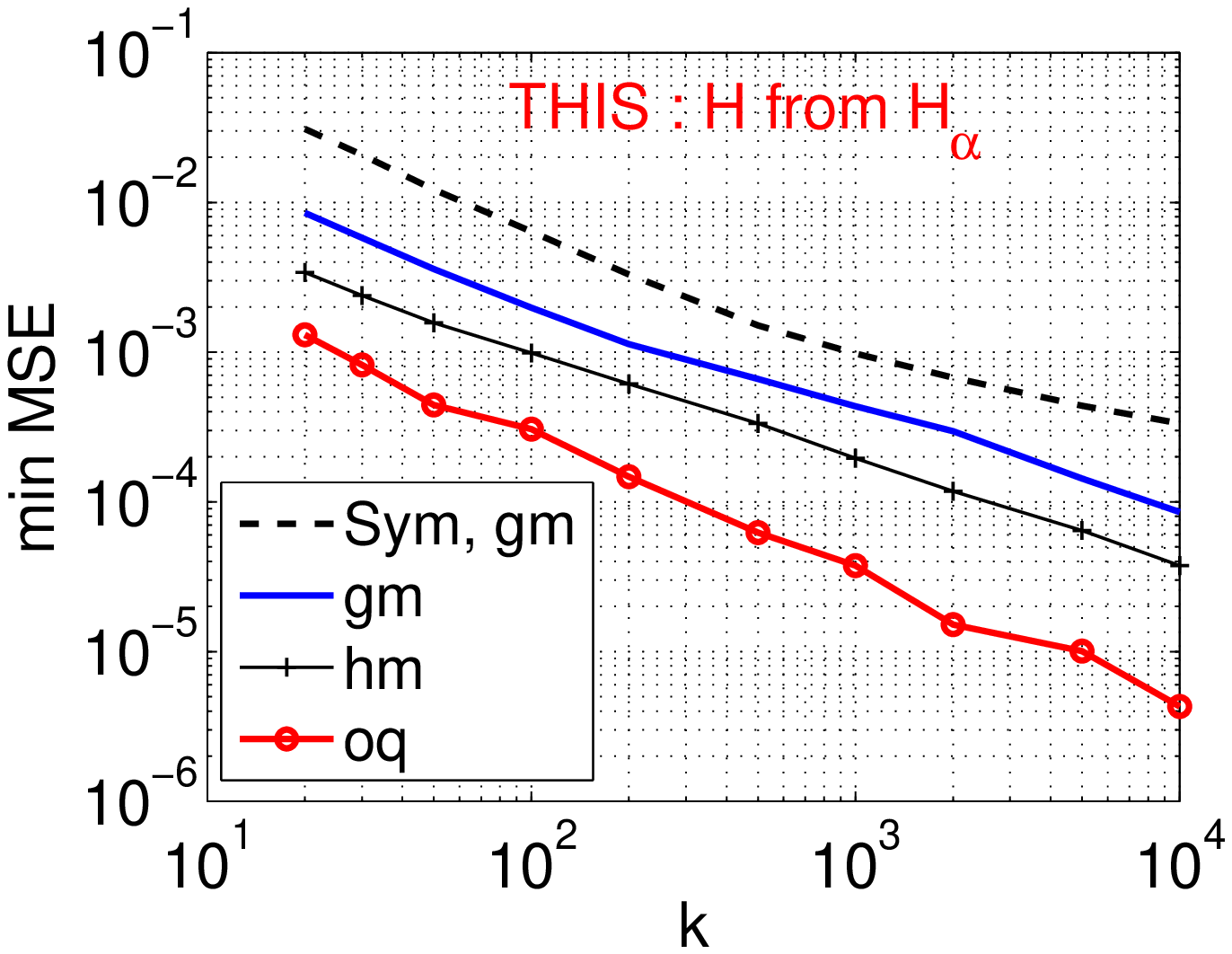}}
}
\mbox{
{\includegraphics[width=1.75in]{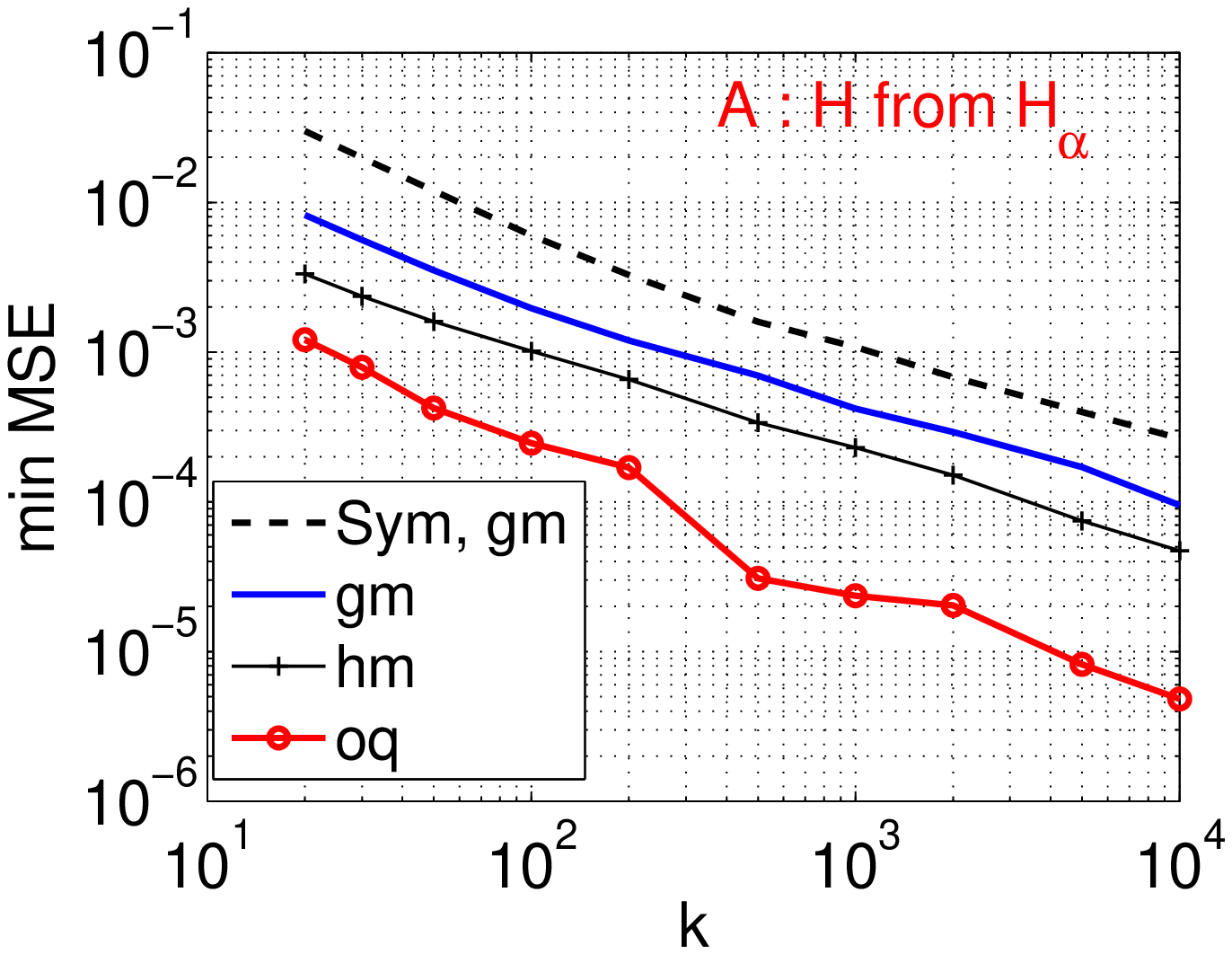}} \hspace{-0.1in}
{\includegraphics[width=1.75in]{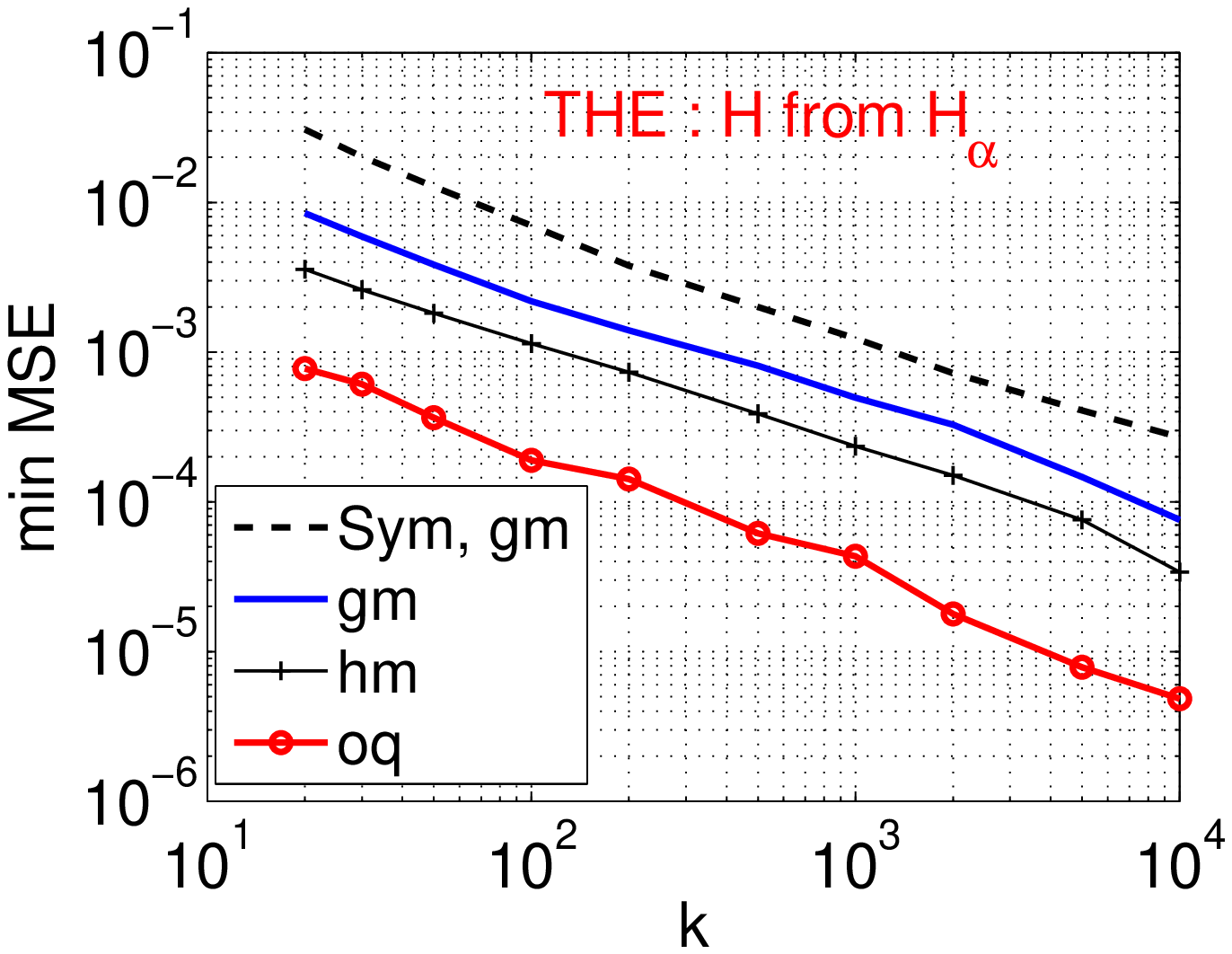}}
}
\end{center}
\vspace{-0.15in}
\caption{Shannon entropy, $H$, estimated from R\'enyi  entropy, $T_\alpha$, for 10 words in Table \ref{tab_data}. Curves are the minimum MSEs at each $k$.  }\label{fig_HR_min}
\end{figure}

Figures \ref{fig_HR_min} presents the minimum MSEs for all 10 words:
\begin{itemize}
\item The optimal quantile estimator is the most accurate. For example, using $k = 20$, the relative MSE is only less than $1\%$ (or even $0.1\%$), which may be already accurate enough for some applications.
\item For every $k$, CC reduces the (minimum) MSE roughly by 20- to 50-fold, compared to {\em symmetric stable random projections}. This is comparing the curves in the vertical direction.
\item To achieve the same accuracy as {\em symmetric stable random projections}, CC requires a much smaller $k$, a reduction by about 50-fold (using the optimal quantile estimator). This is comparing the curves in the horizontal direction.
\item The results are quite similar for all 10 words. While it is boring to present all 10 words, the results deliver a strong hint that the performance of CC and its improvement over {\em symmetric stable random projections} should hold
    universally, not just for these 10 words.
\end{itemize}

\subsubsection{Estimating Shannon Entropy\\ from Tsallis Entropy}

\begin{figure}[h]
\begin{center}\mbox{
{\includegraphics[width=1.75in]{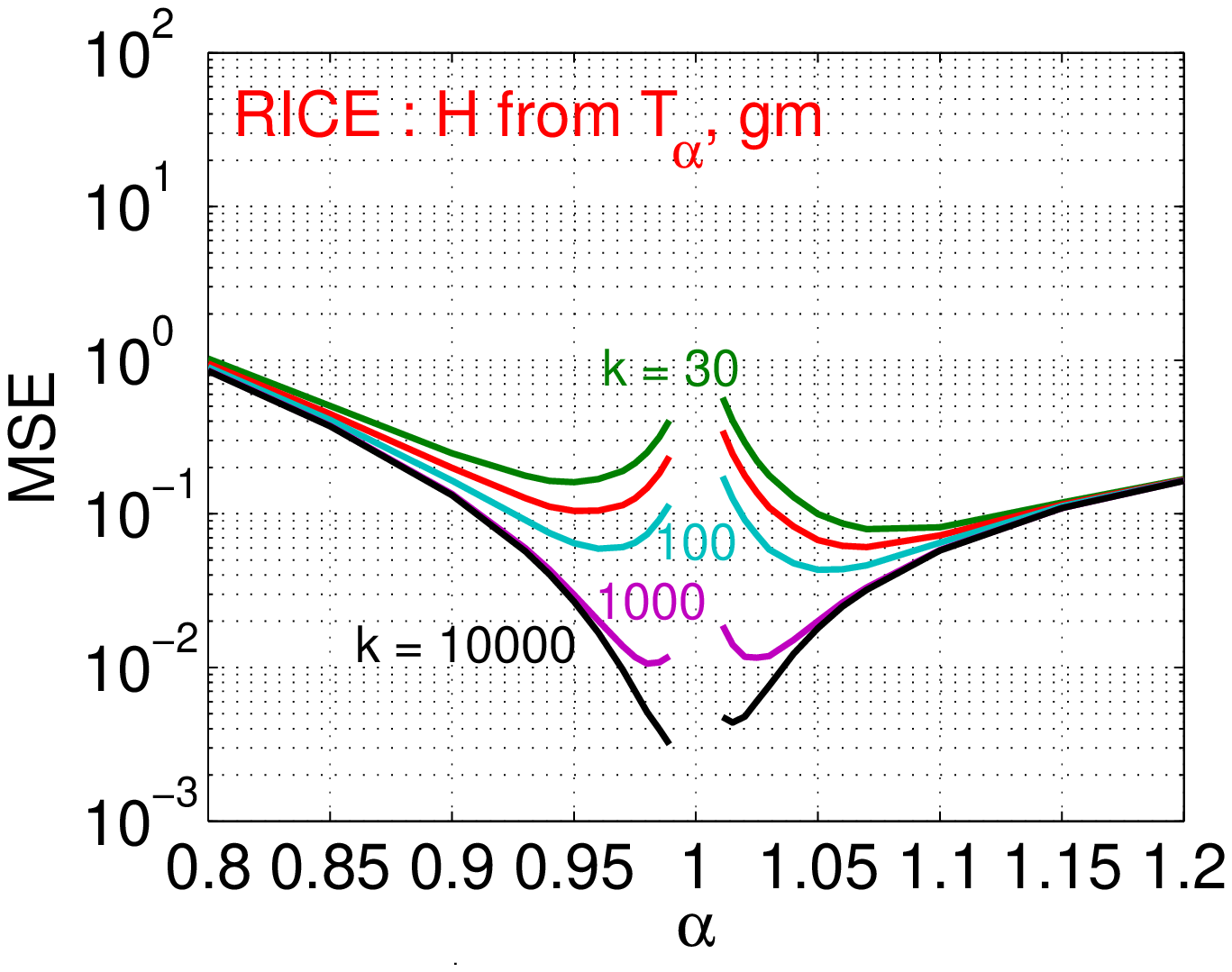}} \hspace{-0.1in}
{\includegraphics[width=1.75in]{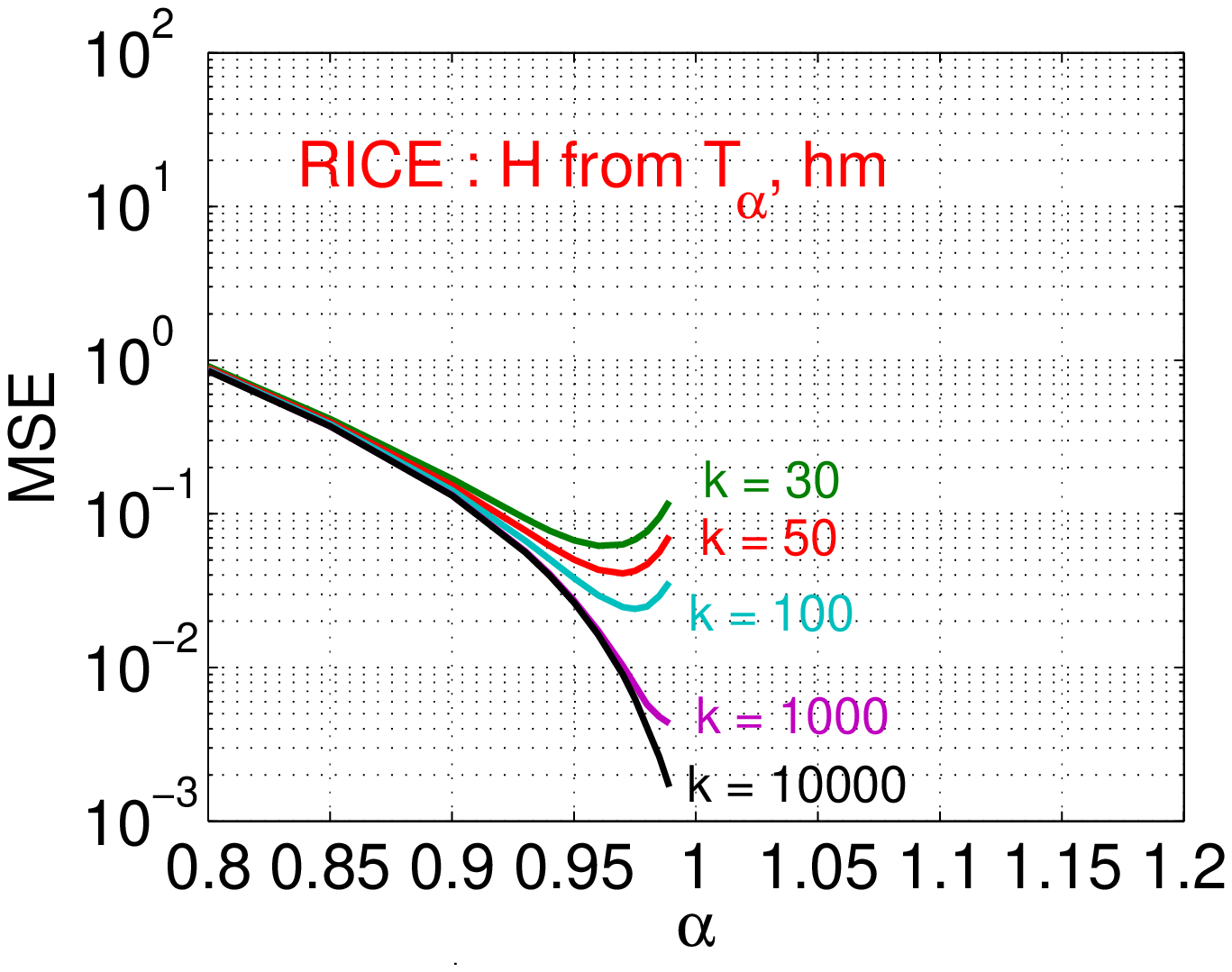}}}
\mbox{
{\includegraphics[width=1.75in]{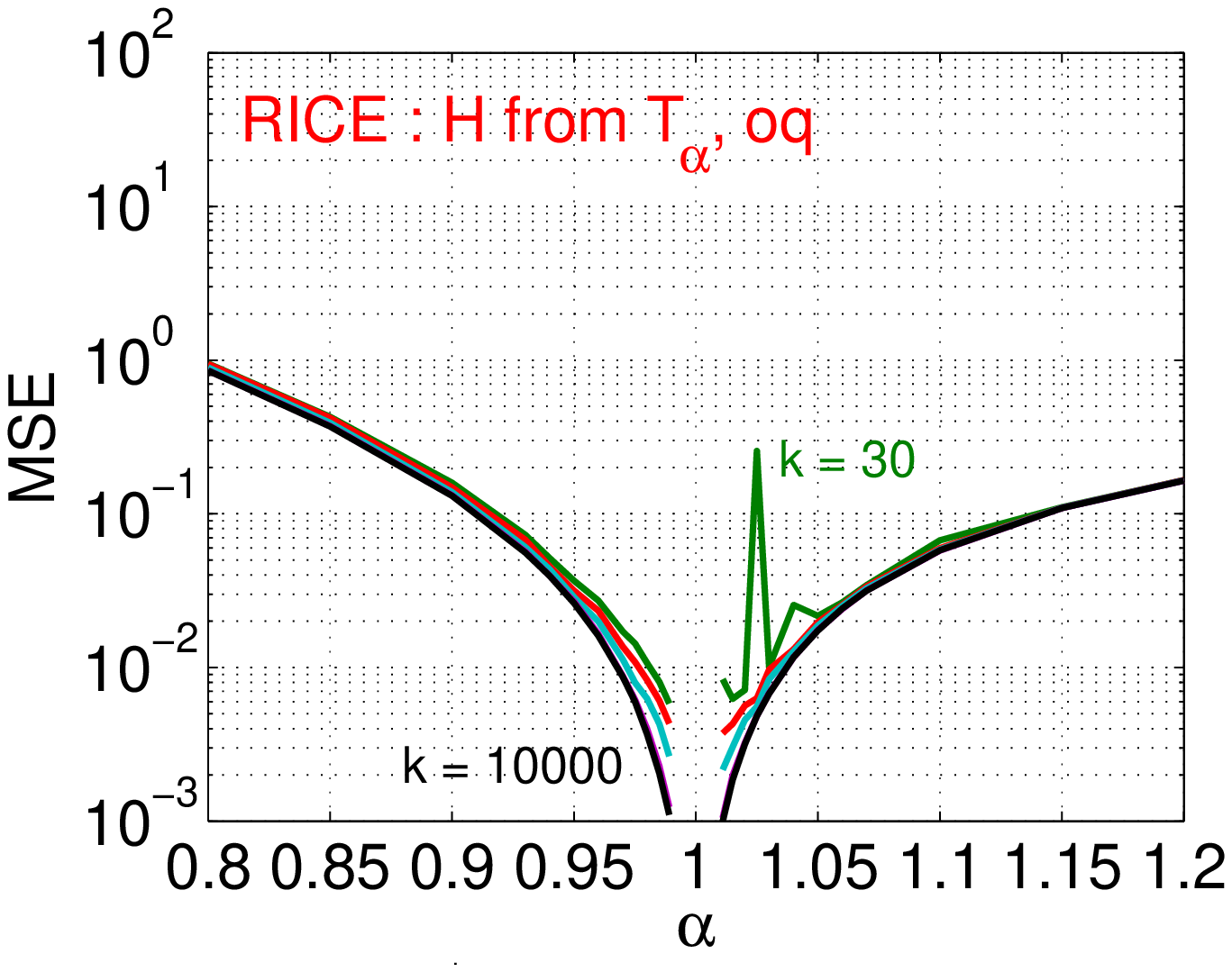}} \hspace{-0.1in}
{\includegraphics[width=1.75in]{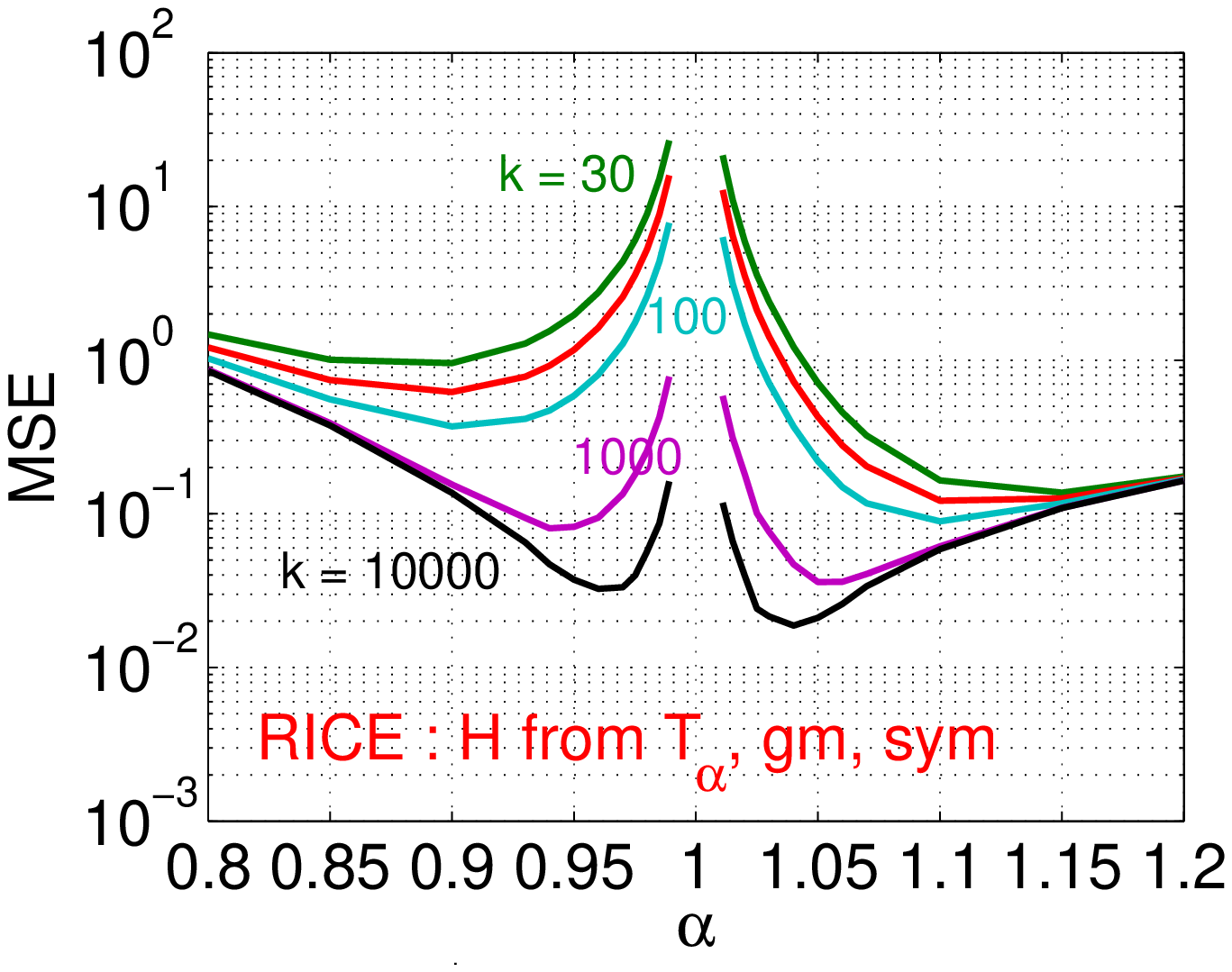}}
}
\end{center}
\vspace{-0.15in}
\caption{ Shannon entropy, $H$, estimated from Tsallis entropy, $H_{\alpha}$, for RICE. Curves are MSEs.}\label{fig_rice_HT}
\end{figure}

Figure \ref{fig_rice_HT} illustrates the MSEs from estimating Shannon entropy using Tsallis entropy, for RICE:
\begin{itemize}
\item Using {\em symmetric stable random projections} with $\alpha=1+\delta$ and very small $|\delta|$ is not a good strategy and not practically feasible. For example, when $|\delta|\approx 0.01$, using $k = 10000$ can only achieve a relative MSE of $10\%$.
\item The effect of the variance-bias trade-off for geometric mean and harmonic mean estimators, is even more significant, because the (intrinsic) bias $T_\alpha - H$ is large, as reported in Table \ref{tab_data}
\item The MSEs of the optimal quantile estimator is not affected much by $k$, because its variance is negligible compared to the (intrinsic) bias.
\end{itemize}

\begin{figure}[h]
\begin{center}\mbox{
{\includegraphics[width=1.75in]{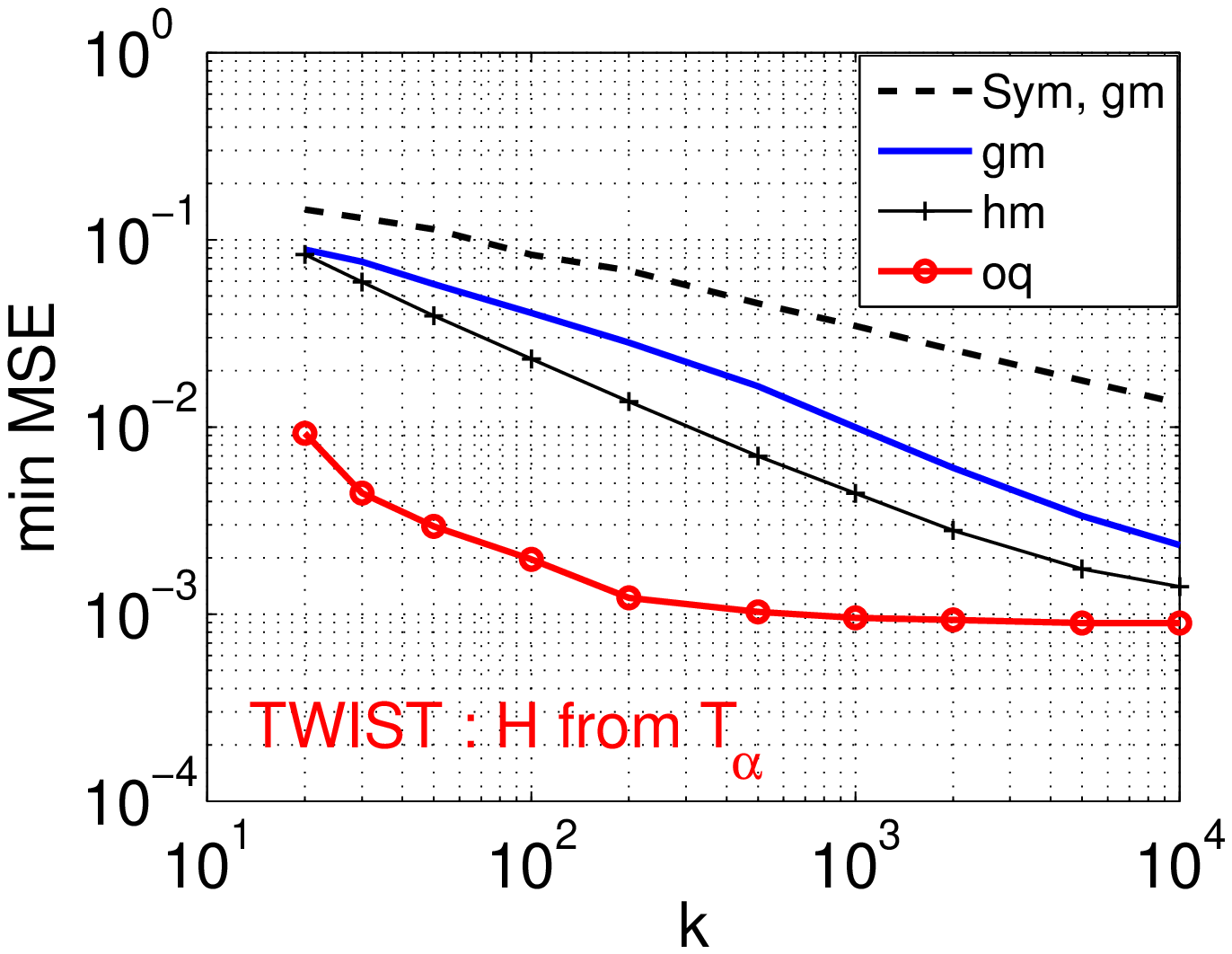}} \hspace{-0.1in}
{\includegraphics[width=1.75in]{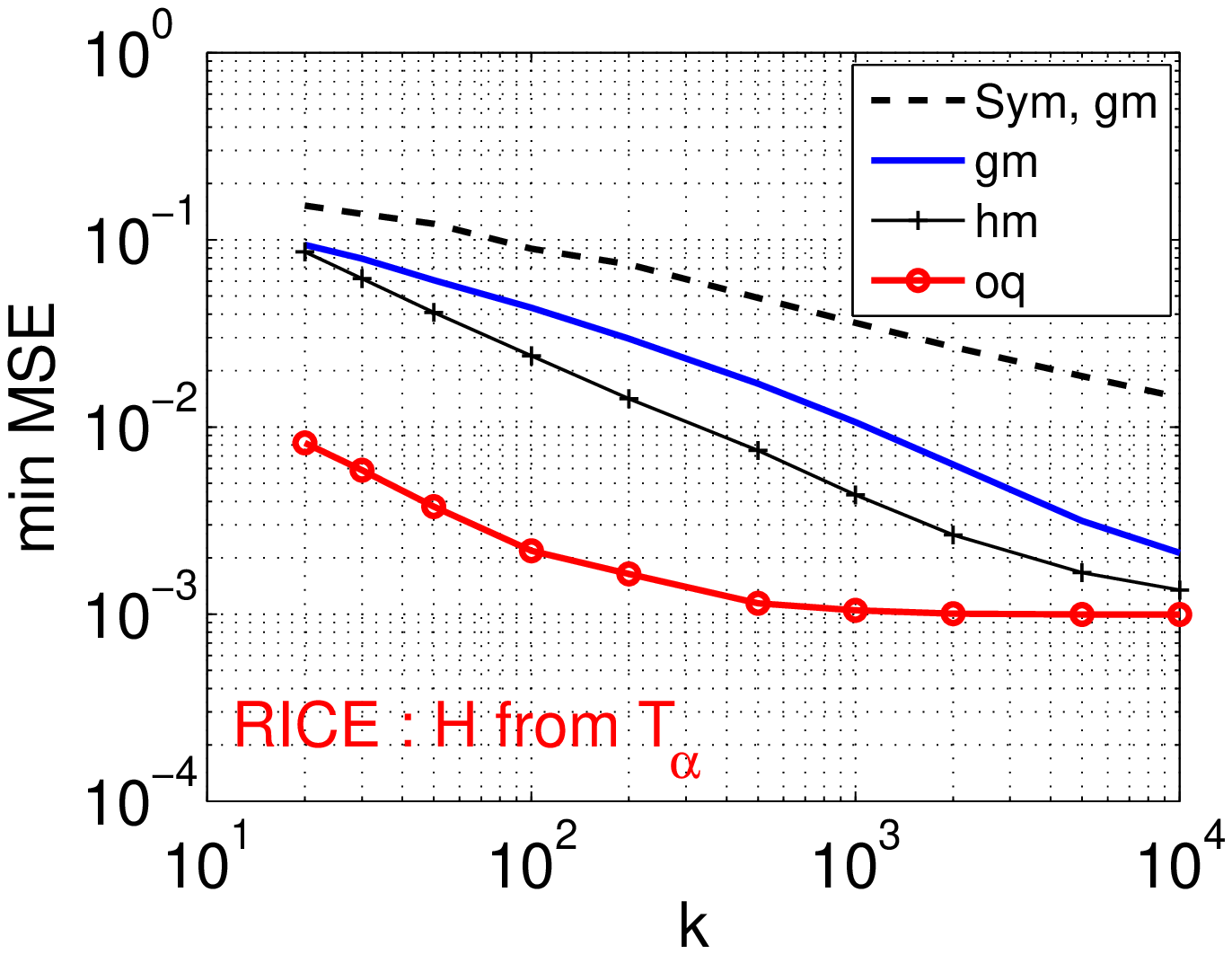}}}
\mbox{
{\includegraphics[width=1.75in]{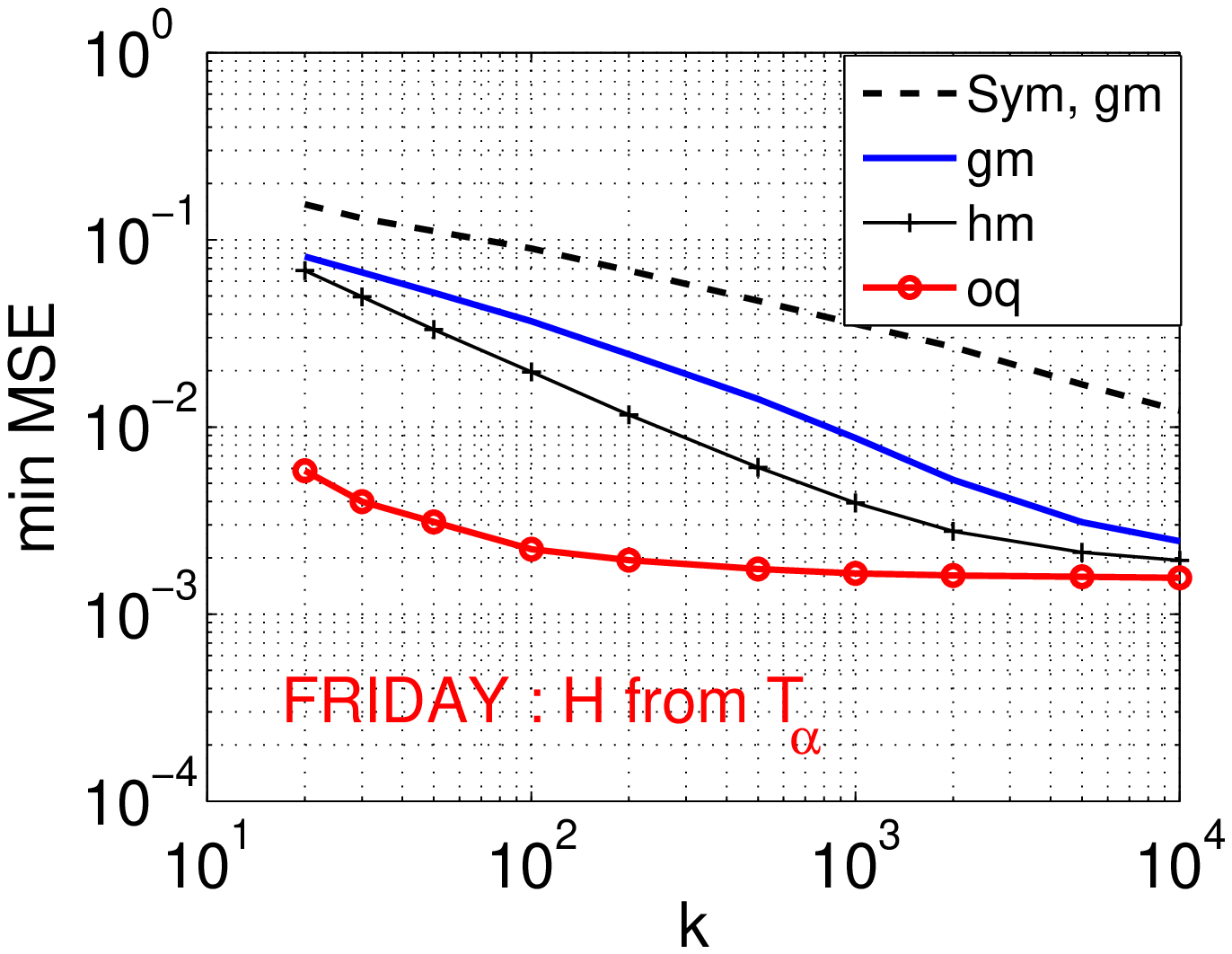}} \hspace{-0.1in}
{\includegraphics[width=1.75in]{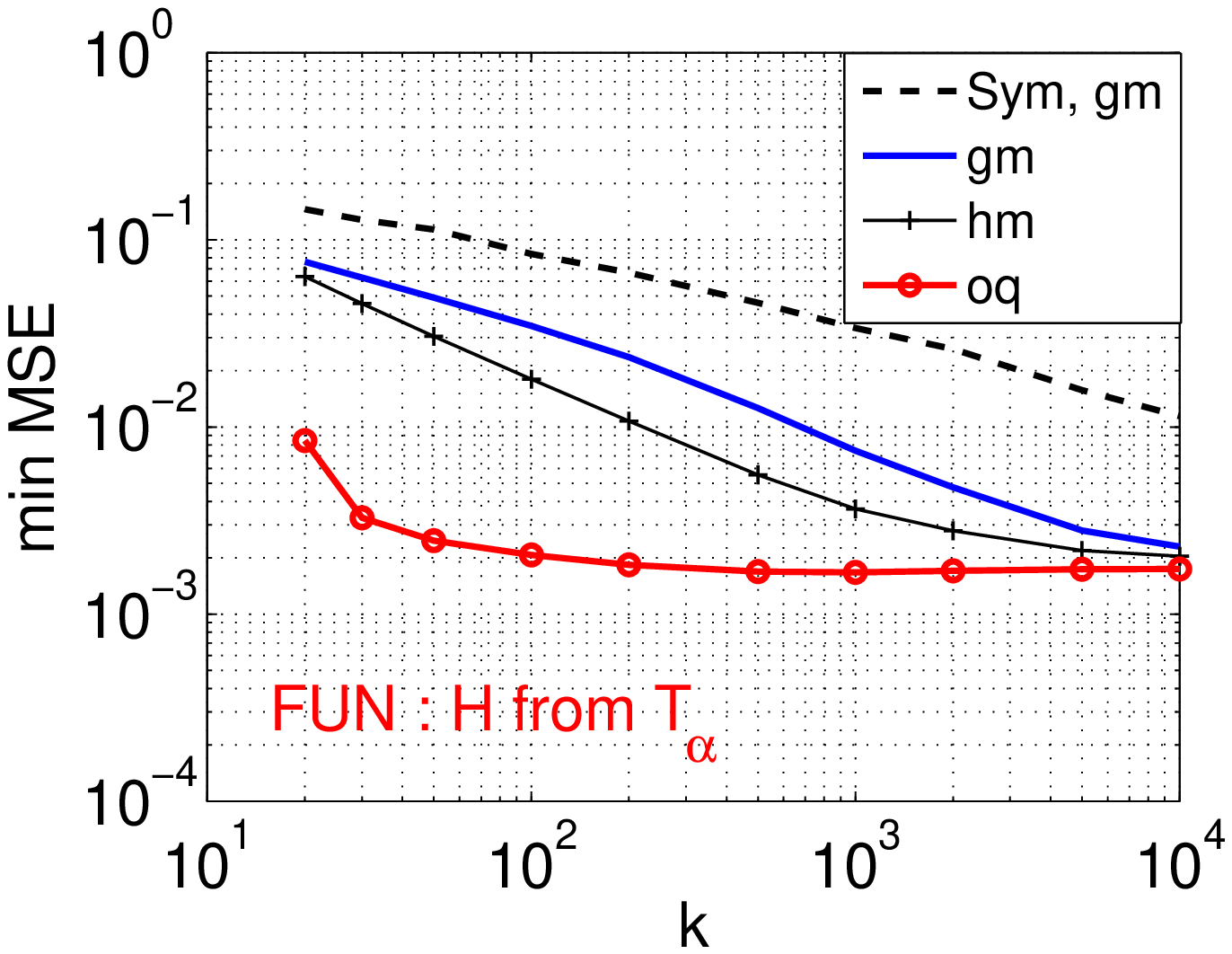}}
}\\
\mbox{
{\includegraphics[width=1.75in]{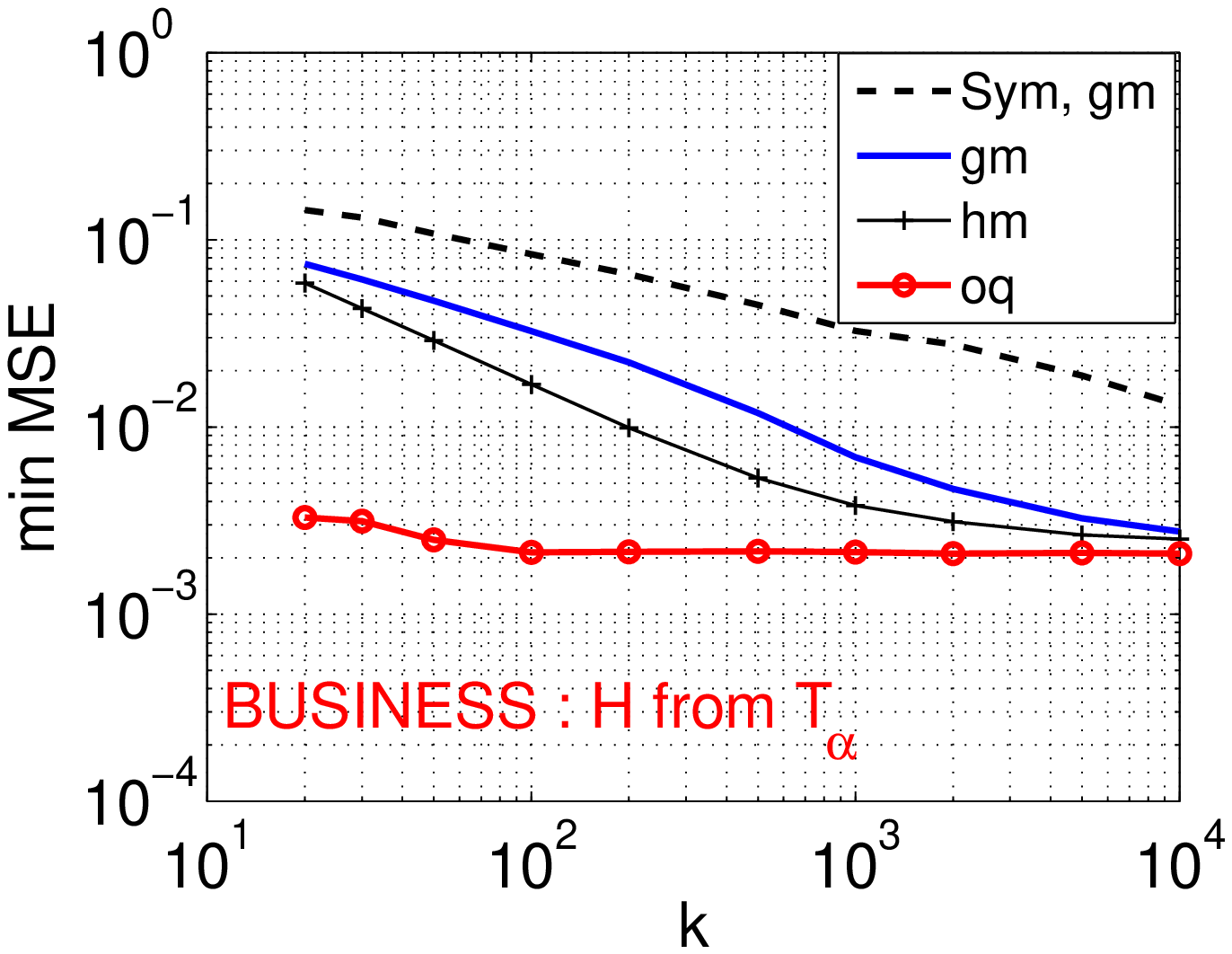}} \hspace{-0.1in}
{\includegraphics[width=1.75in]{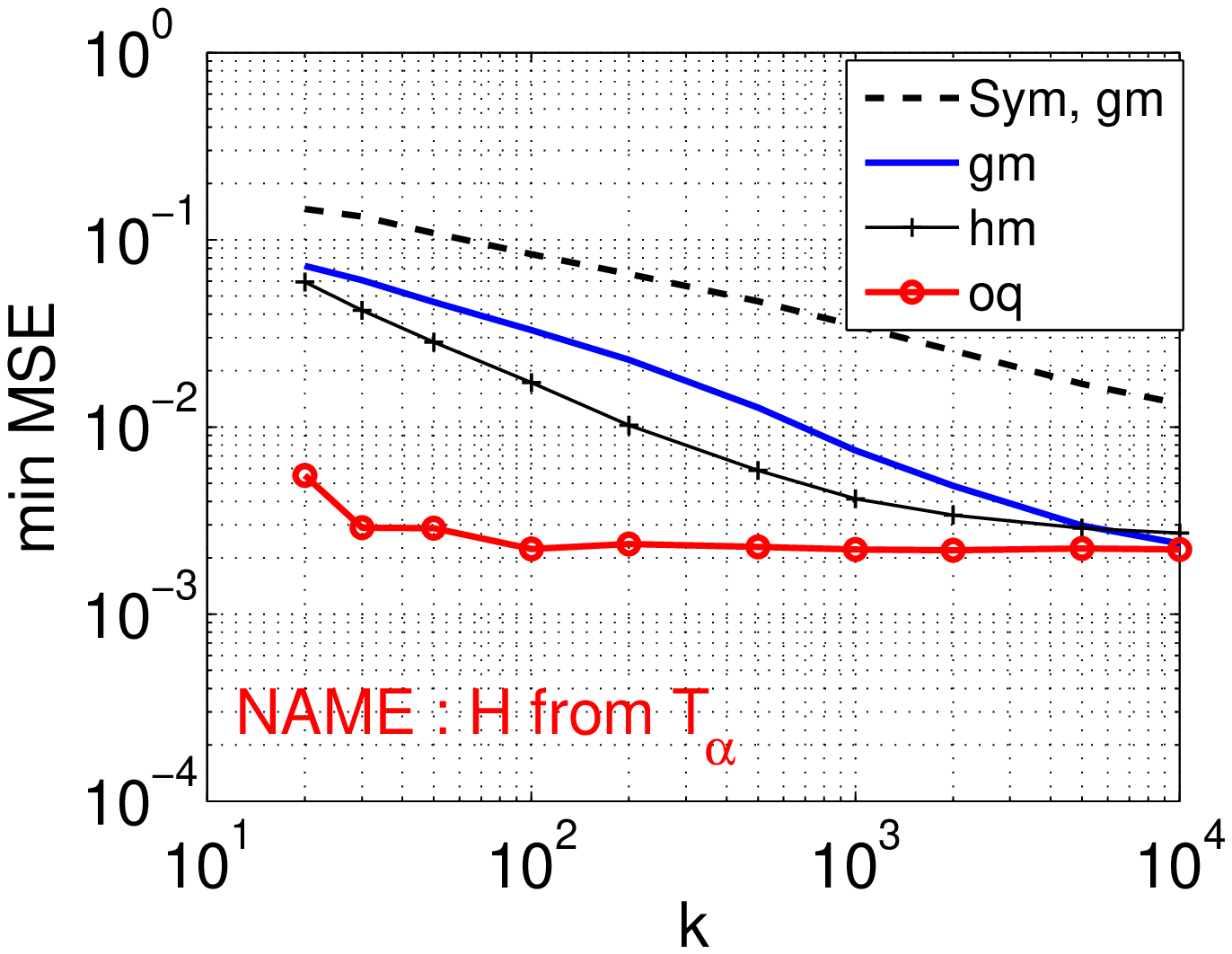}}}
\mbox{
{\includegraphics[width=1.75in]{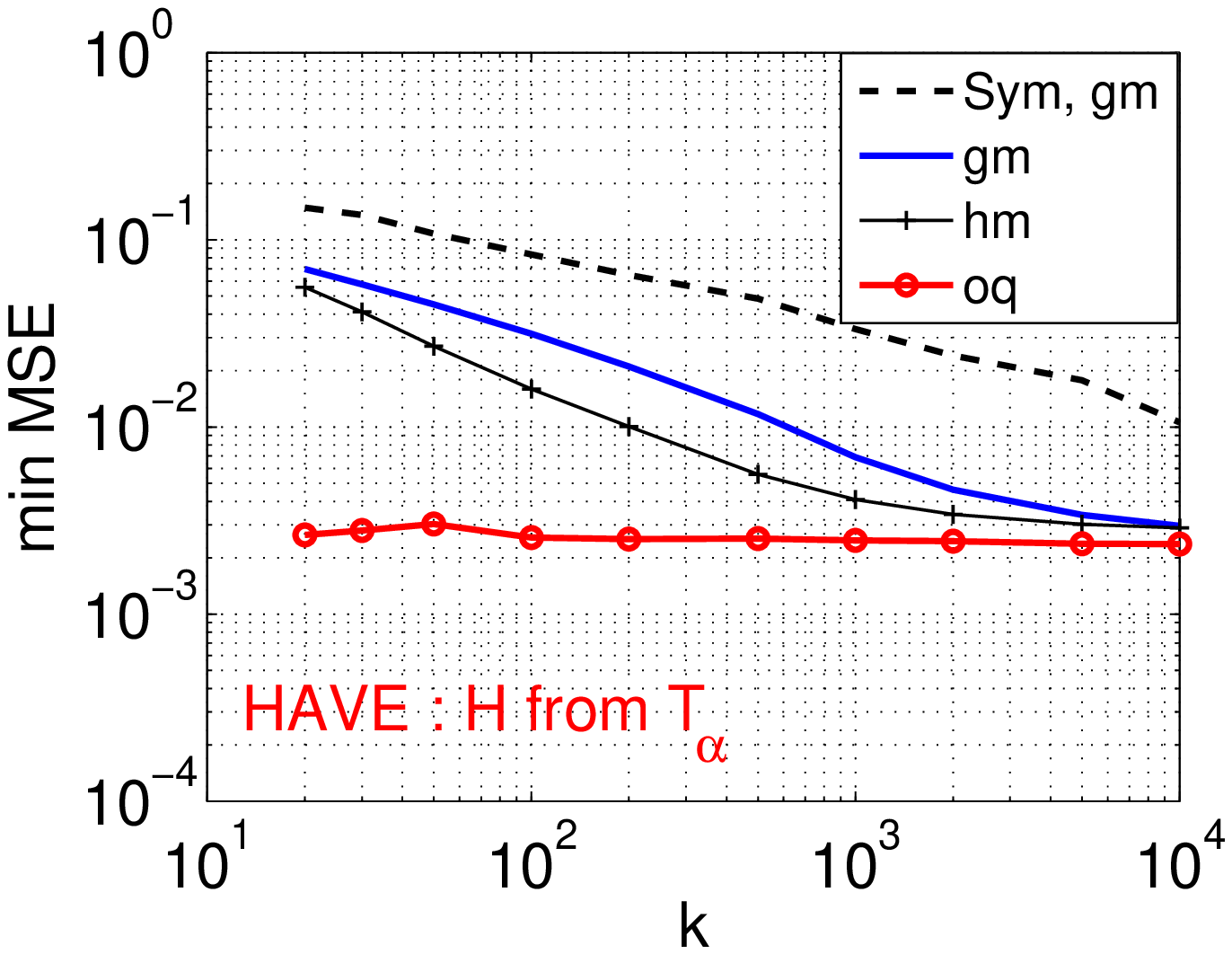}} \hspace{-0.1in}
{\includegraphics[width=1.75in]{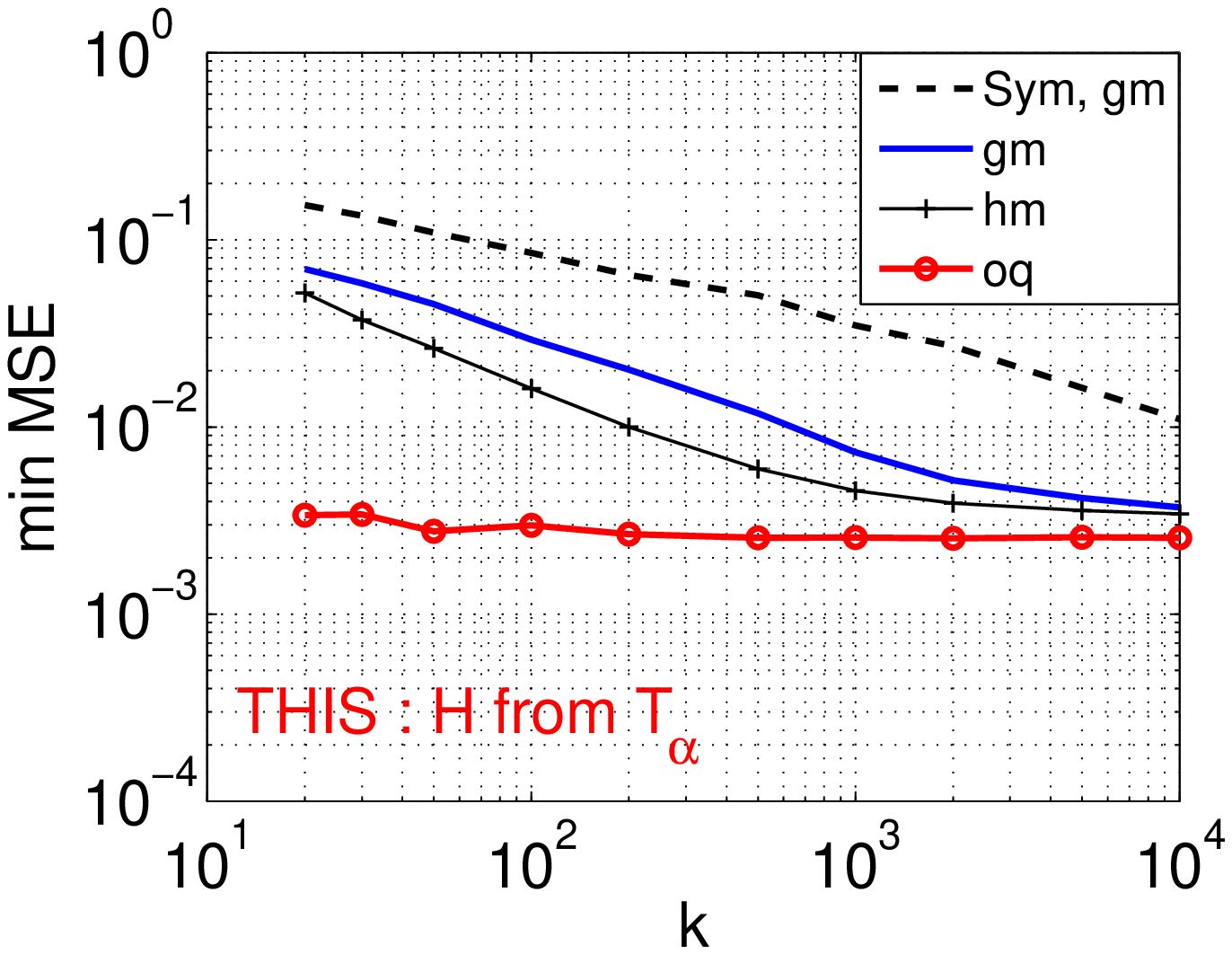}}
}
\mbox{
{\includegraphics[width=1.75in]{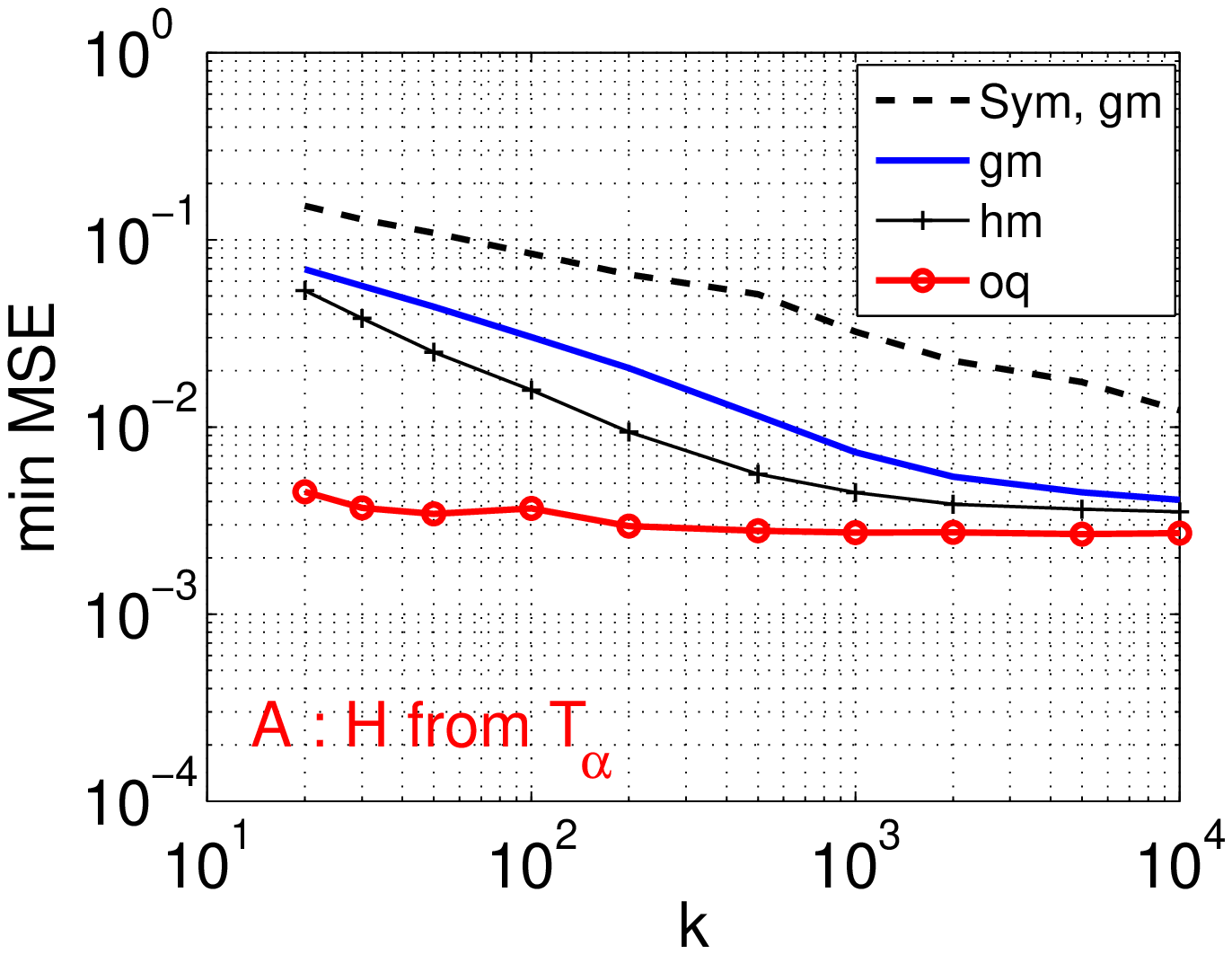}} \hspace{-0.1in}
{\includegraphics[width=1.75in]{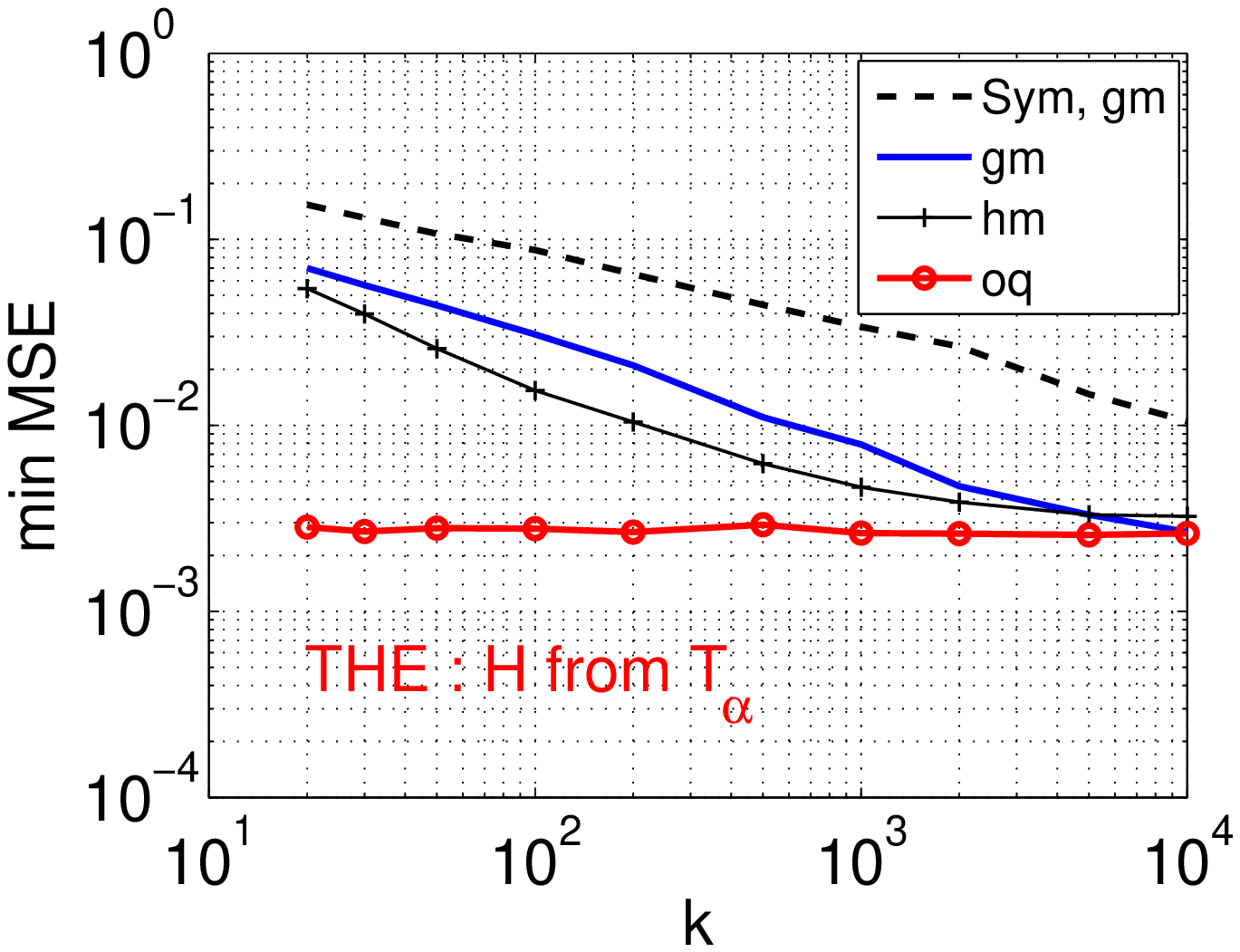}}
}
\end{center}
\vspace{-0.15in}
\caption{Shannon entropy, $H$, estimated from Tsallis entropy, $T_\alpha$, for 10 words in Table \ref{tab_data}. Curves are the minimum MSEs at each $k$. }\label{fig_HT_min}
\end{figure}

Figures \ref{fig_HT_min} presents the minimum MSEs for all 10 words:
\begin{itemize}
\item The optimal quantile estimator is the most accurate.  With $k = 20$, the relative MSE is only less than $1\%$ (or even $0.1\%$).
\item When $k\leq 10^3$, using the optimal quantile estimator, CC reduces  minimum MSEs by roughly 20-  to 50-fold, compared to {\em symmetric stable random projections}. When $k = 10^4$, the reduction is about 5- to 15-fold.
\item Even with $k=10^4$, {\em Symmetric table random projections} can not achieve the same accuracy as CC using the optimal quantile estimator with $k = 20$ only.
\end{itemize}

Again, using the optimal quantile estimator with $\alpha\approx 0.98 ~ 0.99$ would be our recommended procedure for estimating Shannon entropy from Tsallis entropy.

\clearpage
\section{Conclusion}\label{sec_conclusion}

Network data and Web search data are naturally dynamic and can be viewed as data streams. The entropy is an extremely useful summary statistic and has numerous applications, for example, anomaly detection in Web mining and network diagnosis.

Efficiently and accurately computing the entropy in ultra-large and frequently updating data streams, in one-pass, is an active topic of research. A recent trend is to use the $\alpha$th frequency moments with $\alpha\approx 1$ to approximate the entropy. For example, \cite{Article:Harvey_entropy_arXiv08,Proc:Harvey_FOCS08} proposed using the $\alpha = 1+\delta$ frequency moments with very small $|\delta|$ (e.g., $10^{-4}$ or smaller).

For estimating the $\alpha$th frequency moments, the recently proposed {\em Compressed Counting (CC)} dramatically improves the standard data stream algorithm based on {\em symmetric stable random projections}, especially when $\alpha\approx 1$. However, it had never been empirically evaluated before this work.

We experimented with CC to approximate the R\'enyi entropy, the Tsallis entropy, and the Shannon entropy. Some theoretical analysis on the biases and variances was provided. Extensive empirical studies based on some Web crawl data were conducted.

Based on the theoretical and empirical results, important conclusions can be drawn:
\begin{itemize}
\item Compressed Counting (CC) is numerically stable and is capable of providing highly accurate estimates of the $\alpha$th frequency moments. When $\alpha$ is close to 1, the improvements of CC over {\em symmetric stable random projections} in estimating frequency moments is enormous; in fact, the improvements tend to ``infinity'' when $\alpha\rightarrow 1$.
\item When $\alpha$ is close 1,  the optimal quantile estimator for CC is more accurate than the geometric mean and harmonic mean estimators, except when $\alpha>1$ and the sample size $k$ is very small (e.g., $k\leq 20$).
\item It appears not a practical algorithm to approximate the Shannon entropy using {\em symmetric stable random projections} with $\alpha =1+\delta$  and very small $|\delta|$. When we do need to use {\em symmetric stable random projections}, we should take advantage of the variance-bias trade-off by using $\alpha$ away from 1 for achieving smaller mean square errors (MSEs).

\item CC is able to provide highly accurate estimates of the Shannon entropy using either the R\'enyi entropy or the Tsallis entropy. In terms of the best achieable MSEs, the improvements over {\em symmetric stable random projections} can be about 20- to 50-fold.
\item When estimating Shannon entropy from R\'enyi entropy, in order to reach the same accuracy as CC, {\em symmetric stable random projections} would need about 50 times more samples than CC. When estimating Shannon entropy from Tsallis entropy, {\em symmetric stable random projections}  could not reach the same accuracy as CC even with 500 times more samples.
\item The R\'enyi entropy provides a better tool for estimating the Shannon entropy than the Tsallis entropy does.
\item Our recommended procedure for estimating the Shannon entropy is to use CC with the optimal quantile estimator and $\alpha<1$ close 1 (e.g., $0.98\sim 0.99$).
\item Since CC only needs a very small sample to achieve a good accuracy, the processing time of CC will be much reduced, compared to {\em symmetric stable random projections}, if the same level of accuracy is desired.
\end{itemize}

The technique of estimating Shannon entropy using {\em symmetric stable random projections} has been applied with some success in practical applications, such as  network anomaly detection and diagnosis\cite{Proc:Zhao_IMC07}. One major issue reported in \cite{Proc:Zhao_IMC07} (also \cite{Proc:Ganguly_RANDOM07}), is that the required sample size using {\em symmetric stable random projections} could be prohibitive for their real-time applications. Since CC can dramatically reduce the required sample size, we are passionate that  using Compressed Counting for estimating Shannon entropy will be highly practical and beneficial to real-world Web/network/data stream problems.

\section*{Acknowledgement}
This work is supported by Grant NSF DMS-0808864 and a gift from Google. The author would like to thank Jelani Nelson for  helpful communications. The author thanks Kenneth Church.


\end{document}